\documentclass[twocolumn,astrosymb,tighten,trackchanges]{aastex631}

\usepackage{color}
\usepackage{colortbl}
\definecolor{YG}{RGB}{115,80,185}

\graphicspath{{./}{figures2/}}

\begin{document}

\title{Late-time Hubble Space Telescope Observations of AT\,2018cow. II. Evolution of a UV bright Underlying Source 2--4 Yr Post-discovery}

\correspondingauthor{Yuyang Chen}
\email{yuyangf.chen@mail.utoronto.ca}

\author[0000-0002-8804-3501]{Yuyang Chen}
\affiliation{David A. Dunlap Department of Astronomy and Astrophysics, University of Toronto, 50 St. George Street, Toronto M5S 3H4, Canada}
\affiliation{Dunlap Institute for Astronomy and Astrophysics, University of Toronto, 50 St. George Street, Toronto M5S 3H4, Canada}

\author[0000-0001-7081-0082]{Maria R. Drout}
\affiliation{David A. Dunlap Department of Astronomy and Astrophysics, University of Toronto, 50 St. George Street, Toronto M5S 3H4, Canada}
\affiliation{The Observatories of the Carnegie Institution for Science, 813 Santa Barbara St., Pasadena, CA 91101, USA}

\author[0000-0001-6806-0673]{Anthony L. Piro}
\affiliation{The Observatories of the Carnegie Institution for Science, 813 Santa Barbara St., Pasadena, CA 91101, USA}

\author[0000-0002-5740-7747]{Charles~D.~Kilpatrick}
\affiliation{Department of Physics and Astronomy, Northwestern University, Evanston, IL 60208, USA}
\affiliation{Center for Interdisciplinary Exploration and Research in Astrophysics (CIERA), 1800 Sherman Ave, Evanston, IL 60201, USA}

\author[0000-0002-2445-5275]{Ryan~J.~Foley}
\affiliation{Department of Astronomy and Astrophysics, University of California, Santa Cruz, CA 95064, USA}

\author[0000-0002-7559-315X]{C\'{e}sar Rojas-Bravo}
\affiliation{Department of Astronomy and Astrophysics, University of California, Santa Cruz, CA 95064, USA}

\author[0000-0002-0629-8931]{M. R. Magee}
\affiliation{Department of Physics, University of Warwick, Gibbet Hill Road, Coventry CV4 7AL, UK}

\begin{abstract}

In this second of a two-paper series, we present a detailed analysis of three HST observations taken $\sim$2--4 years post-discovery, examining the evolution of a UV-bright underlying source at the precise position of AT\,2018cow. While observations at $\sim$2--3 years post-discovery revealed an exceptionally blue ($L_\nu\propto \nu^{1.99}$) underlying source with relatively stable optical brightness, fading in the NUV was observed at year 4, indicating flattening in the spectrum (to $L_\nu\propto \nu^{1.64}$). The resulting spectral energy distributions can be described by an extremely hot but small blackbody, and the fading may be intrinsic (cooling) or extrinsic (increased absorption). Considering possible scenarios and explanations, we disfavor significant contributions from stellar sources and dust formation based on the observed color and brightness. By comparing the expected power and the observed luminosity, we rule out interaction with the known radio-producing circumstellar material as well as magnetar spin down with $B\sim10^{15}\,\mathrm{G}$ as possible power sources, though we cannot rule out the possible existence of a denser CSM component (e.g., previously ejected hydrogen envelope) or a magnetar with $B\lesssim10^{14}\,\mathrm{G}$. Finally, we find that a highly-inclined precessing accretion disk can reasonably explain the color, brightness, and evolution of the underlying source. However, a major uncertainty in this scenario is the mass of the central black hole (BH), as both stellar-mass and intermediate-mass BHs face notable challenges that cannot be explained by our simple disk model, and further observations and theoretical works are needed to fully constrain the nature of this underlying source.

\end{abstract}

\section{Introduction} \label{sec:introduction}

Over the past decade, a class of peculiar transients has been discovered with brightness similar to supernovae (SNe) but a much shorter timescale, often accompanied by persistent blue color, commonly termed as Fast Blue Optical Transients \citep[FBOT; e.g.,][]{2014ApJ...794...23D,2016ApJ...819....5T,2016ApJ...819...35A,2018NatAs...2..307R,2018MNRAS.481..894P,2020ApJ...894...27T,2020MNRAS.498.2575W,Ho2021}. The much shorter timescale is incompatible with standard SN models powered by radioactive decay and hydrogen recombination (which serve to prolong light curve), and thus alternative models were proposed, most commonly involving power sources such as interaction between ejecta and circumstellar material \citep[CSM;][]{2010ApJ...724.1396O,2016MNRAS.461.3057S,2018NatAs...2..307R,McDowell2018,Kleiser2018CSM,Tolstov2019,Wang2019,Suzuki2020,Wang2020,Karamehmetoglu2021,Maeda2022,Margalit2022a,Margalit2022b,Mor2023,Liu2023,Khatami2023} and energy injection by central engines such as neutron stars \citep[NSs;][]{Yu2015,2017ApJ...850...18H,2017ApJ...851..107W,Wang2022,Liu2022} or black holes \citep[BHs;][]{2015MNRAS.451.2656K,2018NatAs...2..307R,2020ApJ...890L..26K,Tsuna2021,Kremer2021,Fujibayashi2022}. 

In 2018, a particularly intriguing FBOT -- AT\,2018cow -- was discovered by the Asteroid Terrestrial-impact Last Alert System \citep{2018ATel11727....1S,2018ApJ...865L...3P} in the star-forming dwarf galaxy CGCG 137-068 ($z=0.0141$) that rose to a peak brightness $M\sim-22$ similar to superluminous SNe in $\lesssim$\,2 days. Multi-wavelength follow-up observations \citep[e.g.,][]{RiveraSandoval2018,Kuin2019,Ho2019,Margutti2019,Perley2019,Huang2019,Bietenholz2020,Xiang2021,Pasham2021} revealed an array of peculiar properties, including (i) bright mm radio emission associated with mildly relativistic outflow ($v\sim0.1c$) in CSM \citep{Ho2019}, (ii) bright X-ray emission with signs of quasiperiodic oscillation (QPO) indicative of a central engine \citep{RiveraSandoval2018,Kuin2019,Margutti2019,Pasham2021,Zhang2022}, (iii) rapidly-fading optical emission with a persistent blue color and a receding photosphere \citep{Perley2019,Margutti2019,Xiang2021}, (iv) a separate IR component likely associated with dust \citep{Perley2019,Metzger2022dust}, (v) a sudden transition from broad to intermediate-width spectral features $\sim$15--20 days
after peak, suggesting multiple (possibly asymmetric) thermal emission components \citep{Perley2019,Margutti2019,Xiang2021}, and (vi) flashes of high optical polarization that further support an asymmetric configuration \citep[][]{Maund2023}{}{}. 

Following the discovery of AT\,2018cow, several analogs (sometimes referred to as ``Cow-like transients''), CSS161010 \citep[][]{Coppejans2020}, AT\,2018lug \citep[``The Koala'';][]{Ho2020}, AT\,2020xnd \citep[``The Camel'';][]{Perley2021Camel,Bright2022Camel,Ho2022Camel}, and AT\,2020mrf \citep[][]{Yao2022}, have also been discovered with defining characteristics being superluminous optical brightness, rapid timescale, and mildly relativistic outflow accompanied by bright multi-wavelength emissions\footnote{Bright radio emissions were observed for all analogs over the first few hundred days. Bright X-ray emissions were typically observed as well, with the exception of AT\,2018lug which did not have any follow-up X-ray observation (i.e., no confirmed X-ray emission). Also, the X-ray detection of CSS161010 was weaker and relatively less luminous.}. It has been suggested that these luminous FBOTs likely make up an entirely new population of transients distinct from established SNe and typical FBOTs \citep[][]{Ho2021}. Unfortunately, all analogs were discovered either at much larger distances ($z\sim0.1-0.3$) or after the fact, and thus, the most well-studied case at the moment is still AT\,2018cow.

A large variety of models have been proposed to explain AT\,2018cow, typically invoking extreme configurations of CSM \citep[][]{RiveraSandoval2018,Xiang2021,Pellegrino2022} or central engines \citep[][]{Quataert2019,Margutti2019,Fang2019,Mohan2020,Piro2020,Uno2020,Gottlieb2022,Metzger2022} with progenitors ranging from low-mass systems \citep[including white dwarfs;][]{Lyutikov2019,Yu2019,Lyutikov2022} to massive stars \citep[][]{Leung2020,Soker2019,Soker2022,Cohen2023}.

For AT\,2018cow, a Tidal Disruption Event (TDE) by an intermediate-mass BH \citep[IMBH, with $M_{\mathrm{BH}} \lesssim 10^5\,M_\odot$; see review by ][]{Greene2019IMBHReview} or a supermassive BH (SMBH, with $M_{\mathrm{BH}} \gtrsim 10^5-10^6\,M_\odot$) was also proposed initially as a possible explanation \citep[][]{Perley2019,Kuin2019}. However, the TDE hypothesis gradually became less favored given that it is difficult to prove the existence of such a BH at the outskirt of the galaxy where the gas velocity is smoothly varying without any signs of a coincident massive host system \citep[][]{Lyman2020}{}{}, and that a mass limit of $M_{\mathrm{BH}}<850\,M_\odot$ was derived from NICER X-ray quasiperiodic oscillations \citep[QPO;][]{Pasham2021}. The dense CSM around AT\,2018cow needed to explain the bright radio emission and non-detection of radio linear polarization would also be difficult to explain unless the BH was already embedded in a gas-rich environment \citep[][]{Margutti2019,Huang2019}{}{}. The exact nature of AT\,2018cow is still open to debate
and additional constraints are needed to distinguish the viable models.

One way of obtaining additional constraints is through late-time observations, useful for probing the immediate surrounding as well as any fading transient emission. For AT\,2018cow, the Hubble Space Telescsope (HST) was used to acquire six late-time observations, with the first three tracking the fading prompt emission over $\sim$50--60 days post-discovery, and the latest three monitoring the field over $\sim$2, 3, and 4 years post-discovery (in 2020, 2021, and 2022). The HST images taken in 2020 and 2021 were initially examined by \citet{Sun2022}, which led to the discovery of an UV-bright underlying source at the precise location of AT\,2018cow years post-discovery. \citet{Sun2022} found this source to be bluer and brighter than any known stars with a stable optical brightness for over a year, which initially led to the suggestion that it may be a young stellar cluster. We independently discovered this underlying source and requested an additional HST observation to monitor any evolution over time. This resulted in the most recent HST observation taken in 2022. The photometry from this 2022 HST observation was briefly reported (but not used) in an environmental study of AT\,2018cow by \citet{Sun2022new}, which showed fading in the NUV, suggesting the existence of a transient component undergoing spectral evolution.

Such transient component that persists over multiple years post-discovery is an interesting discovery that had never previously been observed for an FBOT. In the case of AT\,2018cow, compared to the rapid fading of $\gtrsim8\,\mathrm{mag}$ in 
the NUV over the first $\sim$60\,days, the underlying source faded extremely slowly in the NUV by only $\sim$0.4--0.5\,mag between 703--1453\,days post-discovery. This (apparent) drastic difference in timescale may imply a transition to a new evolutionary stage, possibly connected to the commonly-proposed power sources, i.e., ejecta-CSM interaction or central engines. For example, ejecta-CSM interaction is well-known to be capable of producing long-lasting emission over multiple years given a sufficient amount of CSM \citep[e.g.,][]{Smith2009}. A remnant central engine can also evolve over long timescales, manifesting at late times as pulsar wind nebulae \citep[e.g.,][]{Metzger2014PWN} or transient accretion disks \citep[e.g.,][]{Strubbe2009TDE}. Therefore, by connecting the underlying source to a specific remnant power source, significant constraints could be placed on the true nature of AT\,2018cow.

In this study, we present a detailed analysis of the UV-bright underlying source at the precise position of AT\,2018cow in the three latest HST observations taken $\sim$2--4 years post-discovery. We examine the highly unusual properties and evolution of this source and use them to place constraints on possible origins and power sources. This is paper II of a two-paper series, while paper I \citep{Chen2023I} focused on the first three HST observations of AT\,2018cow that tracked the fading prompt emission (50--60 days post-discovery).

Note that throughout this paper, we adopt the term \emph{underlying source} to refer to the newly-discovered point source spatially coincident with AT\,2018cow, and as with paper I, we use the term \emph{prompt emission} to refer to the initial evolution of AT\,2018cow over the first two months. 
We opted to use a neutral term for the spatially-coincident object, rather than immediately calling it AT\,2018cow, because of the current ambiguity in (i) the exact classification of this object (e.g., partially or entirely transient) and (ii) the exact physical processes producing the emission and if they are related to the initial evolution of AT\,2018cow (e.g., entirely new processes or the same processes in a new environment). Therefore, we do not use a term that would suggest a particular classification, physical process, or hypothetical phase of AT\,2018cow. This choice of terminology also aligns with our strategy of examining the underlying source as an individual object and associating the hypothetical scenarios with the prompt emission to establish possible links with AT\,2018cow.

The HST observations as well as a new late-time X-ray constraint from \emph{Swift} are described in \ref{sec:observations}. In Section \ref{sec:underlying}, we examine the properties of the underlying source and model the SEDs using simple models. In Section \ref{sec:underinterpret}, we analyze these properties in the context of different physical scenarios and place constraints on the origin of the underlying source. Finally, we summarize the results and overall implications on the nature of AT\,2018cow and the new class of luminous FBOTs in Section \ref{sec:conclusion}.

For our analyses, we adopt a luminosity distance of $d_L = 60\,\mathrm{Mpc}$ for AT\,2018cow \citep[][]{Perley2019,Margutti2019}. We assume an $R_{\mathrm{V}} = 3.1$ Milky Way extinction curve with $E_{\mathrm{B}-\mathrm{V}} = 0.07\,\mathrm{mag}$ \citep[][]{2011ApJ...737..103S} and no internal extinction in the host galaxy of AT\,2018cow. Throughout this study, we refer to $t$ as the rest-frame time after the first discovery date of AT\,2018cow, MJD 58285.441 \citep[][]{2018ATel11727....1S,2018ApJ...865L...3P}{}{}.

\section{Observations and Data Reduction} \label{sec:observations}

\subsection{HST Observations ($t\sim703-1453\,\mathrm{days}$)}

\begin{figure*}
\epsscale{0.825}
\plotone{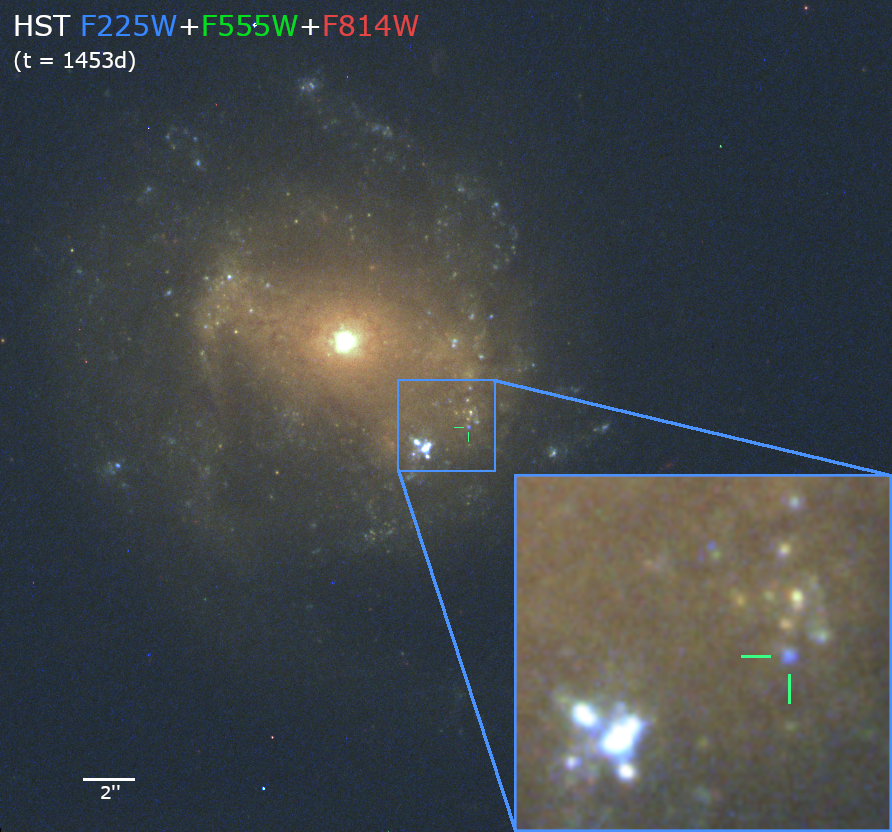}
\caption{Composite HST image of the host galaxy of AT\,2018cow from the F225W (blue), F555W (green), and F814W (red) images taken at $t\simeq1453\,\mathrm{days}$ (around 4 years post-discovery). A crosshair is shown at the position of AT\,2018cow, marking the existence of a spatially-coincident underlying source. The underlying source is much bluer even visually than most other sources in the galaxy. Note that a faint tidal tail can be seen south of the galaxy.
\label{fig:CompImg}}
\end{figure*}

The latest three HST observations of AT\,2018cow monitored the field using WFC3 between $\sim$2--4 years post-discovery. The first observation (PI Levan) was made during 2020-05-29 (MJD 58998 or $t \simeq 703\,\mathrm{days}$), designed to study the local host environment, with images were taken in the F225W, F336W, F555W, F657N, F665N, and F814W bands spanning $\lambda \simeq 2358-8029\,\mathrm{\AA}$. The second observation (PI Filippenko) was made as part of a Snaphot program during 2021-07-25 (MJD 59420 or $t \simeq 1119\,\mathrm{days}$) in the F555W ($\lambda = 5308\,\mathrm{\AA}$) and F814W ($\lambda = 8029\,\mathrm{\AA}$) bands.

We requested the most recent third observation (PI Chen) after independently discovering an underlying source spatially coincident with AT\,2018cow in the first and second observations \citep[subsequently reported in][]{Sun2022}. This observation was made during 2022-06-29 (MJD 59759 or $t \simeq 1453$) in the F225W, F336W, F555W, and F814W bands. A key goal of this observation was to determine whether the underlying source is transient or stable, which was difficult to discern due to the lack of UV spectral coverage in the 2021 Snapshot program and the higher background level present in the optical bands. A composite image from this observation is shown in Figure~\ref{fig:CompImg}. All HST data can be found on the Mikulski Archive for Space Telescopes (MAST): \dataset[10.17909/fmz6-9b21]{http://dx.doi.org/10.17909/fmz6-9b21}.

Data reduction and PSF photometry was performed on all three late-time epochs via the same procedure described in paper I \citep{Chen2023I}. We identified sources using the drizzled WFC3/UVIS F336W frame from $t\simeq50.3\,\mathrm{days}$ as a reference and performed forced photometry at the location of AT\,2018cow, allowing slight recentering in each image. We note that there is some astrometric uncertainty resulting from both our centroid on AT\,2018cow at $t\simeq50.3\,\mathrm{days}$ ($\approx0.002\arcsec$ based on the signal-to-noise of this detection and the PSF size in F336W) and frame-to-frame alignment (0.005--0.049\arcsec). Regarding the spatial coincidence between AT\,2018cow and the underlying source, we also verified by comparing their relative positions from the nucleus of the host galaxy and found an average offset of $\lesssim0.02\arcsec$, consistent with the alignment uncertainty.

\begin{figure*}
\epsscale{1.0}
\plotone{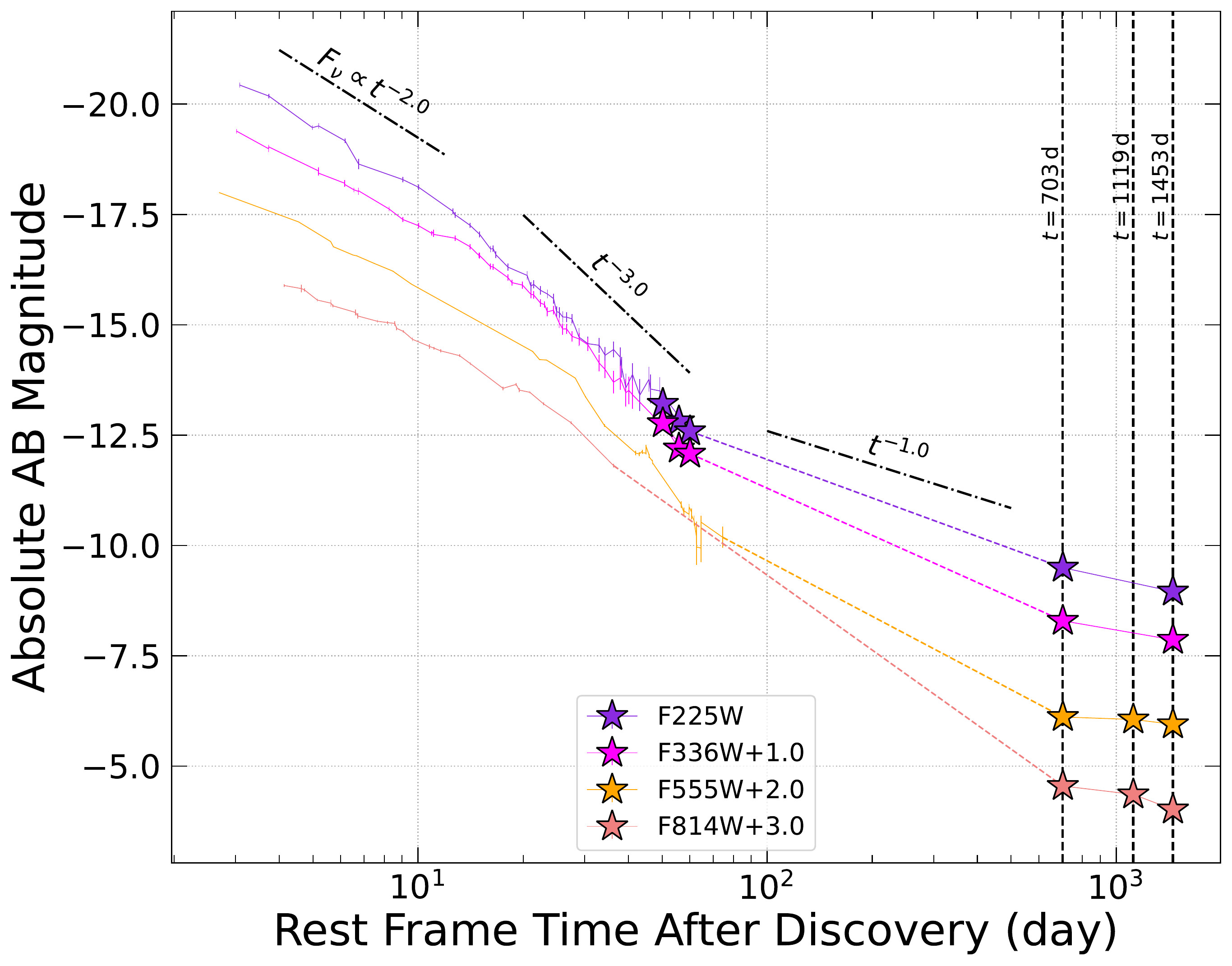}
\caption{Light curves illustrating the evolution of the underlying source from AT\,2018cow. HST photometry are shown as stars, and earlier measurements at similar wavelengths taken from paper I \citep{Chen2023I} are also shown. Offsets are added for visual clarity. Dash-dotted lines corresponding to power law declines of flux density are also added for reference (note these are not best-fit lines). Fading is observed in the F225W and F336W bands for the underlying source, but at a much slower rate than the initial decline of AT\,2018cow.
\label{fig:LC}}
\end{figure*}

Final HST photometry of the underlying source are listed in Table~\ref{tab:HSTmag} and shown in Figure~\ref{fig:LC}. For epochs in which we did not detect any significant ($\geq3\sigma$) emission at the site of AT\,2018cow, we injected artificial stars using built-in methods in {\tt dolphot}. We injected 50,000 stars at magnitudes varying from 20--28\,AB\,mag and estimated the threshold where {\tt dolphot} could detect 99.7\% of all stars at $\geq3\sigma$ significance, which we report as the $3\sigma$ limiting magnitude in \autoref{tab:HSTmag}.

We found that the underlying source was still present in all bands in our 2022 observations. However, we observed fading of $\sim$0.4--0.5\,mag in the UV bands (F225W and F336W), thus confirming the transient nature of the underlying source (see Section \ref{sec:underlying}).

\begin{deluxetable}{cccc}[ht]
\tablecaption{HST photometry of Underlying Source\label{tab:HSTmag}}
\tablewidth{0pt}
\tablehead{
\colhead{Filter} & \multicolumn{3}{c}{Magnitude (AB)}
\\
\colhead{} & \colhead{$703\,\mathrm{d}$} & \colhead{$1119\,\mathrm{d}$} & \colhead{$1453\,\mathrm{d}$} 
}
\startdata
F225W & $24.39(0.11)$ &         &   $24.93(0.11)$   \\
F336W & $24.60(0.08)$ &         &   $25.03(0.07)$   \\
F555W & $25.78(0.10)$ & $25.83(0.17)$ &  $25.94(0.12)$\\
F814W & $26.34(0.20)$ & $26.53(0.24)$ & $26.87(0.33)$\\
F657N & $<23.54$ &        &       \\
F665N & $<23.44$ &        &          
\enddata
\tablecomments{1$\sigma$ errors are given inside the brackets. The $<$ symbol indicates a $3\sigma$ upper limit.}
\end{deluxetable}

\subsubsection{Comparison to \citet{Sun2022,Sun2022new}} \label{subsubsec:compareSun2022}

\begin{figure*}[ht]
\epsscale{1.0}
\plotone{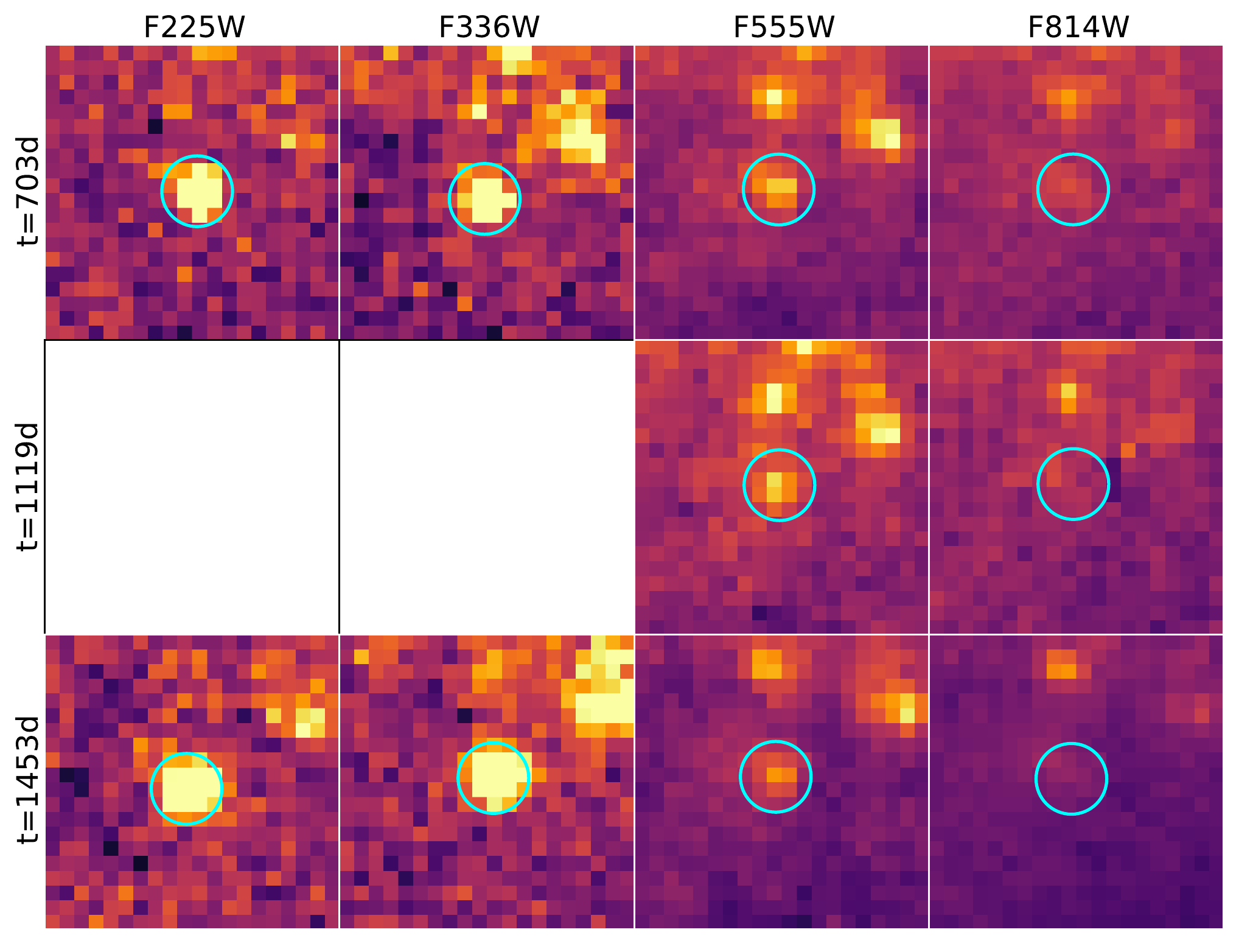}
\caption{Cutout HST images taken at $t\simeq703,1119,1453\,\mathrm{days}$ in the wide bands. Epochs are shown on the left and filter names are shown at the top. Each panel shows a $1\arcsec\times1\arcsec$ box, with a circular aperture ($r=0.12\arcsec$) centered at the position of AT\,2018cow. A UV-bright point source is present at the position of AT\,2018cow in the F225W, F336W, and F555W filters.
\label{fig:UnderImg}}
\end{figure*}

\begin{figure}
\epsscale{1.0}
\plotone{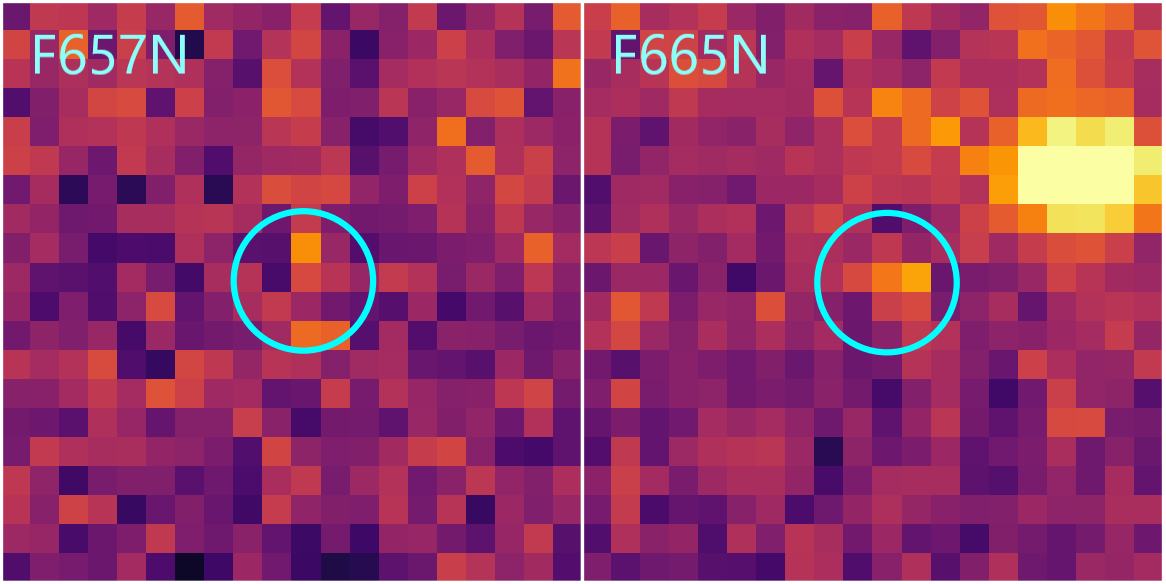}
\caption{Cutout HST images taken at $t\simeq703\,\mathrm{days}$ in the narrow bands (F657N and F665N). A circular aperture ($r=0.12\arcsec$) is centered at the position of AT\,2018cow. \textit{Some} excess emission can be seen in the F665N image.
\label{fig:UnderNarrowImg}}
\end{figure}

Although our HST photometry generally agree with values presented in \citet{Sun2022,Sun2022new} to within about $1\sigma$, there are three discrepancies that we outline in this Section. First, our photometry are almost consistently fainter with larger uncertainties, e.g., our magnitudes in the 2021 Snapshot epoch are $m_{\rm F555W}=25.84\pm0.17$\,AB\,mag and $m_{\rm F814W}=26.53\pm0.24$\,AB\,mag compared with the previously reported $m_{\rm F555W}=25.60\pm0.08$\,AB\,mag and $m_{\rm F814W}=26.39\pm0.24$\,AB\,mag \citep[note that the latter measurements were originally given in Vega mag in][]{Sun2022}. This discrepancy is likely due to the fact that there is extended background emission close to the site of AT\,2018cow. If we instead perform aperture photometry with a radius of 0.12\arcsec, we find that the F555W and F814W measurements are 1 and 0.6\,mag brighter than our PSF photometry, respectively. Moreover, while {\tt dolphot} classifies the F814W source as a point-like object, its crowding is 0.31\,mag, which can be understood as the difference between a PSF and aperture magnitude. We conclude that there is some other source of background emission that is comparable in brightness to the underlying source that {\tt dolphot} deblends from the transient. Differences in {\tt dolphot} parameters between our work and \cite{Sun2022} may therefore contribute to our fainter measurement.

The second discrepancy from \citet{Sun2022} is the F665N observation from 2020, which probes H$\alpha$ emission at the redshift of AT\,2018cow. We report a $3\sigma$ upper limit of 23.44\,mag, while \cite{Sun2022} reported a $4.6\sigma$ detection of 24.47$\pm$0.24\,mag. While there does appear to be some flux inside a 0.12\arcsec\ aperture centered at the site of AT\,2018cow (Figure \ref{fig:UnderNarrowImg}), it is nominally offset from the transient position, leading to a non-detection in our analysis (which utilized forced photometry based on the transient position as described above). This offset is similar to the frame-to-frame alignment uncertainty, which for F665N is on average 0.028\arcsec\ or 8\,pc at the assumed distance to AT\,2018cow. In addition, if we perform aperture photometry with a 0.12\arcsec\ radius, we only find excess emission at a $\approx2.8\sigma$ level in the F665N band. Therefore, it is not clear if the excess emission in the F665N band is (i) associated with the underlying source and 
(ii) significantly above the diffuse background. We discuss the potential excess narrow band emission further in our interpretations in Section \ref{sec:underinterpret}.

The third discrepancy is the F225W measurement at the latest HST epoch ($t\simeq1453\,\mathrm{days}$), for which we report $24.93\pm0.11\,\mathrm{AB}\,\mathrm{mag}$ while \citet{Sun2022new} reported a fainter $25.21\pm0.07\,\mathrm{AB}\,\mathrm{mag}$ (again from Vega mag). This difference is unexpected, especially since the background level is the lowest in this band, and our other measurements at this epoch are generally consistent. We also found that aperture photometry leads to an F225W measurement 0.37\,mag brighter than the PSF photometry, meaning that the issue is unlikely related to excess background. The exact cause of this discrepancy is unclear, and we chose to use our PSF photometry for our analyses. Note that this discrepancy does not significantly impact the results presented in this study because the fading would be even more prominent according to the measurement from \citet{Sun2022new}.

\subsection{Late-time \emph{Swift} XRT Observation ($t\simeq1360\,\mathrm{days}$)}

Finally, to constrain any high energy emission associated with the underlying source, we analyzed two X-ray observations (totaling 8.5\,ks) obtained by the \emph{Swift} X-Ray Telescope \citep[XRT;][]{Burrows2005XRT} on 2022-03-25 and 03-27 (or $t\simeq 1360\,\mathrm{days}$). This corresponds to approximately 93 days prior to the latest HST observations taken on 2022-06-29.
We downloaded the cleaned event files and the exposure images from the Swift archive and extracted the X-ray images using \textsc{xselect}. We then followed the mosaic routine in \textsc{ximage} to combine the images from the two observations.

\begin{figure*}
\epsscale{1.0}
\plotone{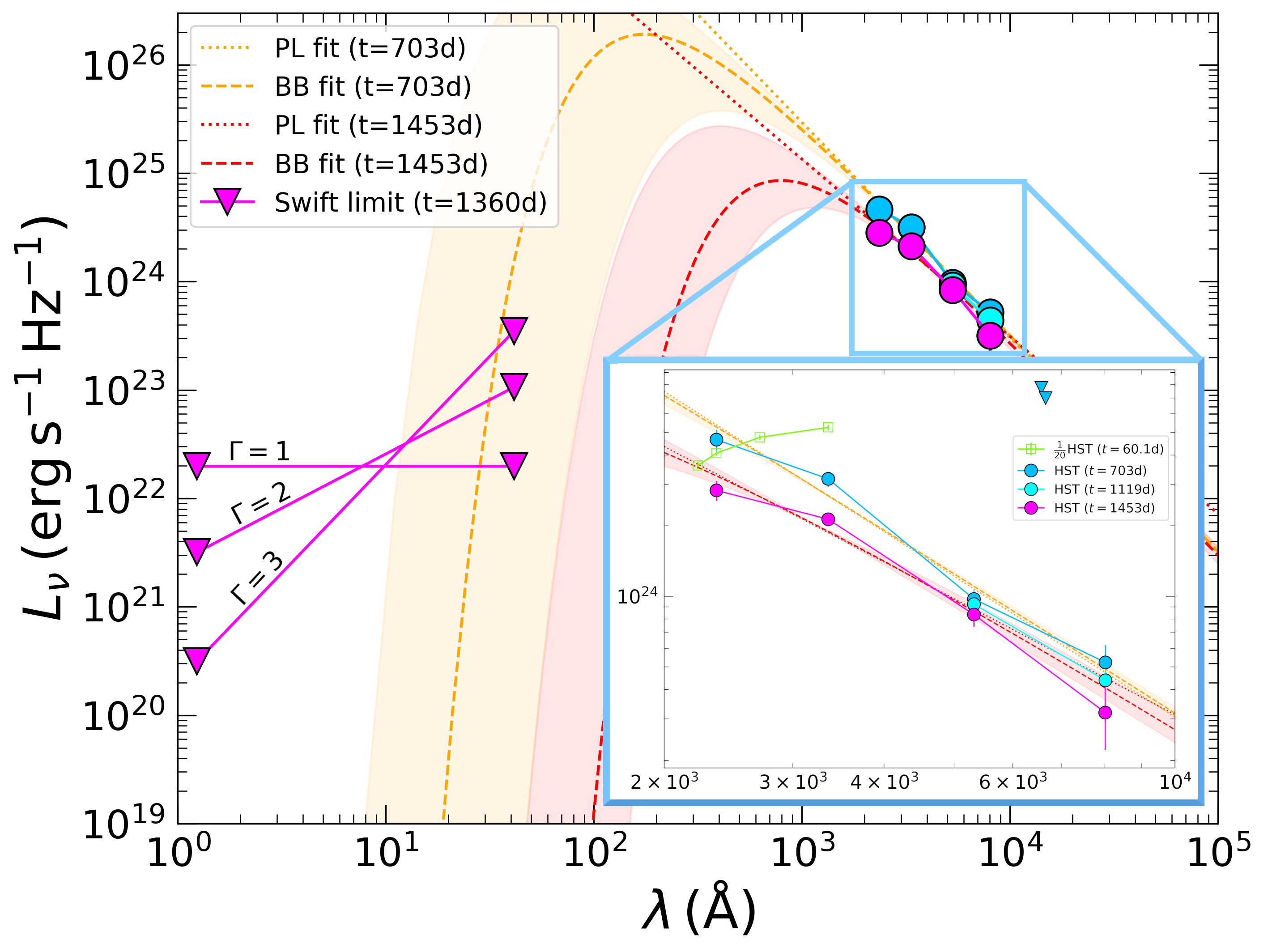}
\caption{The HST SEDs of the underlying source (dereddened from Galactic extinction) and best-fit power laws (dotted lines) and blackbodies (dashed lines), as well as the $3\sigma$ \emph{Swift}-XRT upper limits (downward triangles) assuming a power law and fiducial photon indices $\Gamma$. Shaded regions represent $1\sigma$ uncertainties for the best-fit blackbodies. In the zoomed-in plot, the HST SED of AT\,2018cow at $t\simeq60.1\,\mathrm{days}$ with brightness lowered by a factor of 20 is also shown for comparison. Although the optical emission seems relatively stable, the fading UV emission confirms the transient nature of the underlying source.
\label{fig:underlyingSED}}
\end{figure*}

In the combined image, no X-ray emission was detected at the site of AT\,2018cow. Using the \texttt{sosta} function of \textsc{ximage}, we inferred a $3\sigma$ upper limit count rate of 0.00157\,cts/s at the location of AT\,2018cow. Assuming a power law with a photon index of $\Gamma = 2$ and a galactic neutral hydrogen column density of $N_{\mathrm{H}} = 5\times10^{20}\,\mathrm{cm}^{-2}$ \citep[][]{Margutti2019}, the count rate corresponds to an unabsorbed flux limit of $F_{0.3-10\mathrm{keV}} < 6.27\times10^{-14}\,\mathrm{erg}\,\mathrm{s}^{-1}\,\mathrm{cm}^{-2}$ (or $L_{0.3-10\mathrm{keV}} < 2.70\times10^{40}\,\mathrm{erg}\,\mathrm{s}^{-1}$ at the distance of AT\,2018cow). However, we emphasize that since the origin of the underlying source is unknown, there is significant uncertainty in the X-ray spectral shape. Varying the spectral shape can change this flux limit by more than an order of magnitude. This X-ray non-detection is further assessed in the context of various transient models in Section \ref{sec:underinterpret}.

\section{Evolution of the Underlying Source} \label{sec:underlying}

Cutout HST images of the underlying source are shown in Figure \ref{fig:UnderImg} (wide bands) and Figure \ref{fig:UnderNarrowImg} (narrow bands). The underlying source is detected as a point source in all three epochs, most clearly in the UV bands (F225W and F336W). Compared to its immediate surrounding, the underlying source is the only bright point source in the UV (also see Figure \ref{fig:CompImg}) but is hardly discernable visually from the local diffuse emission in the red optical bands. The light curves of the underlying source are shown in Figure \ref{fig:LC}, revealing fading in the NUV. In this section, we describe the basic properties and evolution of the underlying source. 

\subsection{Basic Properties of the UV-Optical SEDs} \label{subsec:undertransient}

Figure \ref{fig:underlyingSED} shows the SEDs of the underlying source derived from the three HST observations as well as the \emph{Swift}-XRT upper limits at $t\simeq1360\,\mathrm{days}$ (plotted assuming a power-law spectrum with various photon indices~$\Gamma$ for illustrative purposes). The underlying source is detected well above the background in the F225W, F336W, and F555W bands, with signal-to-noise ratios of $\mathrm{SNR}\simeq9-14$. The detection is weaker in the F814W band, with $\mathrm{SNR} = 5.4$ and $3.3$ at $t\simeq703\,\mathrm{days}$ and $1453\,\mathrm{days}$, respectively. In the F657N and F665N narrow bands ($t\simeq703\,\mathrm{days}$), we did not detect emissions above $3\sigma$ (see discussion about F665N band in Section \ref{subsubsec:compareSun2022}). Here we highlight several important properties of the underlying emission.

\emph{SED Shape:} The exceptionally blue color was the most remarkable property of the underlying source, quite distinct from most other sources in the galaxy (visually recognizable in Figure \ref{fig:CompImg}). The SED peak was not directly constrained in any of the HST epochs, suggesting $\lambda_{\mathrm{SED,peak}}\lesssim 2358\,\mathrm{\AA}$, while the lack of an ($0.3-10\,\mathrm{keV}$) X-ray detection shortly before the latest HST epoch implies that the peak must be further in the UV at $t\gtrsim1360\,\mathrm{days}$. At $t\simeq703\,\mathrm{days}$ and $1453\,\mathrm{days}$, we found a dereddened color of $\mathrm{F336W}-\mathrm{F555W}=-1.3\,\mathrm{mag}$ and $-1.0\,\mathrm{mag}$, respectively, which are much bluer than even AT\,2018cow at peak. Finally, there is tentative evidence for a change in the SED slope between the UV and optical bands, with a shallower slope between F225W and F336W and a steeper slope between F336W and F555W relative to a smooth power law (see Figure \ref{fig:underlyingSED}).

\emph{Brightness:} The NUV brightness of the underlying source is also surprisingly bright, with an absolute (dereddened) magnitude of -10.1\,AB\,mag in the F225W filter at $t\simeq703\,\mathrm{days}$. This is only about 20 times fainter than AT\,2018cow at $t\simeq60\,\mathrm{days}$, despite being observed almost two years later (Figure \ref{fig:underlyingSED}). In contrast, we found that the NUV brightness of AT\,2018cow decreased by more than 6000 times over the initial 60 days. In terms of a power law decline, we found $F_{\mathrm{NUV}}\propto t^{k}$ with $k\sim0.7-1.2$ over $t\simeq60-1453\,\mathrm{days}$, much slower compared to the rates of $t^{-2}-t^{-3}$ over the first 60 days (see Figure \ref{fig:LC}). We note that the power law decline of flux density in the NUV does not directly correspond to the decline of bolometric luminosity due to the significant change in color/spectrum. Without assuming the spectral shape, we can calculate the minimum UV-optical luminosity $L_{\mathrm{UVO,min}}$ of the underlying source by directly integrating the SEDs. The values of $L_{\mathrm{UVO,min}}$ at $t\simeq703\,\mathrm{days}$ and $1453\,\mathrm{days}$ are given in Table \ref{tab:UnderProperties}, which are on the order of $10^{39}\,\mathrm{erg}\,\mathrm{s}^{-1}$.

\subsection{Transient Nature of the Underlying Source}\label{subsec:undertransient}

Previous studies \citep[e.g.,][]{Sun2022,Metzger2022} suggested that the relative stability of the optical emission of the underlying source between $t\simeq703\,\mathrm{days}$ and $1119\,\mathrm{days}$ could be an indication of a stable stellar source. However, from the most recent HST observation, we found significant fading in the UV bands between $t\simeq703\,\mathrm{days}$ and $1453\,\mathrm{days}$. Specifically, we found the fading to be $0.54\pm0.15\,\mathrm{mag}$ ($3.5\sigma$) and $0.43\pm0.11\,\mathrm{mag}$ ($3.9\sigma$) in the F225W band and F336W band, respectively. In contrast, we found no significant fading (only $\sim$1$\sigma$) in the optical bands, indicating ongoing spectral evolution of the underlying source.

Note that we have performed additional checks to verify the robustness of the observed fading in the UV. First, we checked the photometry of stable sources in the HST images and found consistent brightnesses between different epochs, meaning that the measured fading of the underlying source was not due to any systematic calibration issue. Second, we performed aperture photometry on the underlying source (using apertures show in Figure \ref{fig:UnderImg}) and recovered similar fading in the UV bands, confirming that the fading was not caused by, e.g., {\tt dolphot} deblending parameters. Together, these confirm the transient nature of the underlying source, which could be intrinsic (i.e., remnant transient of AT\,2018cow) and/or extrinsic (i.e., increased absorption along the line of sight).

\subsection{Modeling of the UV-Optical SEDs} \label{subsec:undermodel}

We performed a set of simple fits to the HST SEDs to further constrain the properties of the underlying source, which we use as a basis for our discussion in Section \ref{sec:underinterpret} when considering specific physical scenarios. Specifically, we characterized (i) the SEDs through power law and blackbody models and (ii) the observed fading through an extinction law.

\subsubsection{Power Law and Blackbody Models}

We fit two models to the HST SEDs: a power law in the form $L_\nu \propto \nu^{\alpha}$ and a blackbody. We performed forward modeling using the Markov Chain Monte Carlo (MCMC) sampler in the Python package \texttt{emcee} \citep[][]{2013PASP..125..306F}. The best-fit parameters and uncertainties were derived from the 50th, 15.9th, and 84.1th percentile of the resulting samples. Note that we did not fit the SED at $t\simeq1119\,\mathrm{days}$ as photometry was only available in two bands. 

The best-fit spectral index $\alpha$, blackbody temperature $T_{\mathrm{BB}}$, blackbody radius $R_{\mathrm{BB}}$, as well as the blackbody luminosity $L_{\mathrm{BB}}$ are given in Table \ref{tab:UnderProperties}. The resulting fits are plotted in Figure \ref{fig:underlyingSED}. It is worth noting that the SED at $t\simeq703\,\mathrm{days}$ was so blue that the spectral index was $\alpha\simeq 2$, the expected value for the Rayleigh-Jeans tail (i.e., $L_\nu\propto\nu^2$). At this epoch, if the emission was blackbody, the blue color would imply a very high temperature ($T_{\mathrm{BB}}>10^5\,\mathrm{K}$), but the brightness would suggest an incredibly small size ($R_\mathrm{BB}\lesssim 15\,R_\odot$). The flattening in the SED at $t\simeq1453\,\mathrm{days}$ can be seen in the decreasing $\alpha$, which could suggest cooling (lower $T_{\mathrm{BB}}$) and expansion (larger $R_{\mathrm{BB}}$) of a blackbody.

These properties of the underlying source are all quite extreme in the context of late-time observations of transients. In particular, if the blackbody characterization is accurate, the derived temperatures are even higher than those of AT\,2018cow, while the radii are no different from the sizes of stars. The derived blackbody luminosities are on the order of $L_{\mathrm{BB}}\sim 10^{40}-10^{43}\,\mathrm{erg}\,\mathrm{s}^{-1}$, similar to those of AT\,2018cow at the second month post-discovery. Note that because the temperatures are so high and the peak wavelengths are further in the UV, the observed luminosities $L_{\mathrm{UVO,min}}$ are only $\sim$0.03--2\% of $L_{\mathrm{BB}}$, meaning that only a tiny fraction of the radiation was actually observed in the NUV-optical bands. Overall, the high temperature and luminosity could be an indication of additional energy injection from a remnant power source of AT\,2018cow. In this case, constraining the nature of the underlying source and the associated power source could be crucial in revealing the exact identity of AT\,2018cow. We further examine possible power sources through additional modeling and discuss the implications in Section \ref{sec:underinterpret}.

\begin{deluxetable}{lcc}
\tablecaption{Basic Properties of the Underlying Source\label{tab:UnderProperties}}
\tablewidth{0pt}
\tablehead{
\colhead{Property} & \colhead{$t\simeq703\,\mathrm{days}$} & \colhead{$t\simeq1453\,\mathrm{days}$}}
\startdata
$L_{\mathrm{UVO,min}}$ $(10^{39}\,\mathrm{erg}\,\mathrm{s}^{-1})$ &  $1.71\pm0.08$  &  $1.18\pm0.06$      \\
$\alpha$ & $1.985^{+0.138}_{-0.137}$ & $1.635^{+0.164}_{-0.167}$   \\
$R_{\mathrm{BB}}$ ($R_\odot$)  & $14.532^{+8.836}_{-5.004}$ &   $29.894^{+9.846}_{-10.388}$    \\   
$T_{\mathrm{BB}}$ ($\mathrm{K}$) & $2.92^{+3.59}_{-1.68}\times10^{5}$  &  $6.42^{+6.12}_{-2.04}\times10^{4}$ \\
$L_{\mathrm{BB}}$ $(\mathrm{erg}\,\mathrm{s}^{-1})$ & $5.27^{+50.68}_{-4.83}\times10^{42}$ &  $5.24^{+27.55}_{-3.24}\times10^{40}$
\enddata
\tablecomments{$L_{\mathrm{UVO,min}}$: integrated luminosity over HST SED; $\alpha$: spectral index; $R_{\mathrm{BB}}$: blackbody radius; $T_{\mathrm{BB}}$: blackbody temperature; $L_{\mathrm{BB}}$: blackbody luminosity. The $1\sigma$ errors of $L_{\mathrm{UVO,min}}$ are from Monte Carlo propagation.
Lower and upper errors of best-fit parameters are from the 15.9th and 84.1th percentile, respectively.}
\end{deluxetable}

\subsubsection{Extinction Law}

\begin{figure}
\epsscale{1.1}
\plotone{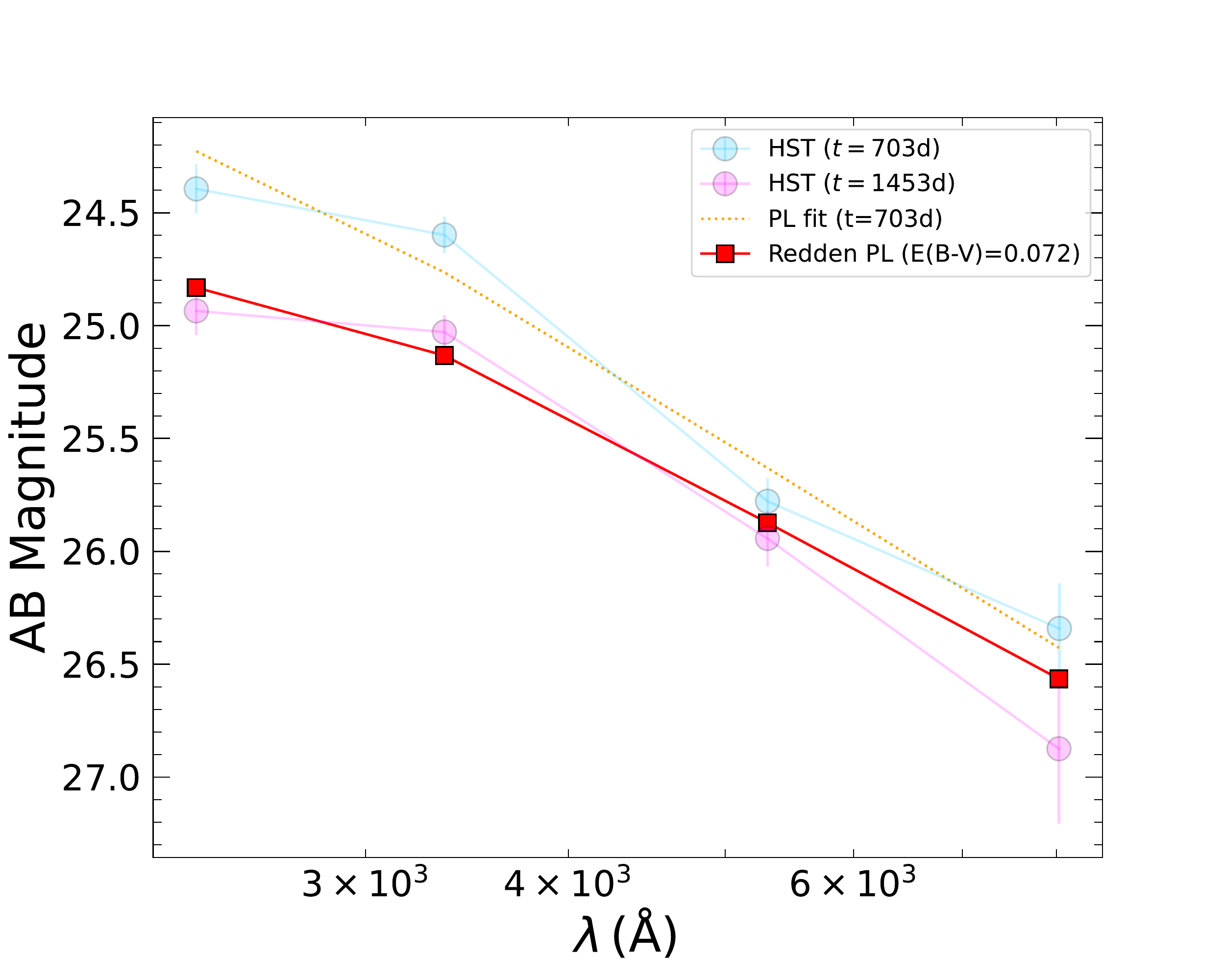}
\caption{The HST SEDs of the underlying source at $t\simeq703\,\mathrm{days}$ and $1453\,\mathrm{days}$, the best-fit power law (PL) at $t\simeq703\,\mathrm{days}$, and the reddened power law with $E_{\mathrm{B}-\mathrm{V}}\simeq0.072$ assuming the Cardelli extinction law. Extinction can account for the observed fading in the UV.
\label{fig:Underdust}}
\end{figure}

We also considered a different scenario where the transient phenomenon was not in the radiation but rather in the extinction that led to preferential dimming at shorter wavelengths. In this case, newly-formed dust grains over $t\simeq703-1453\,\mathrm{days}$ increased the extinction along the line of sight and caused the apparent fading in the UV. To test this case, we assumed the source spectrum to be the best-fit power law at $t\simeq703\,\mathrm{days}$ and reddened the spectrum according to the Cardelli extinction law \citep{1989ApJ...345..245C} with $R_{\mathrm{V}}=3.1$ to fit the observed SED at $t\simeq1453\,\mathrm{days}$ through synthetic photometry. Figure \ref{fig:Underdust} shows the resulting reddened power law with a best-fit color excess of $E_{\mathrm{B}-\mathrm{V}}\simeq0.072$, which matches the observed SED at $t\simeq1453\,\mathrm{days}$. Therefore, dust extinction could be a possible explanation for the observed fading of the underlying source, especially given that dust has been proposed to explain the excess IR of AT\,2018cow \citep[][]{Metzger2022dust}. We further discuss the implications of dust formation in Section \ref{subsec:understardust}. 

\section{Constraints on the Origin of the Underlying Source ($\lowercase{t}\sim703-1453\,\mathrm{\lowercase{days}}$)}\label{sec:underinterpret}

Here, we briefly summarize the properties of the underlying source outlined in Section~\ref{sec:underlying} before discussing implications for its origin. The underlying source was quite bright, with an integrated (minimum) luminosity of $L_{\mathrm{UVO,min}}\sim10^{39}\,\mathrm{erg}\,\mathrm{s}^{-1}$ over $\lambda \sim 2358-8029\,\mathrm{\AA}$. HST photometry of the underlying source showed an exceptionally blue continuum ($\mathrm{F336W}-\mathrm{F555W}=-1.3$) without constraining the peak of the SED ($\lambda_{\mathrm{peak}}\lesssim 2358\,\mathrm{\AA}$). A \emph{Swift}-XRT non-detection at $t\simeq1360\,\mathrm{days}$ suggests that the peak was in the UV during the latest HST observation (Figure \ref{fig:underlyingSED}). The NUV-optical continuum at $t\simeq703\,\mathrm{days}$ matches $L_{\nu}\propto \nu^2$, the Rayleigh-Jeans tail, and can be described by a blackbody with a high temperature $T_{\mathrm{BB}}\gtrsim10^{5}\,\mathrm{K}$ and a small radius $R_{\mathrm{BB}} \lesssim 20\,R_\odot$. Transient nature was also confirmed by the $3.5\sigma$ and $3.9\sigma$ fading in the two NUV bands (F225W and F336W) between $t\simeq703\,\mathrm{days}$ and $1453\,\mathrm{days}$, which flattened the spectrum and could indicate cooling and expansion of a blackbody (see Table \ref{tab:UnderProperties}).

The transition between rapid fading over the first two months to slow fading over a timescale of years is likely associated with a sustained power source. 
However, the exceptionally blue color immediately rules out synchrotron emission and warm dust emission with $T_{\mathrm{dust}}<2000\,\mathrm{K}$ while the brightness also rules out radioactive decay. If we assume that $L_{\mathrm{UVO,min}}\sim10^{39}\,\mathrm{erg}\,\mathrm{s}^{-1}$ at $t\simeq1453\,\mathrm{days}$ was powered by radioactive decay, following the equations in \citet{Afsariardchi2021} and assuming full $\gamma$-ray trapping, we find a completely unphysical $M_{\mathrm{Ni}}\sim74\,M_\odot$.

In this section, focusing on the HST SEDs and the \emph{Swift}-XRT non-detection, we discuss the constraints on the possible power sources of the underlying source. We consider five possible origins of the underlying emission and the observed fading: significant stellar contribution (Section \ref{subsec:understellar}), dust extinction (Section \ref{subsec:understardust}), ejecta-CSM interaction (Section \ref{subsec:underCSM}), a magnetar (Section \ref{subsec:undermag}), and an accreting BH (Section \ref{subsec:underBH}). Note that we assume the fading to be smooth over $t\simeq703-1453\,\mathrm{days}$ and do not consider any sporadic activities (e.g., multiple flares) that could explain the observations because we do not have sufficient temporal coverage to distinguish such a possibility.

\subsection{Significant Stellar Contribution}\label{subsec:understellar}

\begin{figure*}
\epsscale{0.55}
\plotone{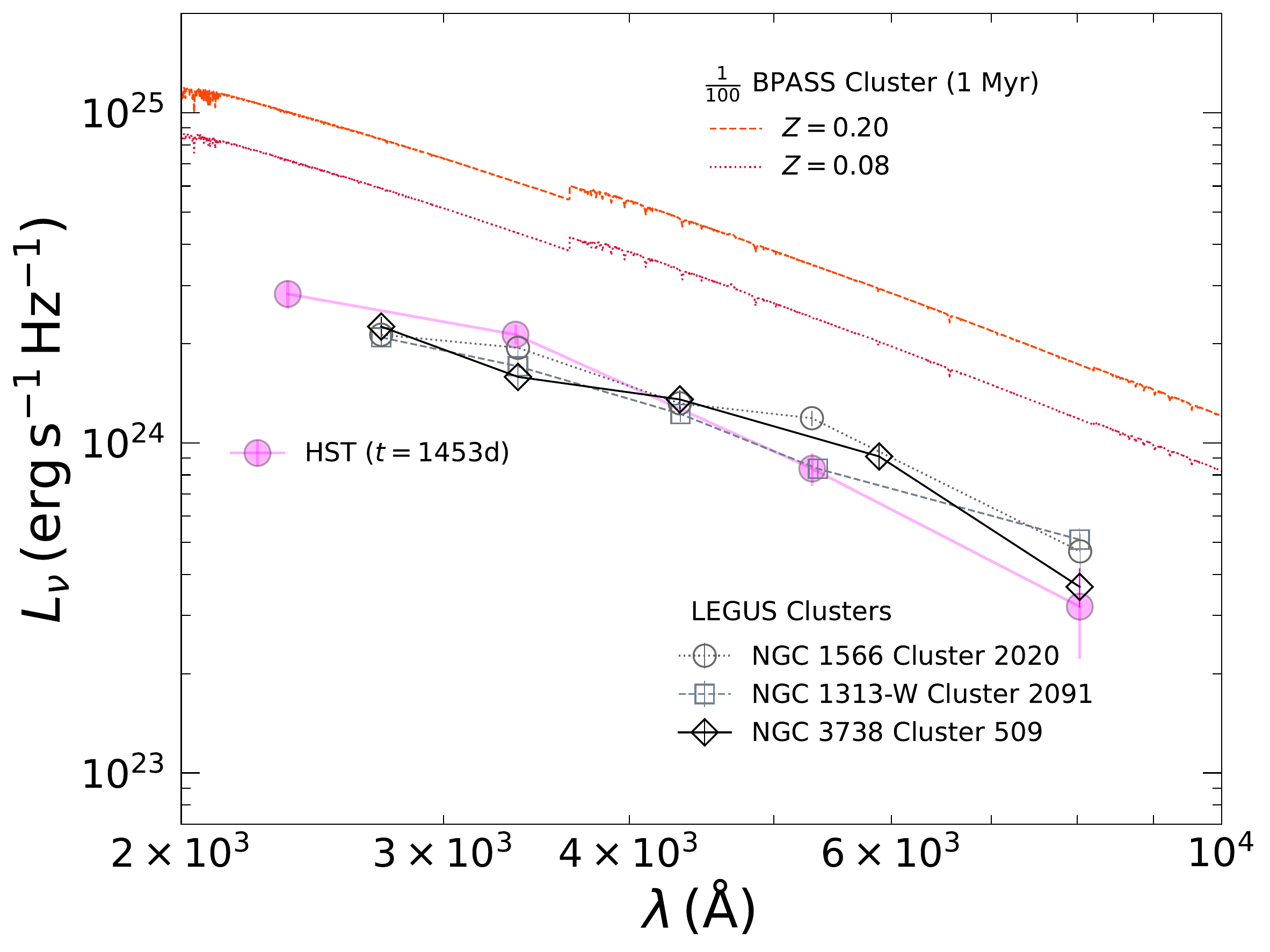}
\plotone{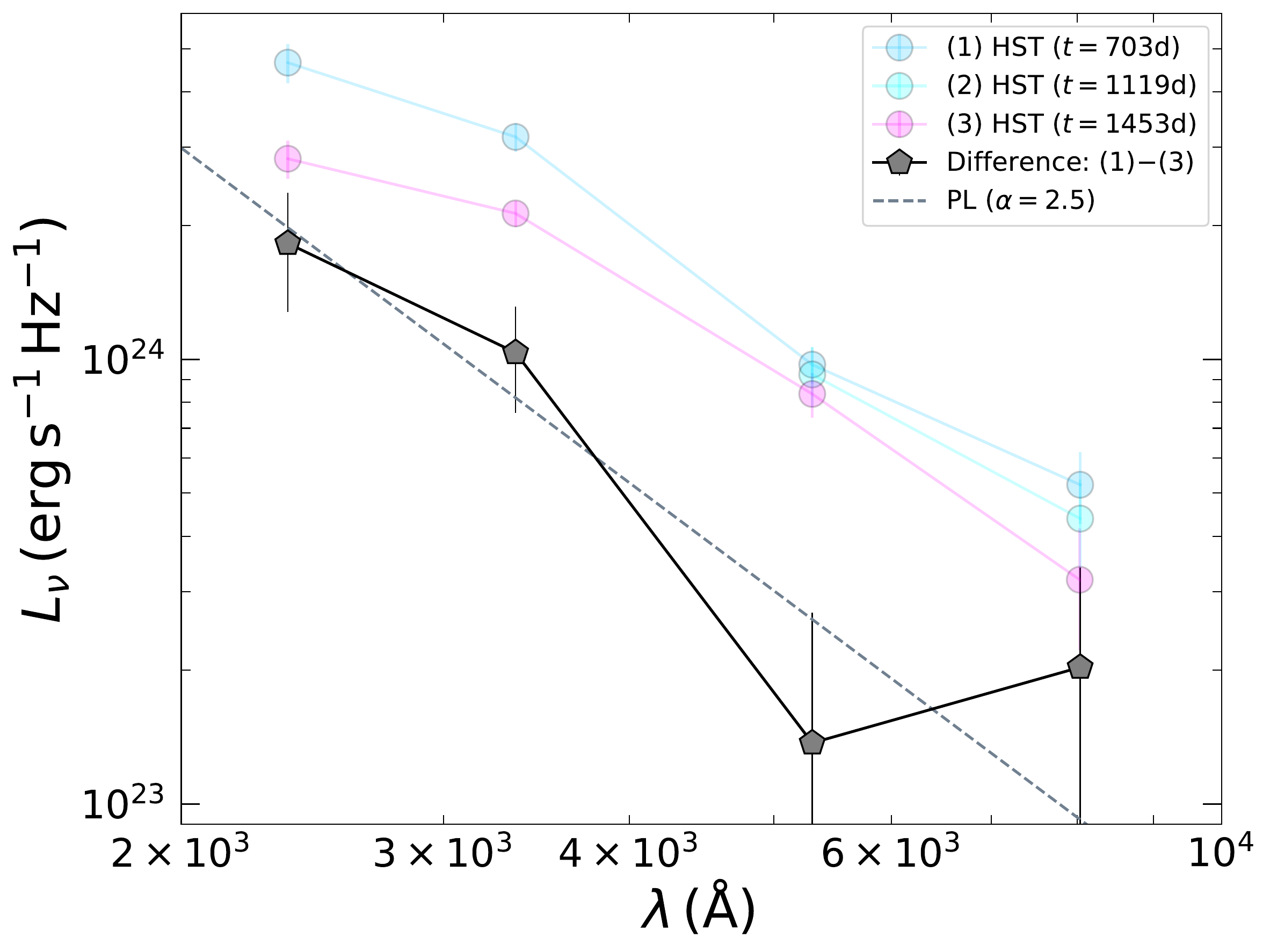}
\caption{Left: Observed (dereddened) HST SED at $t\simeq 1453\,\mathrm{days}$ compared with BPASS clusters and selected LEGUS clusters. The BPASS clusters come from the default model that assumes an initial mass function with a power law slope of $-1.30$ between $0.1-0.5\,M_\odot$ and $-2.35$ above $0.5\,M_\odot$, extending to a maximum mass of $300\,M_\odot$. Right: Observed (dereddened) HST SEDs and the difference SED calculated by taking the difference between the HST SEDs observed at $t\simeq 703\,\mathrm{days}$ and $1453\,\mathrm{days}$. The difference SED corresponds to the transient component at $t\simeq 703\,\mathrm{days}$ if the HST SED at $t\simeq 1453\,\mathrm{days}$ is solely from a stellar source.
\label{fig:Understellar}}
\end{figure*}

Although the fading of the underlying source over $\sim$2--4 years post-discovery is likely associated with some transient phenomenon, there is a possibility that an underlying stellar source contributed to the emission. In particular, the slow fading in the UV and relative stability in the optical might be easier to explain if stellar emission contributed significantly to the HST SED at $t\simeq1453\,\mathrm{days}$. In this case, the actual transient emission would have faded faster, but could not be observed once it became dimmer than the stellar emission. To understand the implications of this scenario, we consider the limiting case where the \emph{entire} HST SED observed at $t\simeq1453\,\mathrm{days}$ was stellar emission. We consider implications for both the stellar and transient emission.

\subsubsection{Implications for the Stellar Population}

First, we note that any stellar emission is unlikely associated with an isolated ``usual'' massive star ($M \lesssim 100\,M_\odot$). This was disfavored by \citet{Sun2022} based on brightness and color. Hypothetically, Very Massive Stars (VMS) with $M \gtrsim 100-200\,M_\odot$, $L\gtrsim10^{39}\,\mathrm{erg}\,\mathrm{s}^{-1}$, and $T_{\mathrm{eff}}\gtrsim10^{4.5}\,\mathrm{K}$ \citep[e.g.,][]{Sabhahit2022} could perhaps explain the color and brightness of the underlying source. However, this would require a surviving single VMS star to be coincident within $\lesssim$6 pc of the location of AT\,2018cow, which is unlikely unless AT\,2018cow also came from a VMS progenitor. The relative isolation of two VMS at the outskirt of the galaxy would be unusual, and although failed explosion is a proposed scenario for FBOTs \citep[e.g.,][]{Kashiyama2015}, a VMS progenitor for AT\,2018cow is still less favored for various reasons, such as the small ejecta mass inferred from the prompt transient emission \citep{Margutti2019} and the possibility that AT\,2018cow comes from an older stellar population that resides in the foreground relative to the nearby star-forming regions \citep{Sun2022new}. 

Thus, a more natural scenario may be that any stellar contribution comes from an underlying star cluster, possibly the host of the progenitor of AT\,2018cow. To investigate the properties of this hypothetical cluster, we compared the HST SED at $t\simeq1453\,\mathrm{days}$ with stellar populations generated by the Binary Population and Spectral Synthesis \citep[BPASS;][]{Eldridge2017,Stanway2018} version 2.2.1 as well as clusters observed by the HST Legacy ExtraGalactic UV Survey \citep[LEGUS;][]{Calzetti2015LEGUS,Adamo2017LEGUScluster}.

The left panel of Figure \ref{fig:Understellar} shows the HST SED at $t\simeq1453\,\mathrm{days}$ and some selected clusters from BPASS and LEGUS. Compared to the observed SED, the youngest BPASS clusters (1\,Myr old) have the best-matching color but overpredict the luminosity by several orders of magnitude. The normalization of the BPASS clusters is based on a cluster with $M_{\mathrm{clus}}=10^6\,M_\odot$, meaning that a lower cluster mass ($M_{\mathrm{clus}}\sim10^3-10^4\,M_\odot$) is required to explain the brightness of the underlying source. 

On the other hand, we were able to identify a number of LEGUS clusters with brightness and color similar to the underlying source. However, these are extremely rare cases (we found less than a dozen searching through the cluster catalogs), highlighting the peculiarity of the brightness and color. In the images, these clusters are seen in blue crowded regions of the galaxy that appear to be associated with young stellar populations and star-formation activity\footnote{Cluster 2020 of NGC\,1566 is inside the spiral arm and beside a very bright concentration of stars. NGC\,1313 and NGC\,3738 are known star-forming galaxies \citep[][]{Larsen1999,Karachentsev2003}}. In the catalogs, they are classified as asymmetric and multi-peaked sources ($\mathrm{Class}=2$ and $3$) with a best-fit age $\sim$few Myr and best-fit $M_{\mathrm{clus}} \sim \mathrm{few}\times10^3\,M_\odot$, consistent with our expectation from the BPASS clusters (however, note that based on the given Q probability, the SED fits are quite poor). Overall, these comparisons suggest that it \emph{is} possible to explain the HST SED at $t\simeq1453\,\mathrm{days}$ as a star cluster (with real examples). However, it would imply a rare young ($\mathrm{age}\sim\mathrm{Myr}$) cluster with a relatively small mass ($M_{\mathrm{clus}}\sim10^{3}\,M_\odot$) and suggest that AT\,2018cow came from a massive progenitor.

We briefly note that we have not considered cases involving long-term evolution of rare stellar merger/interaction events that leave behind compact or stripped stars with some energy source powering a low-mass envelope. An example is the case considered by \citet{Cohen2023} where, after the common envelope jets supernova imposter event that produced AT\,2018cow \citep[][]{Soker2022}{}{}, a NS is inside the red supergiant during a second common envelope phase and influencing the evolution through launching jets. Although these objects could in theory be blue and luminous and become redder over time, their exact spectral evolution is not clear at the moment, and it is uncertain if they can fully explain the underlying source.

\subsubsection{Implications for the Transient Emission}

Next, we consider the implications on the transient emission if the entire HST SED at $t\simeq1453\,\mathrm{days}$ was from a star cluster. In this case, we can subtract the cluster SED at $t\simeq1453\,\mathrm{days}$ from the observed SED at $t\simeq703\,\mathrm{days}$ to obtain a transient SED at $t\simeq703\,\mathrm{days}$. The right panel of Figure \ref{fig:Understellar} shows the transient SED and the best-fit power law. The transient emission in this case is still bright, with a $L_{\mathrm{UVO,min}}=(5.3\pm1.0)\times10^{38}\,\mathrm{erg}\,\mathrm{s}^{-1}$. The transient SED is much bluer now with a best-fit spectral index of $\alpha=2.5$. This extremely hard transient spectrum would be inconsistent with a blackbody and may be significantly challenging to explain with other emission mechanisms. 

In addition, if we consider a scenario where only the optical emission at $t\simeq1453\,\mathrm{days}$ is dominated by a stellar population and the UV is dominated by the transient (e.g., an older stellar population), then the inferred spectral index of the transient emission at $t\simeq703\,\mathrm{days}$ would be \emph{even steeper}. Thus, while a star cluster could have contributed significantly to the HST SED at $t\simeq1453\,\mathrm{days}$, the implied blue color (spectral index $\alpha\gtrsim2.5$) of the transient emission at may be difficult to explain, posing a significant challenge to this scenario.

\subsection{Dust Extinction} \label{subsec:understardust}

As we have shown in Section \ref{subsec:undermodel} and Figure \ref{fig:Underdust}, an increase in extinction could theoretically account for the preferential fading in the UV of the underlying source. Assuming the Cardelli extinction law \citep{1989ApJ...345..245C} with $R_{\mathrm{V}}=3.1$, we found that the fading between $t\simeq703\,\mathrm{days}$ and $1453\,\mathrm{days}$ would correspond to a color excess of $E_{\mathrm{B}-\mathrm{V}}\simeq0.072$. The increase in extinction would imply dust formation over these epochs, likely associated with AT\,2018cow and its surrounding CSM. Here, we examine the properties of a hypothetical dust cloud that could produce the observed fading and discuss the implications of dust formation on the nature of the underlying source.

We can make an order-of-magnitude estimate of the required dust mass column density $\Sigma_{\mathrm{d}}$ from the inferred extinction. Assuming spherical dust grains with radius $a$ and density $\rho$ and uniform size and composition, the optical depth through the dust cloud $\tau_{\lambda}(a)$ can be written as
\begin{equation}
    \tau_{\lambda}(a) = \frac{3Q^{\mathrm{ext}}_\lambda(a)}{4a\rho} \Sigma_{\mathrm{d}},
\end{equation}
where $Q^{\mathrm{ext}}_\lambda$ is the extinction coefficient. The extinction coefficient is given by the sum of the absorption coefficient and scattering coefficient: $Q^{\mathrm{ext}}_\lambda = Q^{\mathrm{abs}}_\lambda + Q^{\mathrm{sca}}_\lambda$. 

The optical depth is related to the extinction in magnitude through $A_{\lambda} = 2.5\log_{10}(e)\tau_{\lambda}$ while the visual extinction in magnitudes is calculated by $A_{\mathrm{V}} = R_{\mathrm{V}}E_{\mathrm{B}-\mathrm{V}}$. Therefore, we can derive a dust mass column density as:
\begin{equation}
    \Sigma_{\mathrm{d}} = \frac{R_{\mathrm{V}}E_{\mathrm{B}-\mathrm{V}}}{2.5\log_{10}(e)}\frac{4a\rho}{3Q^{\mathrm{ext}}_\mathrm{V}}.
\end{equation}

We considered two types of dust, graphite with $\rho=2.26\,\mathrm{g}\,\mathrm{cm}^{-3}$ or silicate with 
$\rho=3.30\,\mathrm{g}\,\mathrm{cm}^{-3}$, and two grain sizes, $a=0.1\,\mu\,\mathrm{m}$ and $1.0\,\mu\,\mathrm{m}$. We use $Q_\lambda^{\mathrm{abs}}$ and $Q_\lambda^{\mathrm{sca}}$ derived in \citet{1984ApJ...285...89D} and \citet{1993ApJ...402..441L}\footnote{Obtained from from Draine’s website \url{https://www.astro.princeton.edu/~draine/dust/dust.diel.html}} and interpolated the coefficients to $\lambda_{\mathrm{V}}=5500\,\mathrm{\AA}$ to obtain $Q^{\mathrm{ext}}_\mathrm{V}$. From these values and assuming $R_{\mathrm{V}}=3.1$ and $E_{\mathrm{B}-\mathrm{V}}=0.072$, we estimated 
\begin{eqnarray}
    \Sigma_{\mathrm{d}}(a=0.1-1.0\,\mu\,\mathrm{m})\sim10^{-5.7}-10^{-4.6}\,\mathrm{g}\,\mathrm{cm}^{-2}\nonumber\\
    \sim10^{-4.9}-10^{-4.4}\,\mathrm{g}\,\mathrm{cm}^{-2}\nonumber
\end{eqnarray}
for graphite (top) and silicate (bottom). 

To convert the column density to a volume density or a mass, we have to consider the size of the dust cloud. For the underlying source, dust formation over $t\simeq703-1453\,\mathrm{days}$ likely occurred in the region where pre-existing dust grains were initially destroyed by AT\,2018cow, but the material cooled over time and started forming new dust grains at these late epochs. The UV radiation of AT\,2018cow would have destroyed the dust grains inside the sublimation radius, where the temperature is higher than the sublimation temperature of the dust grains ($\sim$1100--1500\,K for silicate and $\sim$2000\,K for graphite). If we assume a luminosity of $L\sim4\times10^{44}\,\mathrm{erg}\,\mathrm{s}^{-1}$ (peak of AT\,2018cow), a grain size of $a\sim0.1-1.0\,\mu\mathrm{m}$, and a sublimation temperature of $T_{\mathrm{s}}\sim1100-2000\,\mathrm{K}$, then we can follow the equations in \citet{Metzger2022dust} and obtain a sublimation radius $r_{\mathrm{s}}\sim10^{16.2}-10^{16.4}\,\mathrm{cm}$. The ejecta of AT\,2018cow also could have destroyed the dust, which would have reached $r\sim10^{17.6}-10^{17.9}\,\mathrm{cm}$ by $t=1453\,\mathrm{days}$ assuming a velocity of $v\sim0.1c-0.2c$. Therefore, we can assume that the newly-formed dust cloud has a radius on the order of $R_{\mathrm{d}}\sim10^{16.2}-10^{17.9}\,\mathrm{cm}$.

Note that if the underlying source was at the center of the dust cloud (i.e., associated with AT\,2018cow), its UV emission at $t\simeq1453\,\mathrm{days}$ ($L\sim10^{40}\,\mathrm{erg}\,\mathrm{s}^{-1}$) would create a dust-free cavity with a radius of $r_{\mathrm{s}}\sim10^{15.8}-10^{16.0}\,\mathrm{cm}$. This size is slightly less significant than the order-of-magnitude size we consider here, so we ignore this cavity. Detailed models should consider dust-forming and dust-destroying regions inside an evolving radiation field, which is beyond the scope of this study.

From the column density and physical size, we estimated the dust mass from $M_{\mathrm{d}}\sim\pi R_{\mathrm{d}}^2 \Sigma_{\mathrm{d}}$, which gave
\begin{eqnarray}
    M_{\mathrm{d}}\sim10^{-6.1}-10^{-4.8}\,M_\odot\quad (R_\mathrm{d}\sim10^{16.2}\,\mathrm{cm})\nonumber\\
    \sim10^{-2.7}-10^{-1.4}\,M_\odot\quad (R_\mathrm{d}\sim10^{17.9}\,\mathrm{cm})\nonumber
\end{eqnarray}
assuming $\Sigma_{\mathrm{d}}\sim10^{-5.7}-10^{-4.4}\,\mathrm{g}\,\mathrm{cm}^{-2}$. Note that as shown in paper I \citep{Chen2023I}, the dust mass derived from the excess IR emission of AT\,2018cow was on the order of $M_{\mathrm{d}}\sim10^{-6}-10^{-4}\,M_\odot$, suggesting that the dust mass needed to explain the fading of the underlying source is not unreasonable.

We can also infer a gas density by assuming a dust-to-gas mass density ratio $\rho_{\mathrm{d}}/\rho_{\mathrm{g}}=X_{\mathrm{d}}\sim0.1$. We estimated the gas density from $\rho_{\mathrm{g}}\sim \Sigma_{\mathrm{d}}/R_{\mathrm{d}}X_{\mathrm{d}}$, which gave
\begin{eqnarray}
    \rho_{\mathrm{g}}\sim10^{-20.9}-10^{-19.6}\,\mathrm{g}\,\mathrm{cm}^{-3}\quad (R_\mathrm{d}\sim10^{16.2}\,\mathrm{cm})\nonumber\\
    \sim10^{-22.6}-10^{-21.3}\,\mathrm{g}\,\mathrm{cm}^{-3}\quad (R_\mathrm{d}\sim10^{17.9}\,\mathrm{cm})\nonumber
\end{eqnarray}
assuming $\Sigma_{\mathrm{d}}\sim10^{-5.7}-10^{-4.4}\,\mathrm{g}\,\mathrm{cm}^{-2}$. Interestingly, this gas density is roughly consistent with the radio-producing CSM density at $R\sim10^{16}-10^{17}\,\mathrm{cm}$ (see top panel of Figure \ref{fig:UnderCSM}). However, we note that the exact value of $X_{\mathrm{d}}$ is very uncertain and can depend on properties such as the composition. In particular, if $X_{\mathrm{d}}$ is orders of magnitude lower than 0.1 (e.g., hydrogen-rich gas), the inferred gas could be much denser than the radio-producing CSM. In that case, a possible explanation is that the dust was associated with the dense hydrogen-rich envelope ejected by the progenitor of AT\,2018cow located at $R\gtrsim10^{17.3}\,\mathrm{cm}$, which is outside the range covered by published radio observations. 

Overall, the properties of the hypothetical dust cloud appear reasonable in the context of AT\,2018cow and stellar explosions. However, there are challenges in explaining the actual underlying emission in this scenario. If the fading in UV over $t\simeq703\,\mathrm{days}-1453\,\mathrm{days}$ was purely due to dust extinction (with $E_{\mathrm{B}-\mathrm{V}}\simeq0.072$), then the underlying emission may have been stable with a spectrum similar to \emph{or likely bluer} than the observed HST SED at $t\simeq703\,\mathrm{days}$ (i.e., spectral index $\alpha\gtrsim 2$). Although the stability would suggest a stellar source, if we follow the findings in Section \ref{subsec:understellar}, there are no known stellar sources that can explain such a blue color. Thus, in this scenario, we are left with the possibility of a remnant transient emission that is stable over several years with an extremely blue spectrum ($\alpha\gtrsim 2$), which is still fairly challenging to explain. 

Therefore, if dust was relevant to the underlying source, it is more likely that dust extinction only contributed partially to the fading. This scenario could be very complex, possibly involving some combination of dust, stellar source, and transient source. For this case, the dust and gas densities derived above assuming $E_{\mathrm{B}-\mathrm{V}}\simeq0.072$ should be considered as upper limits. Constraining such a scenario would require additional observations monitoring the evolution of the underlying source and detailed modeling. Deep images through the JWST may also be useful in constraining potential IR echo produced by the hypothetical dust cloud.

\subsection{Ejecta-CSM Interaction}\label{subsec:underCSM}

Ejecta-CSM interaction is known to be capable of producing prolonged emission in interacting SNe such as SN\,2005ip \citep[][]{Smith2009} and SN2010jl \citep[][]{Fransson2014,Jencson2016}, with optical luminosities of $L\gtrsim 10^{41}\,\mathrm{erg}\,\mathrm{s}^{-1}$ over hundreds of days. Although no known interacting SN matches the description of both AT\,2018cow and the underlying source, the existence of CSM around AT\,2018cow may nonetheless contribute significantly to the underlying emission. Here, we model the interaction between the fast ejecta ($v\sim0.1c$) of AT\,2018cow and the known radio-producing CSM to check if this interaction years after the explosion can power the underlying emission.

Our ejecta-CSM interaction model is similar to the method described in Section 5.2 of \citet{Chandra2015} for arbitrary ejecta and CSM density distributions. Under the thin shell approximation \citep[][]{Chevalier1982}, we solve the equation of motion to advance the shock velocity $v_{\mathrm{s}}$ using the Runge-Kutta method. We advance the shock radius $R_{\mathrm{s}}$ and calculate the ejecta and CSM densities, $\rho_{\mathrm{ej}}$ and $\rho_{\mathrm{CSM}}$. At the new time step, for the forward shock (fs) and reverse shock (rs), we calculate the velocities $v_{\mathrm{fs}} = v_{\mathrm{s}}$ and $v_{\mathrm{rs}} = R_{\mathrm{s}}/t - v_{\mathrm{s}}$, temperatures $T_{\mathrm{fs}}$ and $T_\mathrm{rs}$ assuming a strong shock and ion-electron equipartition, and cooling times $t_{\mathrm{c,fs}}$ and $t_{\mathrm{c,rs}}$ \citep[following][]{Nymark2006}. Finally, we calculate the kinetic luminosities $L_{\mathrm{kin,fs}} = 2\pi R_{\mathrm{s}}^2 \rho_{\mathrm{CSM}}v_{\mathrm{fs}}^3$ and $L_{\mathrm{kin,rs}} = 2\pi R_{\mathrm{s}}^2 \rho_{\mathrm{ej}}v_{\mathrm{rs}}^3$ and derive radiated luminosities in the form of $L_{\mathrm{rad}} = \eta L_{\mathrm{kin}}$, where we define a radiation efficiency factor $\eta=t/(t+t_{\mathrm{c}})$.

We assumed power law distributions for the ejecta: $\rho \propto r^{-\delta}$ with $\delta=0$ for the inner region and $\rho \propto r^{-n}$ with $n=12$ for the outer region \citep[][]{Chevalier1994}. From the radio observations of AT\,2018cow \citep[i.e.,][]{Ho2019,Nayana2021}, we adopted a two-component CSM distribution with $\dot{M}=4\times10^{-4}\,M_\odot\,\mathrm{yr}^{-1}$ inside a radius of $R\leq1.7\times10^{16}\,\mathrm{cm}$ and $\dot{M}= 4\times10^{-6}\,M_\odot\,\mathrm{yr}^{-1}$ outside a radius of $R\geq6\times10^{16}\,\mathrm{cm}$, assuming $\rho = \dot{M}/(4\pi R^2v_\mathrm{w})$ with $v_\mathrm{w}=1000\,\mathrm{km}\,\mathrm{s}^{-1}$ for both distributions. A steeper power law was used to connect the two distributions between $R=1.7\times10^{16}\,\mathrm{cm}$ and $6\times10^{16}\,\mathrm{cm}$. For the explosion, we assumed an ejecta mass of $M_{\mathrm{ej}}=0.5\,M_\odot$ and a total energy of $E = 3\times10^{51}\,\mathrm{erg}$, motivated by the short rise time and high peak optical luminosity of AT\,2018cow \citep[][]{Margutti2019}. We also assumed solar composition with a mean molecular weight of $\mu=0.61$.

\begin{figure}
\epsscale{1.0}
\plotone{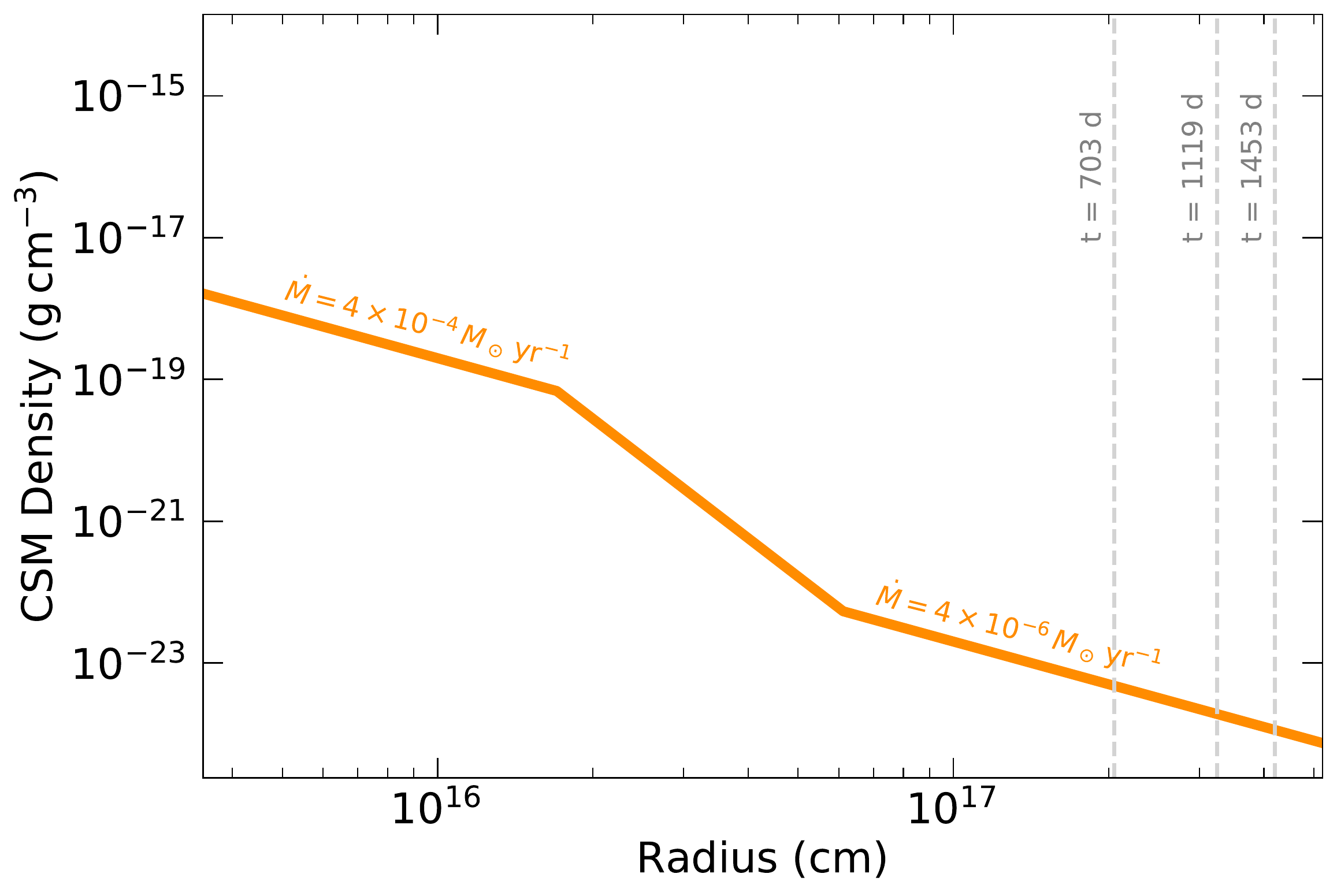}
\plotone{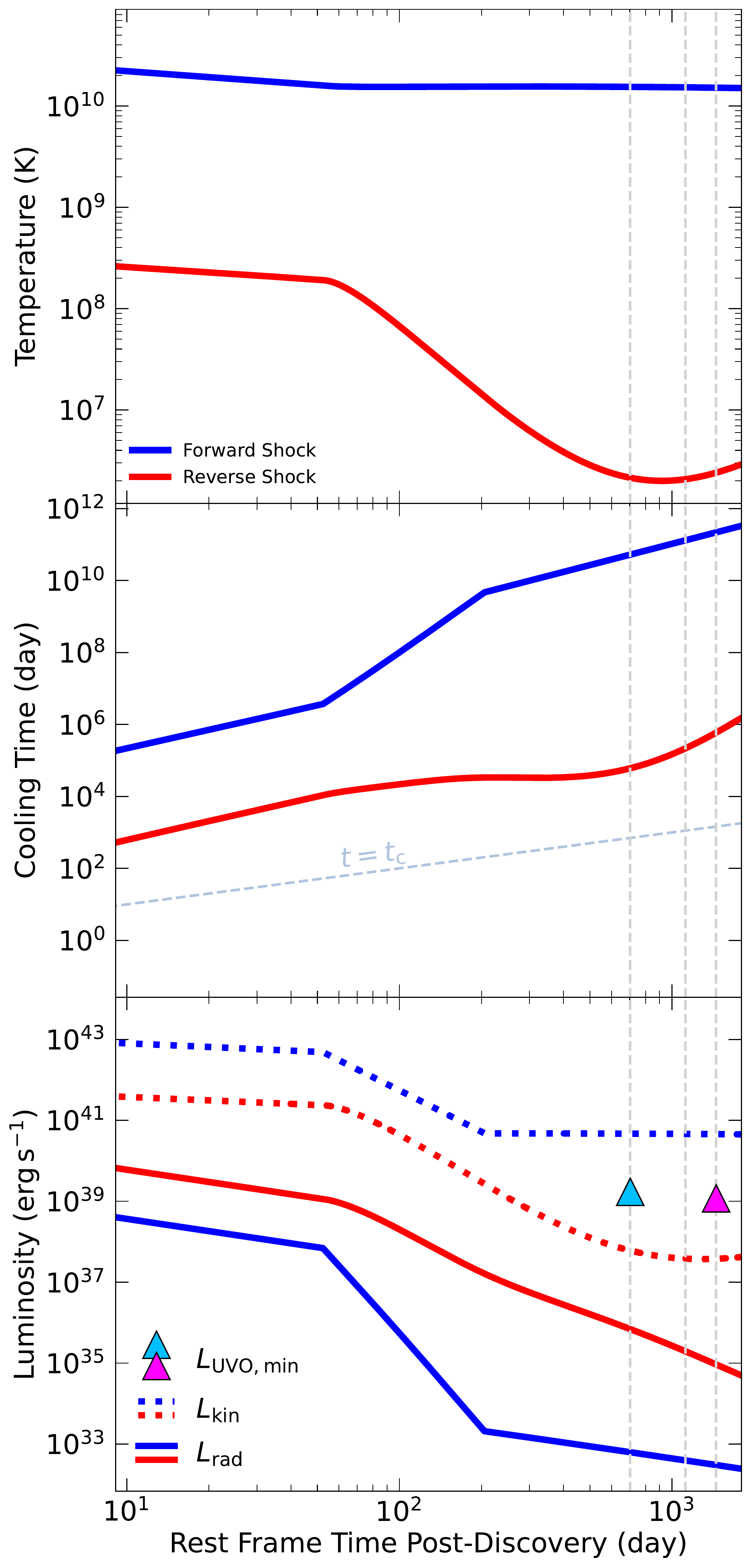}
\caption{CSM density profile used in our ejecta-CSM interaction model (top panels) and the forward and reverse shock properties derived from the model (bottom three panels). See text for details regarding our model. The three HST epochs are marked with gray dashed lines. We also show $L_{\mathrm{UVO,min}}$ (triangles), and a key finding here is that $L_{\mathrm{UVO,min}}$ is orders of magnitude above the $L_{\mathrm{rad}}$.
\label{fig:UnderCSM}}
\end{figure}

Figure \ref{fig:UnderCSM} shows the model CSM density distribution, shock temperatures, cooling times, and kinetic and radiated luminosities. Dashed lines are shown to mark the three HST epochs. Properties for the forward shock and reverse shock are shown in blue and red, respectively. 
The cooling times indicate that both the forward and reverse shocks are adiabatic (i.e., $t_{\mathrm{cool}} \gg t$) from early times because of the high velocity and low ejecta mass. The radiation efficiency $\eta$ is thus very low, and the radiated luminosity is orders of magnitude lower than the kinetic luminosity. 

From our model, the radiated forward shock luminosity and reverse shock luminosity are $L_{\mathrm{rad,fs}} < 10^{33}\,\mathrm{erg}\,\mathrm{s}^{-1}$ and $L_{\mathrm{rad,rs}} < 10^{36}\,\mathrm{erg}\,\mathrm{s}^{-1}$, respectively, which are orders of magnitude lower than the minimum UV-optical luminosity of the underlying source, $L_{\mathrm{UVO,min}} \sim 10^{39}\,\mathrm{erg}\,\mathrm{s}^{-1}$ (bottom panel of Figure \ref{fig:UnderCSM}). The kinetic luminosity of the forward shock at $L_{\mathrm{kin,fs}} \sim 5\times10^{40}\,\mathrm{erg}\,\mathrm{s}^{-1}$ could explain the luminosity of the underlying source if the radiation efficiency is high enough ($\eta \gtrsim 0.02$). However, given the adiabatic nature of the shock, most of the radiation should be in the form of free-free emission in the X-ray and would have difficulty accounting for the observed UV-optical emission. Such a high radiation efficiency would also imply that the interaction should have dominated the fading prompt emission of AT\,2018cow, which was not observed.

Therefore, we conclude that the fast ejecta of AT\,2018cow interacting with the known radio-producing CSM could not account for the observed UV-optical emission of the underlying source. Note that our model does not include the proposed dense equatorial CSM with slow interaction for AT\,2018cow \citep[][]{Margutti2019}. Including such a component may result in an appreciable change but is outside the scope of this study because it would first require a detailed aspherical ejecta-CSM models for AT\,2018cow to constrain the correct CSM profile.

We do not rule out the possibility of ejecta-CSM interaction involving a new CSM component, i.e., the previously ejected hydrogen envelope of the progenitor of AT\,2018cow. Such a phenomenon has been previously observed in cases such as SN\,2014C, where a hydrogen-poor SN transitioned to an interacting hydrogen-rich SN \citep[e.g.,][]{Milisavljevic2015,Margutti2017,Mauerhan2018,Brethauer2022}. However, from the interaction with hydrogen-rich CSM, an expected signature is strong H$\alpha$ emission \citep[e.g.,][]{Fransson2014}. For the underlying source, the detection of H$\alpha$ emission is uncertain (see Section \ref{subsubsec:compareSun2022}). Even assuming an H$\alpha$ luminosity of $L_{\mathrm{H}\alpha}\sim4\times10^{36}\,\mathrm{erg}\,\mathrm{s}^{-1}$ \citep[from][]{Sun2022}, this would still be many orders of magnitude fainter than those typically observed in interacting SNe, e.g., $L_{\mathrm{H}\alpha}\sim10^{39}\,\mathrm{erg}\,\mathrm{s}^{-1}$ for SN\,2014C \citep[][]{Mauerhan2018}. Detailed modeling as well as additional observations could help constrain this hypothesis. In particular, in addition to deeper H$\alpha$ images, radio and X-ray follow-ups may also provide significant constraints on any possible rebrightening from the new interaction, and deep IR observations could also probe warm dust associated with the CSM.

\subsection{Magnetar Spin Down} \label{subsec:undermag}

Studies have suggested that a millisecond magnetar could power the fast-rising luminous optical peak of AT\,2018cow \citep[e.g.,][]{2018ApJ...865L...3P,Margutti2019,Ho2019,Mohan2020,Xiang2021}. Following this hypothesis, the underlying source could be remnant emission powered by such a magnetar.

In the magnetar scenario, a fraction of the spin-down luminosity $L_{\mathrm{sd}}$ is radiated as the observed underlying source, depending on the efficiency. $L_{\mathrm{sd}}$ depends on the rotational energy $E_{\mathrm{sd}}$ and spin-down timescale $t_{\mathrm{sd}}$, which are related to the surface magnetic field strength $B$ and the initial spin period $P_0$ of the magnetar. Therefore, we can place constraints on the magnetar configuration (i.e., $B$ and $P_0$) that would be required to explain the underlying source by comparing $L_{\mathrm{sd}}$ with the observed $L_{\mathrm{UVO,min}}$.

\begin{figure}[t]
\epsscale{1.2}
\plotone{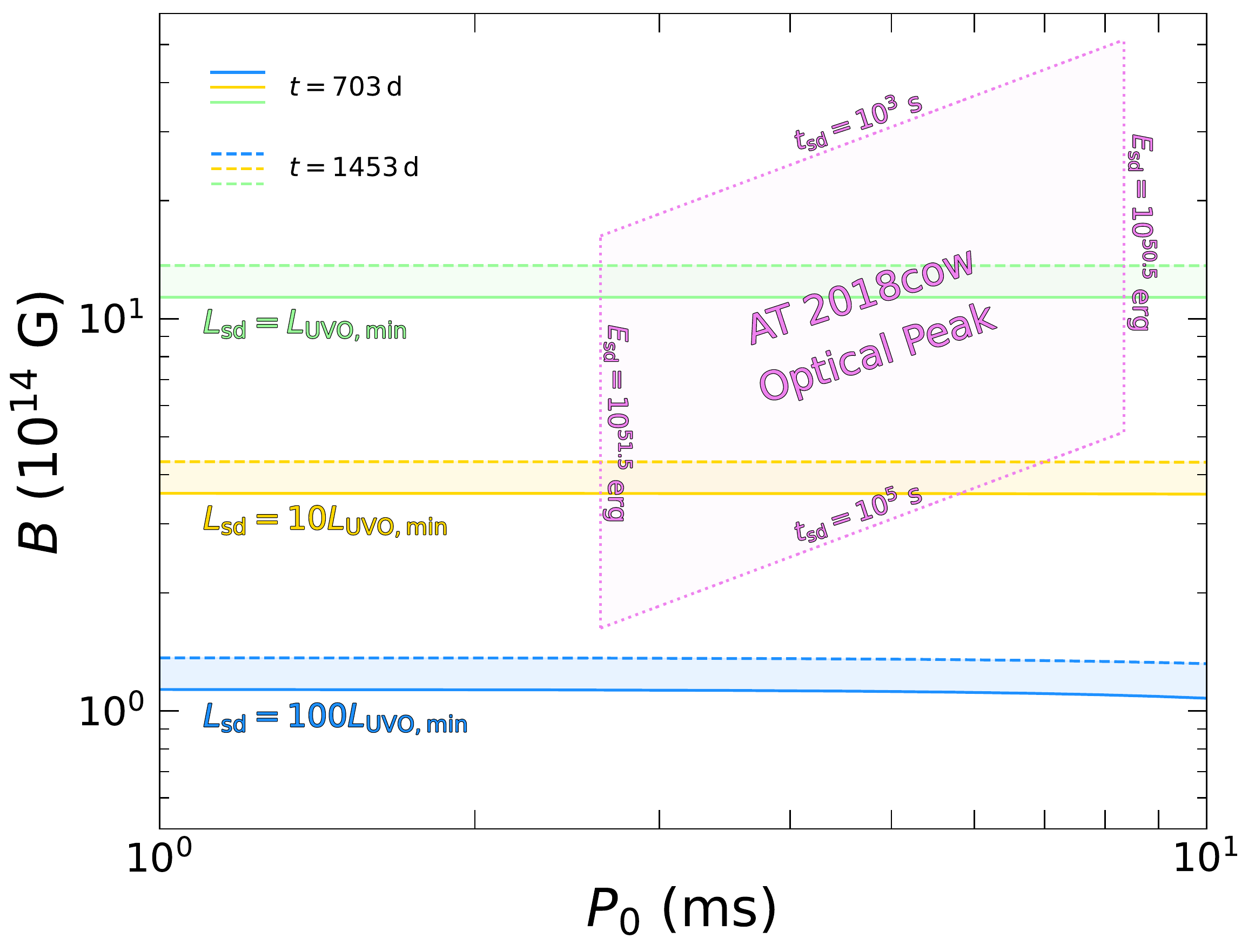}
\caption{Magnetar parameter space with the surface magnetic field strength $B$ and the initial spin period $P_0$. Lines of constant spin-down luminosity $L_{\mathrm{sd}}$ are shown for $L_{\mathrm{sd}}=L_{\mathrm{UVO,min}}$, $10L_{\mathrm{UVO,min}}$, $100L_{\mathrm{UVO,min}}$ (green, yellow, blue). Solid and dashed lines indicate $L_{\mathrm{UVO,min}}$ values derived from $t\simeq703\,\mathrm{days}$ and $1453\,\mathrm{days}$, respectively. A purple region is shown for magnetar parameters that could power the optical peak of AT\,2018cow \citep[][]{Margutti2019}.
\label{fig:magnetar}}
\end{figure}

We calculated $E_{\mathrm{sd}}$, $t_{\mathrm{sd}}$, and $L_{\mathrm{sd}}$ as a function of $B$ and $P_0$ following the formulas in Section 4.3 of \citet{Kasen2017}. We assumed a neutron star with a mass of $M_{\mathrm{ns}}=1.4\,M_\odot$ and a radius of $R_{\mathrm{ns}}=10\,\mathrm{km}$ and assumed a moment of inertia in the form of $I_{\mathrm{ns}} = 2M_{\mathrm{ns}}R_{\mathrm{ns}}^2/5$. Note that for timescales relevant for the underlying source ($t\simeq703--1453\,\mathrm{days}$), $t \gg t_{\mathrm{sd}}$ for large $B$ and $L_{\mathrm{sd}}\approx E_{\mathrm{sd}}t_{\mathrm{sd}}/t^{2} \propto B^{-2}$, i.e., $L_{\mathrm{sd}}$ is actually independent of $P_0$. In Figure \ref{fig:magnetar}, we show the magnetar parameter space ($B$ vs. $P_0$) with lines of constant luminosity for $L_{\mathrm{sd}}=L_{\mathrm{UVO,min}}$, $10L_{\mathrm{UVO,min}}$, $100L_{\mathrm{UVO,min}}$. 

We found that $L_{\mathrm{sd}}=L_{\mathrm{UVO,min}}$ for $B\simeq10^{15}\,\mathrm{G}$, meaning that if the magnetar has an extreme field strength, by these late times, all of its spin-down luminosity is required to power just the \emph{observed} UV-optical luminosity over $\lambda\sim2300-8000\,\mathrm{\AA}$. However, since the spectral peak of the underlying source could be further in the UV ($\lambda_{\mathrm{peak}}\lesssim 2300\,\mathrm{\AA}$), the observed UV-optical luminosity is likely only a small fraction of the total luminosity. Therefore, a magnetar with $B\gtrsim10^{15}\,\mathrm{G}$ cannot power the emission of the underlying source.

Since $L_{\mathrm{sd}}\propto B^{-2}$ at these late times, a larger $L_{\mathrm{sd}}$ can be obtained from a smaller $B$. We found that at $B\simeq10^{14}\,\mathrm{G}$, $L_{\mathrm{sd}}\simeq100L_{\mathrm{UVO,min}}$, the spin-down luminosity is more than two orders of magnitude larger than the observed luminosity. Therefore, a magnetar with $B\lesssim10^{14}\,\mathrm{G}$ could hypothetical have enough energy output to power the emission of the underlying source.

We also show a shaded purple region in Figure \ref{fig:magnetar} to indicate the parameters $B$ and $P_0$ required to produce the energy and timescale associated with the optical peak of AT\,2018cow: $E_{\mathrm{sd}}\sim10^{50.5}-10^{51.5}\,\mathrm{erg}$ and $t_{\mathrm{sd}}\sim10^{3}-10^{5}\,\mathrm{s}$ \citep[ranges from][also see \citealt{2018ApJ...865L...3P} and \citealt{Xiang2021} regarding magnetar parameters]{Margutti2019}. Although a magnetar with $B\sim10^{15}\,\mathrm{G}$ could power the optical peak of AT\,2018cow, we argued above that it does not have sufficient energy output to power the underlying source. On the other hand, a magnetar with a lower field strength $B\sim2-4\times10^{14}\,\mathrm{G}$ may be able to power both emissions if the observed underlying emission $L_{\mathrm{UVO,min}}$ is a at least few percent of the total luminosity.

However, the observed underlying emission could be much less than 1\% of the total luminosity, which would be the case if the emission at $t\simeq703\,\mathrm{days}$ was blackbody (see Section \ref{subsec:undermodel}). In that case, a NS could either power the peak of the AT\,2018cow (with a magnetar) \emph{or} the underlying emission (with a less magnetized pulsar), but not both. A third possibility is that AT\,2018cow did not involve a NS at all. Detailed magnetar models and additional observations of the underlying source are required to further constrain these possibilities.

\subsection{Black Hole Accretion Disk}\label{subsec:underBH}

Many models of AT\,2018cow involve various kinds of winds and jets that could be driven by newborn central engines \citep[e.g.,][]{Piro2020,Gottlieb2022,Metzger2022}. Some studies have also suggested that AT\,2018cow could be a TDE involving an IMBH or a SMBH \citep[][]{Kuin2019,Perley2019}. Following these scenarios, a remnant accretion disk is a natural hypothesis that could explain the slow-evolving, extremely hot and small underlying source. Here, we examine the remnant accretion disk scenario and place constraints on the disk configuration and the central mass by modeling the HST SEDs and the \emph{Swift}-XRT upper limit. Note that in the following sections, although our simple disk model for the underlying source is agnostic to the exact nature of the central object, we refer to the it as a BH. While the central object in principle could be a NS under certain configurations, we prefer a BH in the context of our analysis because (1) much of the mass range we examine (up to $10^{6}\,M_\odot$) are BHs and (2) for masses that can be NSs, the extreme super-Eddington accretion required in our model (up to about $1\,M_\odot\,\mathrm{yr}^{-1}$; see Section \ref{subsubsec:BHMdotMBH} and \ref{subsubsec:summarydisk}) over multiple years would likely provide enough mass to exceed the NS mass limit.

\subsubsection{Multi-Temperature Disk Blackbody Fit}

\begin{figure}
\epsscale{1.2}
\plotone{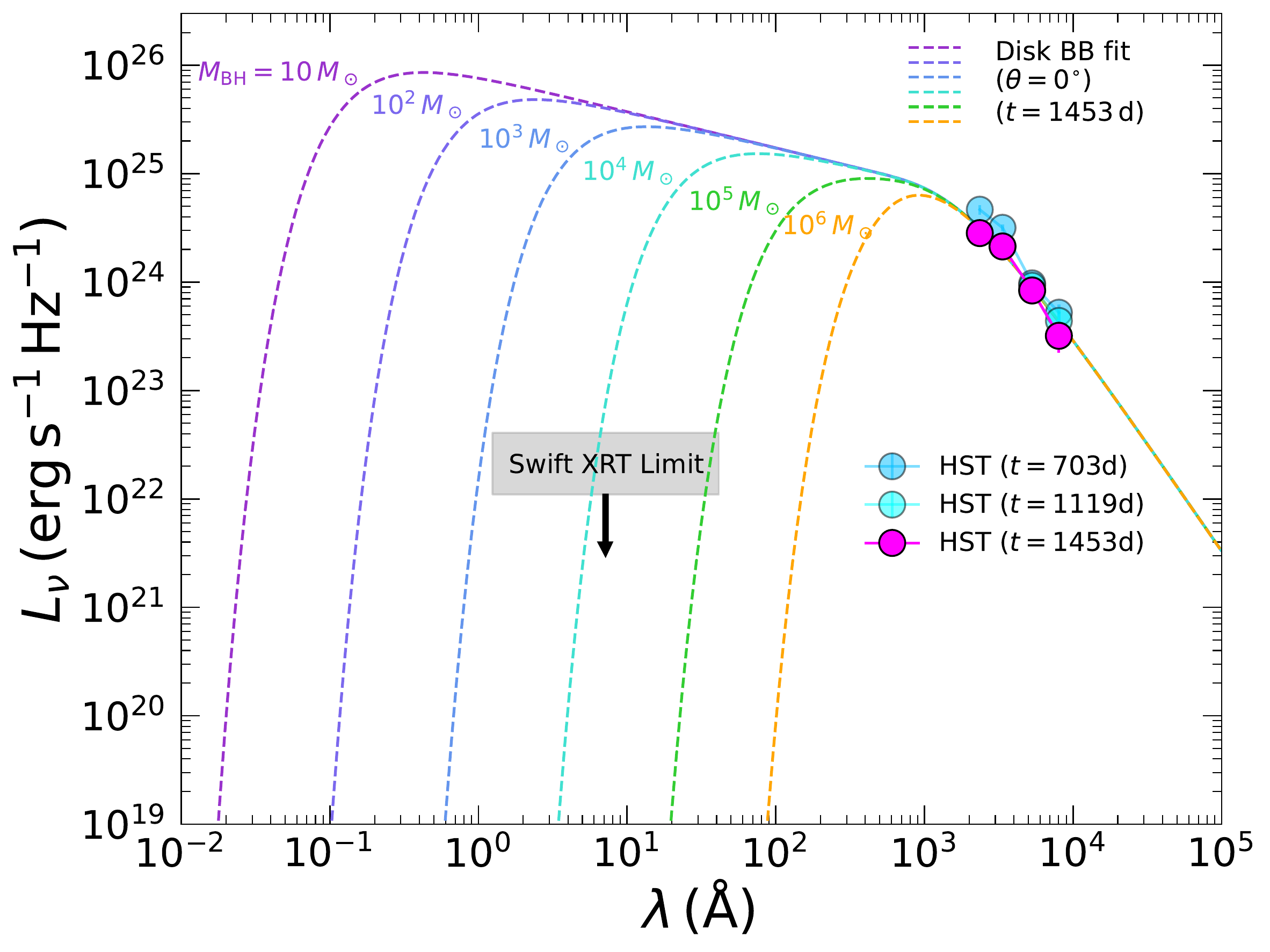}
\caption{Disk blackbody fits to the latest HST SED of the underlying source at $t\simeq1453\,\mathrm{days}$. All three (dereddened) HST SEDs are also shown for reference. The order-of-magnitude estimates of the upper limit $L_\nu$ derived from the \emph{Swift}-XRT non-detection at $t\simeq1360\,\mathrm{days}$ are also shown. The horizontal span of the limit represents the range of $0.3-10\,\mathrm{keV}$, while the vertical span represents the range of $L_{\nu}$ derived from the range of $M_{\mathrm{BH}}$.
\label{fig:diskbbfit22}}
\end{figure}

We fit the latest HST SED of the underlying source ($t\simeq1453\,\mathrm{days}$) using a multi-temperature disk blackbody model \citep[][]{Mitsuda1984,Makishima1986}, which describes a standard geometrically thin accretion disk \citep[][]{ShakuraSunyaev1973,Pringle1981}{}{}. This model assumes a disk with near-Keplerian orbits that is radiatively efficient (cool disk with negligible radial pressure gradient and supersonic orbits) and geometrically thin (small aspect ratio). The disk is also assumed to be optically thick, with each annulus having a local temperature and radiating as a blackbody. We use this simple model as an initial approach to broadly explore the parameter space for the accretion scenario in the context of the underlying source. We note that while the assumptions may be reasonable for sub-Eddington accretion, the thin disk approximation does break down for super-Eddington accretion where the disk is radiatively inefficient and geometrically thick (e.g., \citealt{Paczynski1980,Abramowicz1988}; also see reviews by \citealt{Frank2002,Abramowicz2013}). This caveat becomes relevant for lower BH masses when modeling the underlying source (see Section \ref{subsubsec:BHMdotMBH} and \ref{subsubsec:summarydisk}).

We assumed a temperature profile as a function of radius of $T(r) = (3G\dot{M}M_{\mathrm{BH}}/8\pi\sigma r^3)^{1/4}$, where $G$ is the gravitational constant, $\sigma$ is the Stefan-Boltzmann constant, $\dot{M}$ is the mass accretion rate, and $M_{\mathrm{BH}}$ is the central BH mass. The model disk spectrum is a superposition of the multi-temperature blackbody components, where superposition produces $L_\nu \propto \nu^{1/3}$ with a high energy turnover to the Wien's tail $L_\nu \propto e^{-h\nu/kT}$ associated with the hotter inner edge of the disk and a low energy turnover to the Rayleigh-Jeans tail $L_\nu \propto \nu^2$ associated with the colder outer edge of the disk. 

Since the HST SED at $t\simeq1453\,\mathrm{days}$ was close to the Rayleigh-Jeans tail, the emission would mostly be from the outer edge of the disk, and the fit cannot constrain the inner edge radius (and thus the BH mass). Therefore, we considered a range of BH masses $M_{\mathrm{BH}}=1-10^6\,M_\odot$ and fiducial values of the inclination angle $\theta=0^{\circ},40^{\circ},80^{\circ}$. From this, we derived a range of $\dot{M}$ and $T_{\mathrm{out}} = T(R_{\mathrm{out}})$ from the fits, where $R_{\mathrm{out}}$ is the outer edge radius. Note that we do not fit the SED at $t\simeq703\,\mathrm{days}$ because it is essentially at the Rayleigh-Jeans tail, meaning that even the outer edge radius cannot be well-constrained.

We also utilized the \emph{Swift}-XRT upper limit from $t\simeq1360\,\mathrm{days}$ to place additional constraints on the inner region of the disk which, depending on the configuration, could be hot enough to produce bright X-ray emission. These constraints can have important implications for the BH mass because brighter X-ray emissions are expected from a smaller central mass with a smaller disk inner edge radius (and vice versa). With the upper limit count rate (0.00157\,cts/s), we used the \textsc{xspec} \texttt{fakeit} function and the built-in \texttt{tbabs$\times$diskbb} model to derive the corresponding upper limit $\dot{M}$ and $L_{\mathrm{0.3-10keV}}$ considering the same range of $M_{\mathrm{BH}}$ and fiducial values of $\theta$. For \texttt{fakeit}, we used the standard XRT response files: \texttt{swxs6\_20010101v001.arf} and \texttt{swxpc0to12s6\_20130101v014.rmf}. For \texttt{tbabs}, we assumed a $N_{\mathrm{H}}=0.05\times10^{22}\,\mathrm{cm}^{-2}$, similar to AT\,2018cow \citep[][]{Margutti2019}. For \texttt{diskbb}, we assumed the inner edge radius to be the innermost stable circular orbit for a non-rotating BH and derived the apparent radius following \citet{Kubota1998}.

\begin{figure*}
\gridline{\fig{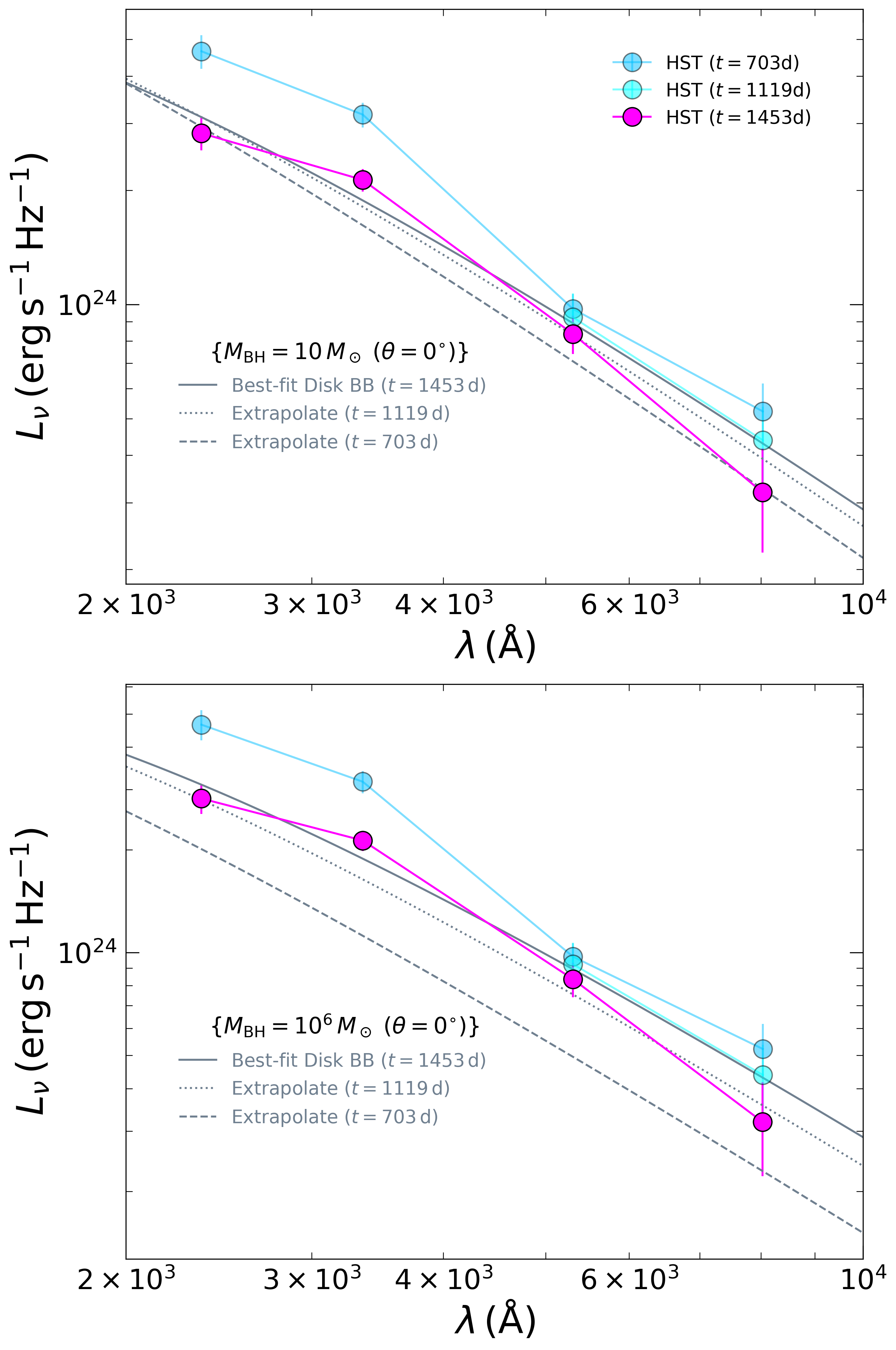}{0.48\textwidth}{(a) Extrapolations with $\theta$ as a fixed parameter}
          \fig{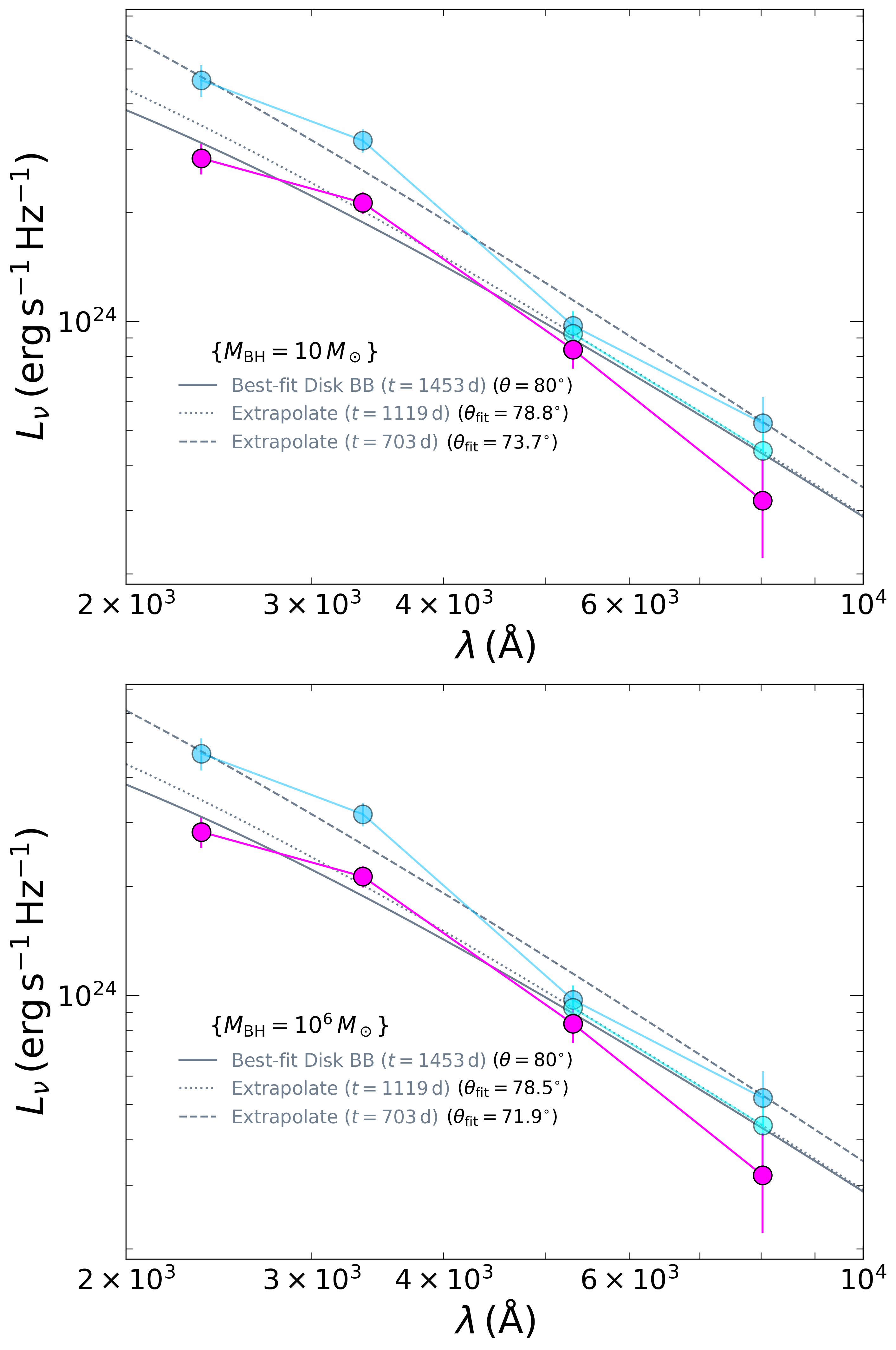}{0.48\textwidth}{(b) Extrapolations with $\theta$ as a free parameter}}
\caption{Extrapolated disk blackbody models at $t\simeq703\,\mathrm{days}$ (dashed lines) and $t\simeq1119\,\mathrm{days}$ (dotted lines) based on the best-fit models at $t\simeq1453\,\mathrm{days}$ (solid lines). The observed (dereddened) HST SEDs are also shown for reference. Top and bottom panels show models for $M_{\mathrm{BH}}=10\,M_\odot$ and $M_{\mathrm{BH}}=10^6\,M_\odot$, respectively. The left panels assume $\theta=0^{\circ}$ for the best-fit models with $\theta$ fixed for the extrapolations. The right panels assume $\theta=80^{\circ}$ for the best-fit models with $\theta$ being a free parameter for the extrapolations and $\theta_{\mathrm{fit}}$ being the best-fit value that match the observed amplitude.}
\label{fig:diskbbExtp}
\end{figure*}

\begin{figure*}
\epsscale{1.0}
\plotone{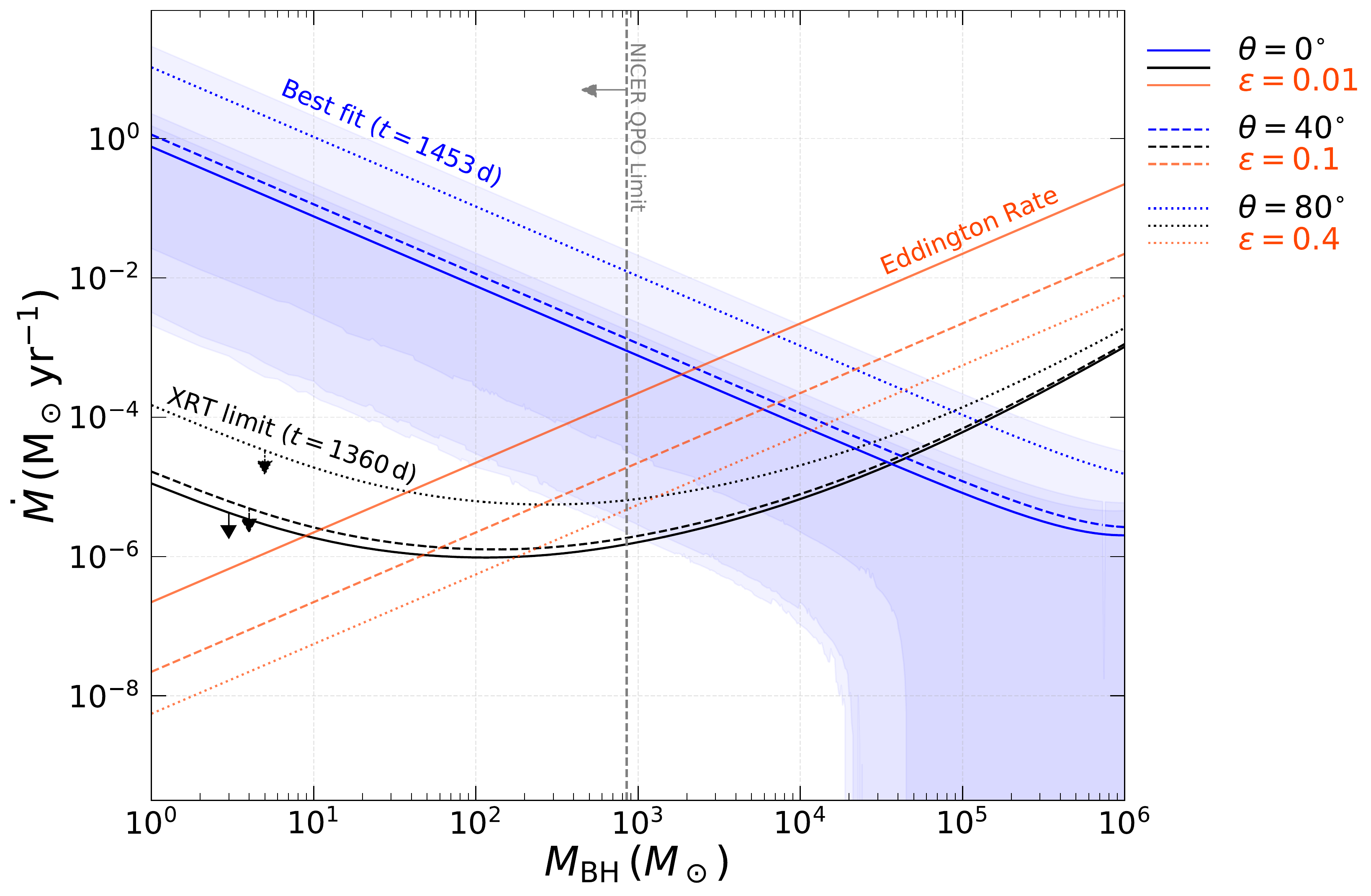}
\caption{Parameter space $\dot{M}$ vs. $M_{\mathrm{BH}}$ for the disk blackbody model representing an accreting BH scenario. $\dot{M}$ values derived from fitting HST SED at $t\simeq1453\,\mathrm{days}$ are shown in blue (errors shown as shaded regions) while values derived from the \emph{Swift}-XRT upper limit are shown in black. Three fiducial $\theta$ values were considered: $0^{\circ}$ (solid lines), $40^{\circ}$ (dashed lines), and $80^{\circ}$ (dotted lines). The Eddington accretion rates are also shown in orange considering three fiducial values of the radiative efficiency $\epsilon$: 0.01 (solid lines), 0.1 (dashed lines), and 0.4 (dotted lines). A vertical dashed line is shown to indicate the limiting $M_{\mathrm{BH}}<850\,M_\odot$ derived from the NICER QPO \citep[][]{Pasham2021}.
\label{fig:diskbbMdotMBH}}
\end{figure*}

In Figure \ref{fig:diskbbfit22}, we show example disk blackbody fits to the latest HST SED assuming $\theta=0^{\circ}$ with $M_{\mathrm{BH}}$ spanning six orders of magnitude. We also show a shaded region indicating the upper limits derived from the \emph{Swift}-XRT non-detection. Note that the limits shown on this plot are for illustrative purposes, calculated assuming $L_{\nu} = L_{\mathrm{0.3-10keV}}/\nu$ with $h\nu = 5.15\,\mathrm{keV}$. 

The disk blackbody can fit the latest HST SED reasonably well for the entire range of $M_{\mathrm{BH}}$. As expected, models with different $M_{\mathrm{BH}}$ are indistinguishable in the optical but predict entirely different X-ray brightness. In particular, the best-fit models predict that the inner regions of the disk around a stellar-mass BH should produce extremely bright X-ray emission: $L_\nu\sim10^{25}-10^{26}\,\mathrm{erg}\,\mathrm{s}^{-1}\,\mathrm{Hz}^{-1}$ over $0.3-10\,\mathrm{keV}$ or $\nu L_\nu \sim 10^{43}-10^{44}\,\mathrm{erg}\,\mathrm{s}^{-1}$ at $5.15\,\mathrm{keV}$. This prediction contradicts the \emph{Swift}-XRT non-detection, which sets an upper limit at $L_\nu \lesssim 10^{22}\,\mathrm{erg}\,\mathrm{s}^{-1}\,\mathrm{Hz}^{-1}$ over $0.3-10\,\mathrm{keV}$. The \emph{Swift}-XRT non-detection is more consistent with best-fit models that involve IMBHs or SMBHs with much larger inner edge radii that predict much fainter X-ray emission. We discuss this further in Section \ref{subsubsec:BHMdotMBH}.

\subsubsection{Predicted Evolution of a Remnant Accretion Disk}

With the best-fit models, we checked to see if the predicted evolution of an accretion disk could explain the evolution of the observed SEDs. Specifically, we followed the predictions of $R_{\mathrm{out}} \propto t^{2/3}$ and $\dot{M}\propto t^{-4/3}$ from self-similar solutions \citep[e.g.,][]{Metzger2008} and extrapolated the best-fit $R_{\mathrm{out}}$ and $\dot{M}$ to the earlier HST epochs, $t=703\,\mathrm{days}$ and $1119\,\mathrm{days}$. We built the extrapolated disk blackbody models based on these parameters.

We show some example comparisons between extrapolated disk blackbody models and observed SEDs in Figure \ref{fig:diskbbExtp}. The best-fit models at $t=1453\,\mathrm{days}$ are shown as solid lines while the extrapolated models are shown as dotted lines at $t=1119\,\mathrm{days}$ and dashed lines at $t=703\,\mathrm{days}$. We show two cases at the two ends of the mass range, $M_{\mathrm{BH}}=10\,M_\odot$ and $M_{\mathrm{BH}}=10^6\,M_\odot$, for comparison at the top and bottom panels, respectively. We perform two sets of fits: one where the inclination angle of the disk is fixed between epochs (examples shown in the left panels with $\theta=0^{\circ}$) and one where we allow the angle to vary between epochs which changes the model normalization (examples shown in the right panels; starting from $\theta=80^{\circ}$), a situation that might correspond to a precessing disk. Fitting the inclination angle yields the required precession to explain the SED amplitude at each epoch.

When $\theta$ is fixed, the extrapolation predicts a lower NUV-optical brightness at earlier epochs (left panels of Figure \ref{fig:diskbbExtp}) because the smaller $R_{\mathrm{out}}$ shifts the turnover (i.e., the outer edge blackbody) to a shorter wavelength, which has a greater effect than the larger $\dot{M}$. Although we only show two cases in Figure \ref{fig:diskbbExtp}, we found this behavior to hold regardless of $M_{\mathrm{BH}}$ and $\theta$. This prediction contradicts the observed HST SEDs, which show a brighter NUV-optical emission at earlier epochs. Therefore, pure extrapolation of $R_{\mathrm{out}}$ and $\dot{M}$ cannot explain the evolution of the HST SEDs of the underlying source.

On the other hand, if $\theta$ can be varied, the extrapolated models can actually match the observed SEDs quite well if the disk was less inclined (by $\sim$2$^{\circ}$--8$^{\circ}$) at the earlier epochs for the case of $\theta=80^{\circ}$ at $t=1453\,\mathrm{days}$ (right panels of Figure \ref{fig:diskbbExtp}). In particular, the color of the extrapolated models are very consistent with the observed color at $t\simeq703\,\mathrm{days}$. Note that the scenario with varying $\theta$ only works for high inclination angles because in this case, a small change in $\theta$ can cause an appreciable change in $\cos \theta$ to change the model amplitude. A high inclination angle is an interesting point in the context of AT\,2018cow because this was also proposed to explain properties such as the receding photosphere and the asymmetric line profiles \citep[][]{Margutti2019}.

\subsubsection{Comparison of Fit Parameters} \label{subsubsec:BHMdotMBH}

In Figure \ref{fig:diskbbMdotMBH}, we show the parameter space $\dot{M}$ vs. $M_{\mathrm{BH}}$ with all $\dot{M}$ values derived in our analyses: from the best-fit models to the latest HST SED (blue lines) and the \emph{Swift}-XRT upper limit (black lines) considering $\theta=0^{\circ},40^{\circ},80^{\circ}$ (solid, dashed, dotted). Additionally, we show the Eddington accretion rate $\dot{M}_{\mathrm{Edd}} = L_{\mathrm{Edd}}/\epsilon c^2$ (orange lines) considering fiducial values for the radiative efficiency $\epsilon=0.01,0.1,0.4$ (solid, dashed, dotted). For comparison, a vertical dashed line is also shown to indicate the limit $M_{\mathrm{BH}}<850\,M_\odot$ derived from the NICER QPO \citep[][]{Pasham2021}. 

This parameter space highlights the discrepancy between the optical and X-ray in the disk blackbody model for stellar-mass BHs. For $M_{\mathrm{BH}}\lesssim100\,M_\odot$, extremely high (super-Eddington) $\dot{M}\sim10^{-2}-10\,M_\odot\,\mathrm{yr}^{-1}$ is required to explain the optical emission, while the \emph{Swift}-XRT non-detection places a limit at $\dot{M}\lesssim 10^{-4}-10^{-6}\,M_\odot\,\mathrm{yr}^{-1}$. The orders of magnitude differences and the required super-Eddington accretion rates are significant challenges to the disk blackbody model if AT\,2018cow involved a stellar-mass BH. This discrepancy disappears at $M_{\mathrm{BH}}\gtrsim10^4\,M_\odot$, the IMBH and SMBH range, where (sub-Eddington) $\dot{M}\sim10^{-4}-10^{-6}\,M_\odot\,\mathrm{yr}^{-1}$ is required to explain the optical emission. Although this finding is consistent with the scenario that AT\,2018cow was a TDE involving an IMBH or a SMBH, this interpretation contradicts the limit $M_{\mathrm{BH}}<850\,M_\odot$ derived by \citep[][]{Pasham2021} from NICER QPOs.

Additional absorption may be another way of explaining the lack of X-ray. We tested this case by multiplying the original \textsc{xspec} model by \texttt{phabs} and found the required $N_{\mathrm{H}}$ such that the best-fit $\dot{M}$ from fitting the HST SED can produce the upper limit \emph{Swift}-XRT count rate (0.00157\,cts/s). For $M_{\mathrm{BH}}\sim10-10^3\,M_\odot$, we found that significant absorption with $N_{\mathrm{H}}\gtrsim 10^{24}-10^{25}\,\mathrm{cm}^{-2}$ is required explain the lack of X-ray. Such high absorption has been observed for some Galactic X-ray binaries \citep[e.g.,][]{Matt2003}{}{} and Compton-thick active galactic nuclei \citep[e.g.,][]{Ricci2015,Marchesi2018}{}{}. We do note that our estimated $N_{\mathrm{H}}$ is highly uncertain since the X-ray spectral shape and emitting mechanism are essentially unconstrained.

\subsubsection{Summary of Implications For Remnant Disk} \label{subsubsec:summarydisk}

Here we summarize our main findings about the viability of the accretion disk model for the underlying source and implications for various transient models.
\begin{itemize}
    \item An accretion disk can naturally explain the high temperature and small size of the underlying source, and a disk blackbody can reasonably fit the observed HST SED at $t=1453\,\mathrm{days}$.
    \item The predicted evolution of an accretion disk (i.e., $R_{\mathrm{out}} \propto t^{2/3}$ and $\dot{M}\propto t^{-4/3}$) is consistent with the color evolution of the underlying source, but inconsistent with the brightness/amplitude evolution. We found that if the inclination angle is large (close to edge-on) at $t=1453\,\mathrm{days}$, then a precessing disk with an increasing inclination angle can explain the brightness/amplitude evolution.
    \item For stellar-mass BHs ($M_{\mathrm{BH}}\lesssim100\,M_\odot$), a super-Eddington accretion rate is required to explain the optical emission but cannot explain the lack of X-ray emission. For IMBHs and SMBHs ($M_{\mathrm{BH}}\gtrsim10^{4}\,M_\odot$), both the optical emission and the lack of X-ray emission can be explained by sub-Eddington accretion rate, but the central mass would violate the limit derived from the NICER QPO, $M_{\mathrm{BH}}<850\,M_\odot$ \citep[][]{Pasham2021}.
\end{itemize}

In the context of transient models, precessing accretion disks are often considered in TDE frameworks where the disk angular momentum vector is misaligned with the BH spin axis \citep[e.g.,][]{Shen2014,Bonnerot2016,Hayasaki2016,Liska2018}, and the precession is due to the Lense-Thirring effect. Intriguingly, the same effect was proposed to explain the zero time lag between the optical and X-ray QPO in AT\,2018cow \citep[][]{Li2022}. Recent TDE simulations have also shown that high inclination angles can lead to significant X-ray reprocessing and bright UV-optical emission \citep[][]{Dai2018,Thomsen2022}. Late-time UV-optical emission have also been observed for TDEs \citep[][]{VanVelzen2019}{}{}, interpreted as emission from viscously spreading accretion disks. In this context, the precessing highly-inclined accretion disk scenario appears to be compatible with the observational constraints of the underlying source.

In terms of the precessing timescale, as an example, if we consider the Bardeen--Petterson effect \citep[][]{BardeenPetterson1975,Hatchett1981,Nelson2000,Fragile2001}{}{} and derive the alignment timescale \citep[][]{Scheuer1996}{}{}, we find
\begin{eqnarray}\label{eq:BPtime}
t_{\mathrm{align}} \approx 147\,\alpha_{0.1}^{1/3} \delta_{0.5}^{2/3} a_{*,0.1}^{2/3}\qquad\qquad\qquad\quad\nonumber \\
\times\left( \frac{M_\mathrm{BH}}{10\,M_\odot} \right) \left( \frac{\dot{M}}{10^{-2}M_\odot\,\mathrm{yr}^{-1}} \right)^{-1}\,\mathrm{yr}
\end{eqnarray}
where $\alpha_{0.1}=\alpha/0.1$ with $\alpha$ being the viscosity parameter, $\delta_{0.5} = \delta/0.5$ with $\delta$ being the disk aspect ratio, and $a_{*,0.1} = a_{*}/0.1$ with $a_*$ being the dimensionless specific angular momentum of the BH. The rate of precession is given by $2\pi/t_{\mathrm{align}}$, which following the equation above can be a few degrees per year for $M_\mathrm{BH}\sim10\,M_\odot$ and $\dot{M}\sim10^{-2}M_\odot\,\mathrm{yr}^{-1}$. However, note that the assumed values for the dimensionless parameters are uncertain.

Perhaps the largest uncertainty in the accretion disk scenario for the underlying source is the mass of the BH involved, which has significant implications on the progenitor system of AT\,2018cow. A stellar-mass BH would likely imply a stellar progenitor for AT\,2018cow (core collapse of single star or binary merger with a BH) but would have to maintain super-Eddington accretion for many years to explain the observed underlying UV-bright emission and not produce bright X-ray emission. We note that if we follow the wind-reprocessed framework for prompt emission of AT\,2018cow as discussed in paper I \citep{Chen2023I} and assume that the accretion rate is on the order of $\dot{M}\sim10^{-1}-10^{1}\,M_\odot\,\mathrm{yr}^{-1}$ over the first two months post-discovery, then a constant decline of $\dot{M}\propto t^{-4/3}$ would predict an accretion rate on the order of $\dot{M}\sim0.5-1.5\times10^{-3}\,M_\odot\,\mathrm{yr}^{-1}$ at $t=1453\,\mathrm{days}$. This is inconsistent with the accretion rates derived from our analysis for stellar-mass BHs using a disk blackbody model ($\dot{M}\sim 0.1-1\,M_\odot\,\mathrm{yr}^{-1}$; see Figure~\ref{fig:diskbbMdotMBH}). We also note that while the stellar-mass BH scenario may produce significant precession over a few years (Equation \ref{eq:BPtime}), it is unclear if core collapse can naturally lead to a misalignment between the remnant disk and the spin of the compact object in the first place to cause the precession.

However, since the thin disk approximation does break down for super-Eddington accretion, the disk blackbody model is likely not appropriate for accretion around stellar-mass BHs and our derived parameters may not be entirely accurate. Properly accounting for the super-Eddington accretion may also help explain the lack of X-ray emission through mechanisms such as the obscuration by a geometrically thick disk or dense outflow \citep[][]{Done2007}. For example, \citet{Metzger2022} constructed a model to explain AT\,2018cow involving super-Eddington accretion from a delayed binary merger between a Wolf-Rayet star and a BH or NS. They predicted the late-time evolution of the accretion disk \citep[Section 2.3.3 in][]{Metzger2022}{}{} with the accretion rate decreasing with radius as a power law due to outflow carrying away angular momentum, which may lower the inner accretion rate by a few orders of magnitude and explain the lack of X-ray. We note that \citet{Metzger2022} also predicted late-time thermal emission from the outer accretion disk, and while their model radius and temperature appear similar to those observed for the underlying source, their model luminosity (being the Eddington luminosity $L_{\mathrm{Edd}}\sim10^{38}M_{\mathrm{BH}}\,\mathrm{erg}\,\mathrm{s}^{-1}$) would require $M_{\mathrm{BH}}\gtrsim10^{4}\,M_\odot$ (i.e., an IMBH) to match the derived blackbody luminosity for the underlying source at $t\simeq703\,\mathrm{days}$ (Table \ref{tab:UnderProperties}). Overall, additional theoretical works are needed to constrain the viability of the accreting stellar-mass BH scenario as an explanation of the underlying source.

On the other hand, a geometrically thin accretion disk around an IMBH or a SMBH with sub-Eddington accretion rate ($\dot{M}\lesssim 10^{-2}\,M_\odot\,\mathrm{yr}^{-1}$) can reasonably explain the UV-bright underlying source. These accretion rate implied for an IMBH would also be roughly consistent with the wind-reprocessed framework for the prompt emission of AT\,2018cow that follow $\dot{M}\propto t^{-4/3}$, though the accretion rate may not be enough to cause significant precession (Equation \ref{eq:BPtime}). The IMBH or SMBH scenario would imply that AT\,2018cow was a TDE. Previous studies have disfavored the TDE hypothesis for various reasons, most notably the difficulty in explaining the dense CSM \citep[][]{Margutti2019,Huang2019}{}{}, the existence of such BH at the outskirt of the galaxy without any signs of a coincident massive host system \citep[][]{Lyman2020}{}{}, and the mass limit $M_{\mathrm{BH}}<850\,M_\odot$ derived from the NICER QPO \citep[][]{Pasham2021}. However, some evidence may suggest otherwise. For example, environmental studies \citep[][]{Roychowdhury2019,Lyman2020} have noted that a faint tidal tail in the host galaxy (also see Figure \ref{fig:CompImg}) that traces star-forming activities around AT\,2018cow could be evidence of recent dynamical interaction. One may speculate that there could be a chance that a straggling IMBH or SMBH was left behind from this dynamical interaction. Another example is the low-frequency X-ray QPO discovered by \citet{Zhang2022} from XMM-Newton and \emph{Swift} observations. \citet{Zhang2022} suggested that the low-frequency QPO is more consistent with IMBHs and SMBHs, and the NICER QPO limit could be relaxed by introducing the IMBH or SMBH in a compact binary. Therefore, the TDE scenario for AT\,2018cow may be worth revisiting along with the hypothesis that the underlying source was an accreting IMBH or SMBH.

\section{Summary \& Conclusion}\label{sec:conclusion}

In this study, we examined the UV-bright transient underlying source at the precise position of AT\,2018cow revealed by the three HST observations taken $\sim$2--4 years post-discovery ($t\simeq703,\,1119,\,1453\,\mathrm{days}$). The HST observation at $t\simeq1453\,\mathrm{days}$, which we requested after independently discovering the underlying source, showed significant fading in the UV bands relative to the observations at $t\simeq703$ days (Figure \ref{fig:underlyingSED}). This establishes the transient nature of the source, which could be cause either by an intrinsic (i.e., emission associated with AT\,2018cow) and/or extrinsic (i.e., increased absorption along the line of sight) effect. 

The underlying source is bright ($L_{\mathrm{UVO,min}}\sim10^{39}\,\mathrm{erg}\,\mathrm{s}^{-1}$) and exceptionally blue ($\mathrm{F336W}-\mathrm{F555W}=-1.3$) with an unconstrained peak further in the UV ($\lambda_{\mathrm{peak}}\lesssim 2358\,\mathrm{\AA}$). The blue spectrum at $t\simeq703\,\mathrm{days}$ can be described by a spectral index of $\alpha=1.99$ (similar to the Rayleigh-Jeans tail) or by a blackbody with a high temperature ($T\gtrsim10^{5}\,\mathrm{K}$) and a small radius ($R \lesssim 20\,R_\odot$). A flatter spectrum at $t\simeq1453\,\mathrm{days}$ with $\alpha=1.66$ could be an indication of of cooling (to $T\sim6\times10^4\,\mathrm{K}$) and expansion (to $R\sim30\,R_\odot$) of a blackbody, or an increase in extinction (with a color excess of $E_{\mathrm{B}-\mathrm{V}}\simeq0.072$) assuming the Cardelli extinction law \citep{1989ApJ...345..245C} with $R_{\mathrm{V}}=3.1$.

We considered five origins of the properties and evolution of this peculiar UV-bright underlying source: (i) significant contribution from a star cluster, (ii) increased extinction from newly-formed dust along the line of sight (iii) ejecta-CSM interaction, (iv) magnetar spin down, and (v) a remnant accretion disk around a BH. We disfavored significant contribution from a star cluster based on comparisons with BPASS and LEGUS clusters because for this scenario to work, the underlying source has to contain both an extremely young cluster \emph{and} a transient source bluer than the Rayleigh-Jeans tail. We found that although dust formation appears reasonable in the context of AT\,2018cow, the fading was unlikely purely due to dust extinction because of the extremely blue color already observed. 

We additionally ruled out ejecta-CSM interaction involving the known radio-producing CSM from modeling the expected radiation, and magnetar spin down with $B\sim10^{15}\,\mathrm{G}$ based on the energy output. However, we cannot rule out ejecta-CSM interaction involving a denser CSM component (e.g., a previously ejected hydrogen-rich envelope) or magnetar spin down with $B\lesssim10^{14}\,\mathrm{G}$, and additional modeling would be required to constrain these possibilities. 

Finally, we found that a precessing accretion disk at a high inclination angle can reasonably explain the color, brightness, and evolution of the HST SEDs. However, a major uncertainty is the type of BH at the center of the accretion disk. A stellar-mass BH would require super-Eddington accretion over multiple years with an accretion rate possibly declining slower than the predicted $\dot{M}\propto t^{-4/3}$ and additional mechanisms to explain the lack of X-ray emission. On the other hand, while an IMBH or a SMBH could naturally explain the lack of associate X-ray emission with an inferred accretion rate consistent with the wind-reprocessed framework for AT\,2018cow \citep[paper I;][]{Chen2023I} that follows $\dot{M}\propto t^{-4/3}$, this would appear to violate the limit of $M_{\mathrm{BH}}<850\,M_\odot$ from the NICER QPO \citep[][]{Pasham2021} and its existence at the location of AT\,2018cow would still be difficult to explain.

Putting together all the pieces, including results from paper I \citep{Chen2023I}, we find that central engine and ejecta-CSM interaction are still the preferred power sources that can coherently explain both the luminous FBOT AT\,2018cow and the remnant UV-bright slow-evolving transient underlying source. However, we note that for ejecta-CSM interaction to fully explain the observations, multiple CSM components are necessary: (I) a dense CSM shell ($\dot{M}\sim1\,M_\odot\,\mathrm{yr}^{-1}$) is required to power the fast-rising luminous peak of AT\,2018cow at $t\sim1\,\mathrm{day}$ \citep[][]{Xiang2021,Pellegrino2022}, (II) dense aspherical CSM (unknown density and distribution) is required to sustain the optically thick rapidly-fading prompt emission over $t\sim20-60\,\mathrm{days}$, (III) relatively less-dense CSM ($\dot{M}\sim10^{-6}-10^{-4}\,M_\odot\,\mathrm{yr}^{-1}$) is required to power the radio emission up to $t\sim600\,\mathrm{days}$ \citep[][]{Ho2019,Nayana2021}, and (IV) dense and extended CSM (unknown density and distribution) farther away from the transient ($R\gtrsim10^{17}\,\mathrm{cm}$), likely the previously ejected hydrogen envelope, is required to power the UV-bright slow-fading transient over $t\sim 700-1500\,\mathrm{days}$. While these CSM components could hypothetically exist, all but the radio-producing CSM are poorly constrained both observationally and theoretically. Additional theoretical works, and perhaps follow-up observations for the underlying source, can help constrain or rule out possible CSM components.

In the context of our analyses, we favor the accreting BH scenario because some of the expected phenomena from this scenario can reasonably explain the observations of AT\,2018cow and the underlying source. As shown in paper I \citep{Chen2023I}, the fading prompt emission and the associated peculiar thermal properties can be explained by continuous wind outflow driven by an accreting central engine, and as argued in this paper, the evolution of the remnant accretion disk can naturally give rise to the underlying source. This would support the hypothesis that AT\,2018cow and the class of luminous FBOTs may form entirely new class of BH transients powered predominantly by accretion. However, there are challenges faced by this interpretation, such as the requirement of disk precession and the uncertainty in the mass of the BH involved. These challenges could either be counterarguments to this hypothesis or interesting constraints that may contain new information on AT\,2018cow and BH transients. Late-time evolution of transient accretion disks could be an interesting topic for future theoretical studies not only to examine the underlying source of AT\,2018cow but also to explore potential transient phenomena related to accreting BHs and potentially reveal new classes of UV-bright BH transients similar to the underlying source.

Our studies highlight the importance of late-time observations, which for AT\,2018cow, provided significant constraints on the late thermal properties \citep[paper I;][]{Chen2023I} and led to the discovery of an unprecedented underlying transient for an FBOT years post-discovery (this paper). These merits justify similar late-time observations and monitoring for nearby peculiar transients through powerful telescopes such as the HST and JWST. For new discoveries of nearby ``Cow-like transients'', late-time observations will be crucial in probing remnant emission possibly associated with ejecta-CSM interaction or an accretion disk and if the host galaxy experienced similar dynamical interaction that could leave behind a straggling IMBH or SMBH.

Lastly, we mention that our study demonstrates the need for next-generation UV telescopes because a UV-bright transient such as the underlying source, which was not explicitly predicted by previous studies for AT\,2018cow, was barely recognized through the HST. Without NUV monitoring from the HST, the underlying transient source would have been completely missed based on the optical and (the lack of) X-ray emission. Next-generation UV telescopes will be crucial for expanding the transient phase space and exploring potential new UV transients such as the underlying source of AT\,2018cow.

\begin{acknowledgments}

We thank the anonymous referee for comments and suggestions that improved this manuscript. Y.C. thanks Christopher D. Matzner for helpful discussions.

Y.C. acknowledges support from the Natural
Sciences and Engineering Research Council of Canada (NSERC) Canada Graduate Scholarships -- Doctoral Program. M.R.D. acknowledges support from NSERC through grant RGPIN-2019-06186, the Canada Research Chairs Program, the Canadian Institute for Advanced Research (CIFAR), and the Dunlap Institute at the University of Toronto. C.D.K. is supported in part by a Center for Interdisciplinary
Exploration and Research in Astrophysics (CIERA) postdoctoral fellowship.

Parts of this research is based on observations made with the NASA/ESA Hubble Space Telescope obtained from the Space Telescope Science Institute, which is operated by the Association of Universities for Research in Astronomy, Inc., under NASA contract NAS 5–26555. The HST observations are associated with programs 15974, 16179, 16925 and can be accessed via \dataset[10.17909/fmz6-9b21]{http://dx.doi.org/10.17909/fmz6-9b21}. We acknowledge the use of public data from the Swift data archive. 

\end{acknowledgments}

\vspace{5mm}
\facilities{HST (UVIS), Swift (XRT)}

\software{Astropy \citep{2013A&A...558A..33A,2018AJ....156..123A}, \texttt{emcee} \citep{2013PASP..125..306F}, \textsc{xspec} \citep[][]{Arnaud1996XSPEC}, {\tt dolphot} \citep{dolphot}, \texttt{hotpants} \citep{becker2015}, HEASoft, Hoki \citep[][]{Stevance2020hoki}, SciPy \citep{2020scipy}, photutils \citep{photutils}.}

\bibliography{ref}{}

\begin{thebibliography}{}
\expandafter\ifx\csname natexlab\endcsname\relax\def\natexlab#1{#1}\fi
\providecommand{\url}[1]{\href{#1}{#1}}
\providecommand{\dodoi}[1]{doi:~\href{http://doi.org/#1}{\nolinkurl{#1}}}
\providecommand{\doeprint}[1]{\href{http://ascl.net/#1}{\nolinkurl{http://ascl.net/#1}}}
\providecommand{\doarXiv}[1]{\href{https://arxiv.org/abs/#1}{\nolinkurl{https://arxiv.org/abs/#1}}}

\bibitem[{{Abramowicz} {et~al.}(1988){Abramowicz}, {Czerny}, {Lasota}, \&
  {Szuszkiewicz}}]{Abramowicz1988}
{Abramowicz}, M.~A., {Czerny}, B., {Lasota}, J.~P., \& {Szuszkiewicz}, E. 1988,
  \apj, 332, 646, \dodoi{10.1086/166683}

\bibitem[{{Abramowicz} \& {Fragile}(2013)}]{Abramowicz2013}
{Abramowicz}, M.~A., \& {Fragile}, P.~C. 2013, Living Reviews in Relativity,
  16, 1, \dodoi{10.12942/lrr-2013-1}

\bibitem[{{Adamo} {et~al.}(2017){Adamo}, {Ryon}, {Messa}, {Kim}, {Grasha},
  {Cook}, {Calzetti}, {Lee}, {Whitmore}, {Elmegreen}, {Ubeda}, {Smith},
  {Bright}, {Runnholm}, {Andrews}, {Fumagalli}, {Gouliermis}, {Kahre}, {Nair},
  {Thilker}, {Walterbos}, {Wofford}, {Aloisi}, {Ashworth}, {Brown}, {Chandar},
  {Christian}, {Cignoni}, {Clayton}, {Dale}, {de Mink}, {Dobbs}, {Elmegreen},
  {Evans}, {Gallagher}, {Grebel}, {Herrero}, {Hunter}, {Johnson}, {Kennicutt},
  {Krumholz}, {Lennon}, {Levay}, {Martin}, {Nota}, {{\"O}stlin}, {Pellerin},
  {Prieto}, {Regan}, {Sabbi}, {Sacchi}, {Schaerer}, {Schiminovich}, {Shabani},
  {Tosi}, {Van Dyk}, \& {Zackrisson}}]{Adamo2017LEGUScluster}
{Adamo}, A., {Ryon}, J.~E., {Messa}, M., {et~al.} 2017, \apj, 841, 131,
  \dodoi{10.3847/1538-4357/aa7132}

\bibitem[{{Afsariardchi} {et~al.}(2021){Afsariardchi}, {Drout}, {Khatami},
  {Matzner}, {Moon}, \& {Ni}}]{Afsariardchi2021}
{Afsariardchi}, N., {Drout}, M.~R., {Khatami}, D.~K., {et~al.} 2021, \apj, 918,
  89, \dodoi{10.3847/1538-4357/ac0aeb}

\bibitem[{{Arcavi} {et~al.}(2016){Arcavi}, {Wolf}, {Howell}, {Bildsten},
  {Leloudas}, {Hardin}, {Prajs}, {Perley}, {Svirski}, {Gal-Yam}, {Katz},
  {McCully}, {Cenko}, {Lidman}, {Sullivan}, {Valenti}, {Astier}, {Balland},
  {Carlberg}, {Conley}, {Fouchez}, {Guy}, {Pain}, {Palanque-Delabrouille},
  {Perrett}, {Pritchet}, {Regnault}, {Rich}, \&
  {Ruhlmann-Kleider}}]{2016ApJ...819...35A}
{Arcavi}, I., {Wolf}, W.~M., {Howell}, D.~A., {et~al.} 2016, \apj, 819, 35,
  \dodoi{10.3847/0004-637X/819/1/35}

\bibitem[{{Arnaud}(1996)}]{Arnaud1996XSPEC}
{Arnaud}, K.~A. 1996, in Astronomical Society of the Pacific Conference Series,
  Vol. 101, Astronomical Data Analysis Software and Systems V, ed. G.~H.
  {Jacoby} \& J.~{Barnes}, 17

\bibitem[{{Astropy Collaboration} {et~al.}(2013){Astropy Collaboration},
  {Robitaille}, {Tollerud}, {Greenfield}, {Droettboom}, {Bray}, {Aldcroft},
  {Davis}, {Ginsburg}, {Price-Whelan}, {Kerzendorf}, {Conley}, {Crighton},
  {Barbary}, {Muna}, {Ferguson}, {Grollier}, {Parikh}, {Nair}, {Unther},
  {Deil}, {Woillez}, {Conseil}, {Kramer}, {Turner}, {Singer}, {Fox}, {Weaver},
  {Zabalza}, {Edwards}, {Azalee Bostroem}, {Burke}, {Casey}, {Crawford},
  {Dencheva}, {Ely}, {Jenness}, {Labrie}, {Lim}, {Pierfederici}, {Pontzen},
  {Ptak}, {Refsdal}, {Servillat}, \& {Streicher}}]{2013A&A...558A..33A}
{Astropy Collaboration}, {Robitaille}, T.~P., {Tollerud}, E.~J., {et~al.} 2013,
  \aap, 558, A33, \dodoi{10.1051/0004-6361/201322068}

\bibitem[{{Astropy Collaboration} {et~al.}(2018){Astropy Collaboration},
  {Price-Whelan}, {Sip{\H{o}}cz}, {G{\"u}nther}, {Lim}, {Crawford}, {Conseil},
  {Shupe}, {Craig}, {Dencheva}, {Ginsburg}, {VanderPlas}, {Bradley},
  {P{\'e}rez-Su{\'a}rez}, {de Val-Borro}, {Aldcroft}, {Cruz}, {Robitaille},
  {Tollerud}, {Ardelean}, {Babej}, {Bach}, {Bachetti}, {Bakanov}, {Bamford},
  {Barentsen}, {Barmby}, {Baumbach}, {Berry}, {Biscani}, {Boquien}, {Bostroem},
  {Bouma}, {Brammer}, {Bray}, {Breytenbach}, {Buddelmeijer}, {Burke},
  {Calderone}, {Cano Rodr{\'\i}guez}, {Cara}, {Cardoso}, {Cheedella}, {Copin},
  {Corrales}, {Crichton}, {D'Avella}, {Deil}, {Depagne}, {Dietrich}, {Donath},
  {Droettboom}, {Earl}, {Erben}, {Fabbro}, {Ferreira}, {Finethy}, {Fox},
  {Garrison}, {Gibbons}, {Goldstein}, {Gommers}, {Greco}, {Greenfield},
  {Groener}, {Grollier}, {Hagen}, {Hirst}, {Homeier}, {Horton}, {Hosseinzadeh},
  {Hu}, {Hunkeler}, {Ivezi{\'c}}, {Jain}, {Jenness}, {Kanarek}, {Kendrew},
  {Kern}, {Kerzendorf}, {Khvalko}, {King}, {Kirkby}, {Kulkarni}, {Kumar},
  {Lee}, {Lenz}, {Littlefair}, {Ma}, {Macleod}, {Mastropietro}, {McCully},
  {Montagnac}, {Morris}, {Mueller}, {Mumford}, {Muna}, {Murphy}, {Nelson},
  {Nguyen}, {Ninan}, {N{\"o}the}, {Ogaz}, {Oh}, {Parejko}, {Parley}, {Pascual},
  {Patil}, {Patil}, {Plunkett}, {Prochaska}, {Rastogi}, {Reddy Janga},
  {Sabater}, {Sakurikar}, {Seifert}, {Sherbert}, {Sherwood-Taylor}, {Shih},
  {Sick}, {Silbiger}, {Singanamalla}, {Singer}, {Sladen}, {Sooley},
  {Sornarajah}, {Streicher}, {Teuben}, {Thomas}, {Tremblay}, {Turner},
  {Terr{\'o}n}, {van Kerkwijk}, {de la Vega}, {Watkins}, {Weaver}, {Whitmore},
  {Woillez}, {Zabalza}, \& {Astropy Contributors}}]{2018AJ....156..123A}
{Astropy Collaboration}, {Price-Whelan}, A.~M., {Sip{\H{o}}cz}, B.~M., {et~al.}
  2018, \aj, 156, 123, \dodoi{10.3847/1538-3881/aabc4f}

\bibitem[{{Bardeen} \& {Petterson}(1975)}]{BardeenPetterson1975}
{Bardeen}, J.~M., \& {Petterson}, J.~A. 1975, \apjl, 195, L65,
  \dodoi{10.1086/181711}

\bibitem[{{Becker}(2015)}]{becker2015}
{Becker}, A. 2015, {HOTPANTS: High Order Transform of PSF ANd Template
  Subtraction}.
\newblock \doeprint{1504.004}

\bibitem[{{Bietenholz} {et~al.}(2020){Bietenholz}, {Margutti}, {Coppejans},
  {Alexander}, {Argo}, {Bartel}, {Eftekhari}, {Milisavljevic}, {Terreran}, \&
  {Berger}}]{Bietenholz2020}
{Bietenholz}, M.~F., {Margutti}, R., {Coppejans}, D., {et~al.} 2020, \mnras,
  491, 4735, \dodoi{10.1093/mnras/stz3249}

\bibitem[{{Bonnerot} {et~al.}(2016){Bonnerot}, {Rossi}, {Lodato}, \&
  {Price}}]{Bonnerot2016}
{Bonnerot}, C., {Rossi}, E.~M., {Lodato}, G., \& {Price}, D.~J. 2016, \mnras,
  455, 2253, \dodoi{10.1093/mnras/stv2411}

\bibitem[{Bradley {et~al.}(2020)Bradley, Sip{\H o}cz, Robitaille, Tollerud,
  Vin{\'{\i}}cius, Deil, Barbary, Wilson, Busko, G{\"u}nther, Cara, Conseil,
  Bostroem, Droettboom, Bray, Bratholm, Lim, Barentsen, Craig, Pascual, Perren,
  Greco, Donath, de~Val-Borro, Kerzendorf, Bach, Weaver, D'Eugenio, Souchereau,
  \& Ferreira}]{photutils}
Bradley, L., Sip{\H o}cz, B., Robitaille, T., {et~al.} 2020, astropy/photutils:
  1.0.0, 1.0.0,  Zenodo, \dodoi{10.5281/zenodo.4044744}

\bibitem[{{Brethauer} {et~al.}(2022){Brethauer}, {Margutti}, {Milisavljevic},
  {Bietenholz}, {Chornock}, {Coppejans}, {Colle}, {Hajela}, {Terreran},
  {Vargas}, {DeMarchi}, {Harris}, {Jacobson-Gal{\'a}n}, {Kamble}, {Patnaude},
  \& {Stroh}}]{Brethauer2022}
{Brethauer}, D., {Margutti}, R., {Milisavljevic}, D., {et~al.} 2022, \apj, 939,
  105, \dodoi{10.3847/1538-4357/ac8b14}

\bibitem[{{Bright} {et~al.}(2022){Bright}, {Margutti}, {Matthews}, {Brethauer},
  {Coppejans}, {Wieringa}, {Metzger}, {DeMarchi}, {Laskar}, {Romero},
  {Alexander}, {Horesh}, {Migliori}, {Chornock}, {Berger}, {Bietenholz},
  {Devlin}, {Dicker}, {Jacobson-Gal{\'a}n}, {Mason}, {Milisavljevic}, {Motta},
  {Mroczkowski}, {Ramirez-Ruiz}, {Rhodes}, {Sarazin}, {Sfaradi}, \&
  {Sievers}}]{Bright2022Camel}
{Bright}, J.~S., {Margutti}, R., {Matthews}, D., {et~al.} 2022, \apj, 926, 112,
  \dodoi{10.3847/1538-4357/ac4506}

\bibitem[{{Burrows} {et~al.}(2005){Burrows}, {Hill}, {Nousek}, {Kennea},
  {Wells}, {Osborne}, {Abbey}, {Beardmore}, {Mukerjee}, {Short}, {Chincarini},
  {Campana}, {Citterio}, {Moretti}, {Pagani}, {Tagliaferri}, {Giommi},
  {Capalbi}, {Tamburelli}, {Angelini}, {Cusumano}, {Br{\"a}uninger}, {Burkert},
  \& {Hartner}}]{Burrows2005XRT}
{Burrows}, D.~N., {Hill}, J.~E., {Nousek}, J.~A., {et~al.} 2005, \ssr, 120,
  165, \dodoi{10.1007/s11214-005-5097-2}

\bibitem[{{Calzetti} {et~al.}(2015){Calzetti}, {Lee}, {Sabbi}, {Adamo},
  {Smith}, {Andrews}, {Ubeda}, {Bright}, {Thilker}, {Aloisi}, {Brown},
  {Chandar}, {Christian}, {Cignoni}, {Clayton}, {da Silva}, {de Mink}, {Dobbs},
  {Elmegreen}, {Elmegreen}, {Evans}, {Fumagalli}, {Gallagher}, {Gouliermis},
  {Grebel}, {Herrero}, {Hunter}, {Johnson}, {Kennicutt}, {Kim}, {Krumholz},
  {Lennon}, {Levay}, {Martin}, {Nair}, {Nota}, {{\"O}stlin}, {Pellerin},
  {Prieto}, {Regan}, {Ryon}, {Schaerer}, {Schiminovich}, {Tosi}, {Van Dyk},
  {Walterbos}, {Whitmore}, \& {Wofford}}]{Calzetti2015LEGUS}
{Calzetti}, D., {Lee}, J.~C., {Sabbi}, E., {et~al.} 2015, \aj, 149, 51,
  \dodoi{10.1088/0004-6256/149/2/51}

\bibitem[{{Cardelli} {et~al.}(1989){Cardelli}, {Clayton}, \&
  {Mathis}}]{1989ApJ...345..245C}
{Cardelli}, J.~A., {Clayton}, G.~C., \& {Mathis}, J.~S. 1989, \apj, 345, 245,
  \dodoi{10.1086/167900}

\bibitem[{{Chandra} {et~al.}(2015){Chandra}, {Chevalier}, {Chugai}, {Fransson},
  \& {Soderberg}}]{Chandra2015}
{Chandra}, P., {Chevalier}, R.~A., {Chugai}, N., {Fransson}, C., \&
  {Soderberg}, A.~M. 2015, \apj, 810, 32, \dodoi{10.1088/0004-637X/810/1/32}

\bibitem[{{Chen} {et~al.}(2023){Chen}, {Drout}, {Piro}, {Kilpatrick}, {Foley},
  {Rojas-Bravo}, {Taggart}, {Siebert}, \& {Magee}}]{Chen2023I}
{Chen}, Y., {Drout}, M.~R., {Piro}, A.~L., {et~al.} 2023, arXiv e-prints,
  arXiv:2303.03500, \dodoi{10.48550/arXiv.2303.03500}

\bibitem[{{Chevalier}(1982)}]{Chevalier1982}
{Chevalier}, R.~A. 1982, \apj, 259, 302, \dodoi{10.1086/160167}

\bibitem[{{Chevalier} \& {Fransson}(1994)}]{Chevalier1994}
{Chevalier}, R.~A., \& {Fransson}, C. 1994, \apj, 420, 268,
  \dodoi{10.1086/173557}

\bibitem[{{Cohen} \& {Soker}(2023)}]{Cohen2023}
{Cohen}, T., \& {Soker}, N. 2023, \mnras, 522, 885,
  \dodoi{10.1093/mnras/stad1015}

\bibitem[{{Coppejans} {et~al.}(2020){Coppejans}, {Margutti}, {Terreran},
  {Nayana}, {Coughlin}, {Laskar}, {Alexander}, {Bietenholz}, {Caprioli},
  {Chandra}, {Drout}, {Frederiks}, {Frohmaier}, {Hurley}, {Kochanek},
  {MacLeod}, {Meisner}, {Nugent}, {Ridnaia}, {Sand}, {Svinkin}, {Ward}, {Yang},
  {Baldeschi}, {Chilingarian}, {Dong}, {Esquivia}, {Fong}, {Guidorzi},
  {Lundqvist}, {Milisavljevic}, {Paterson}, {Reichart}, {Shappee}, {Stroh},
  {Valenti}, {Zauderer}, \& {Zhang}}]{Coppejans2020}
{Coppejans}, D.~L., {Margutti}, R., {Terreran}, G., {et~al.} 2020, \apjl, 895,
  L23, \dodoi{10.3847/2041-8213/ab8cc7}

\bibitem[{{Dai} {et~al.}(2018){Dai}, {McKinney}, {Roth}, {Ramirez-Ruiz}, \&
  {Miller}}]{Dai2018}
{Dai}, L., {McKinney}, J.~C., {Roth}, N., {Ramirez-Ruiz}, E., \& {Miller},
  M.~C. 2018, \apjl, 859, L20, \dodoi{10.3847/2041-8213/aab429}

\bibitem[{{Dolphin}(2016)}]{dolphot}
{Dolphin}, A. 2016, {DOLPHOT: Stellar photometry}, Astrophysics Source Code
  Library, record ascl:1608.013.
\newblock \doeprint{1608.013}

\bibitem[{{Done} {et~al.}(2007){Done}, {Gierli{\'n}ski}, \&
  {Kubota}}]{Done2007}
{Done}, C., {Gierli{\'n}ski}, M., \& {Kubota}, A. 2007, \aapr, 15, 1,
  \dodoi{10.1007/s00159-007-0006-1}

\bibitem[{{Draine} \& {Lee}(1984)}]{1984ApJ...285...89D}
{Draine}, B.~T., \& {Lee}, H.~M. 1984, \apj, 285, 89, \dodoi{10.1086/162480}

\bibitem[{{Drout} {et~al.}(2014){Drout}, {Chornock}, {Soderberg}, {Sanders},
  {McKinnon}, {Rest}, {Foley}, {Milisavljevic}, {Margutti}, {Berger},
  {Calkins}, {Fong}, {Gezari}, {Huber}, {Kankare}, {Kirshner}, {Leibler},
  {Lunnan}, {Mattila}, {Marion}, {Narayan}, {Riess}, {Roth}, {Scolnic},
  {Smartt}, {Tonry}, {Burgett}, {Chambers}, {Hodapp}, {Jedicke}, {Kaiser},
  {Magnier}, {Metcalfe}, {Morgan}, {Price}, \& {Waters}}]{2014ApJ...794...23D}
{Drout}, M.~R., {Chornock}, R., {Soderberg}, A.~M., {et~al.} 2014, \apj, 794,
  23, \dodoi{10.1088/0004-637X/794/1/23}

\bibitem[{{Eldridge} {et~al.}(2017){Eldridge}, {Stanway}, {Xiao}, {McClelland},
  {Taylor}, {Ng}, {Greis}, \& {Bray}}]{Eldridge2017}
{Eldridge}, J.~J., {Stanway}, E.~R., {Xiao}, L., {et~al.} 2017, \pasa, 34,
  e058, \dodoi{10.1017/pasa.2017.51}

\bibitem[{{Fang} {et~al.}(2019){Fang}, {Metzger}, {Murase}, {Bartos}, \&
  {Kotera}}]{Fang2019}
{Fang}, K., {Metzger}, B.~D., {Murase}, K., {Bartos}, I., \& {Kotera}, K. 2019,
  \apj, 878, 34, \dodoi{10.3847/1538-4357/ab1b72}

\bibitem[{{Foreman-Mackey} {et~al.}(2013){Foreman-Mackey}, {Hogg}, {Lang}, \&
  {Goodman}}]{2013PASP..125..306F}
{Foreman-Mackey}, D., {Hogg}, D.~W., {Lang}, D., \& {Goodman}, J. 2013, \pasp,
  125, 306, \dodoi{10.1086/670067}

\bibitem[{{Fragile} {et~al.}(2001){Fragile}, {Mathews}, \&
  {Wilson}}]{Fragile2001}
{Fragile}, P.~C., {Mathews}, G.~J., \& {Wilson}, J.~R. 2001, \apj, 553, 955,
  \dodoi{10.1086/320990}

\bibitem[{{Frank} {et~al.}(2002){Frank}, {King}, \& {Raine}}]{Frank2002}
{Frank}, J., {King}, A., \& {Raine}, D.~J. 2002, {Accretion Power in
  Astrophysics: Third Edition}

\bibitem[{{Fransson} {et~al.}(2014){Fransson}, {Ergon}, {Challis}, {Chevalier},
  {France}, {Kirshner}, {Marion}, {Milisavljevic}, {Smith}, {Bufano},
  {Friedman}, {Kangas}, {Larsson}, {Mattila}, {Benetti}, {Chornock}, {Czekala},
  {Soderberg}, \& {Sollerman}}]{Fransson2014}
{Fransson}, C., {Ergon}, M., {Challis}, P.~J., {et~al.} 2014, \apj, 797, 118,
  \dodoi{10.1088/0004-637X/797/2/118}

\bibitem[{{Fujibayashi} {et~al.}(2022){Fujibayashi}, {Sekiguchi}, {Shibata}, \&
  {Wanajo}}]{Fujibayashi2022}
{Fujibayashi}, S., {Sekiguchi}, Y., {Shibata}, M., \& {Wanajo}, S. 2022, arXiv
  e-prints, arXiv:2212.03958, \dodoi{10.48550/arXiv.2212.03958}

\bibitem[{{Gottlieb} {et~al.}(2022){Gottlieb}, {Tchekhovskoy}, \&
  {Margutti}}]{Gottlieb2022}
{Gottlieb}, O., {Tchekhovskoy}, A., \& {Margutti}, R. 2022, \mnras, 513, 3810,
  \dodoi{10.1093/mnras/stac910}

\bibitem[{{Greene} {et~al.}(2020){Greene}, {Strader}, \&
  {Ho}}]{Greene2019IMBHReview}
{Greene}, J.~E., {Strader}, J., \& {Ho}, L.~C. 2020, \araa, 58, 257,
  \dodoi{10.1146/annurev-astro-032620-021835}

\bibitem[{{Hatchett} {et~al.}(1981){Hatchett}, {Begelman}, \&
  {Sarazin}}]{Hatchett1981}
{Hatchett}, S.~P., {Begelman}, M.~C., \& {Sarazin}, C.~L. 1981, \apj, 247, 677,
  \dodoi{10.1086/159079}

\bibitem[{{Hayasaki} {et~al.}(2016){Hayasaki}, {Stone}, \&
  {Loeb}}]{Hayasaki2016}
{Hayasaki}, K., {Stone}, N., \& {Loeb}, A. 2016, \mnras, 461, 3760,
  \dodoi{10.1093/mnras/stw1387}

\bibitem[{{Ho} {et~al.}(2019){Ho}, {Phinney}, {Ravi}, {Kulkarni}, {Petitpas},
  {Emonts}, {Bhalerao}, {Blundell}, {Cenko}, {Dobie}, {Howie}, {Kamraj},
  {Kasliwal}, {Murphy}, {Perley}, {Sridharan}, \& {Yoon}}]{Ho2019}
{Ho}, A. Y.~Q., {Phinney}, E.~S., {Ravi}, V., {et~al.} 2019, \apj, 871, 73,
  \dodoi{10.3847/1538-4357/aaf473}

\bibitem[{{Ho} {et~al.}(2020){Ho}, {Perley}, {Kulkarni}, {Dong}, {De},
  {Chandra}, {Andreoni}, {Bellm}, {Burdge}, {Coughlin}, {Dekany}, {Feeney},
  {Frederiks}, {Fremling}, {Golkhou}, {Graham}, {Hale}, {Helou}, {Horesh},
  {Kasliwal}, {Laher}, {Masci}, {Miller}, {Porter}, {Ridnaia}, {Rusholme},
  {Shupe}, {Soumagnac}, \& {Svinkin}}]{Ho2020}
{Ho}, A. Y.~Q., {Perley}, D.~A., {Kulkarni}, S.~R., {et~al.} 2020, \apj, 895,
  49, \dodoi{10.3847/1538-4357/ab8bcf}

\bibitem[{{Ho} {et~al.}(2022){Ho}, {Margalit}, {Bremer}, {Perley}, {Yao},
  {Dobie}, {Kaplan}, {O'Brien}, {Petitpas}, \& {Zic}}]{Ho2022Camel}
{Ho}, A. Y.~Q., {Margalit}, B., {Bremer}, M., {et~al.} 2022, \apj, 932, 116,
  \dodoi{10.3847/1538-4357/ac4e97}

\bibitem[{{Ho} {et~al.}(2023){Ho}, {Perley}, {Gal-Yam}, {Lunnan}, {Sollerman},
  {Schulze}, {Das}, {Dobie}, {Yao}, {Fremling}, {Adams}, {Anand}, {Andreoni},
  {Bellm}, {Bruch}, {Burdge}, {Castro-Tirado}, {Dahiwale}, {De}, {Dekany},
  {Drake}, {Duev}, {Graham}, {Helou}, {Kaplan}, {Karambelkar}, {Kasliwal},
  {Kool}, {Kulkarni}, {Mahabal}, {Medford}, {Miller}, {Nordin}, {Ofek},
  {Petitpas}, {Riddle}, {Sharma}, {Smith}, {Stewart}, {Taggart}, {Tartaglia},
  {Tzanidakis}, \& {Winters}}]{Ho2021}
{Ho}, A. Y.~Q., {Perley}, D.~A., {Gal-Yam}, A., {et~al.} 2023, \apj, 949, 120,
  \dodoi{10.3847/1538-4357/acc533}

\bibitem[{{Hotokezaka} {et~al.}(2017){Hotokezaka}, {Kashiyama}, \&
  {Murase}}]{2017ApJ...850...18H}
{Hotokezaka}, K., {Kashiyama}, K., \& {Murase}, K. 2017, \apj, 850, 18,
  \dodoi{10.3847/1538-4357/aa8c7d}

\bibitem[{{Huang} {et~al.}(2019){Huang}, {Shimoda}, {Urata}, {Toma}, {Yamaoka},
  {Asada}, {Nagai}, {Takahashi}, {Petitpas}, \& {Tashiro}}]{Huang2019}
{Huang}, K., {Shimoda}, J., {Urata}, Y., {et~al.} 2019, \apjl, 878, L25,
  \dodoi{10.3847/2041-8213/ab23fd}

\bibitem[{{Jencson} {et~al.}(2016){Jencson}, {Prieto}, {Kochanek}, {Shappee},
  {Stanek}, \& {Pogge}}]{Jencson2016}
{Jencson}, J.~E., {Prieto}, J.~L., {Kochanek}, C.~S., {et~al.} 2016, \mnras,
  456, 2622, \dodoi{10.1093/mnras/stv2795}

\bibitem[{{Karachentsev} {et~al.}(2003){Karachentsev}, {Sharina}, {Dolphin},
  {Grebel}, {Geisler}, {Guhathakurta}, {Hodge}, {Karachentseva}, {Sarajedini},
  \& {Seitzer}}]{Karachentsev2003}
{Karachentsev}, I.~D., {Sharina}, M.~E., {Dolphin}, A.~E., {et~al.} 2003, \aap,
  398, 467, \dodoi{10.1051/0004-6361:20021598}

\bibitem[{{Karamehmetoglu} {et~al.}(2021){Karamehmetoglu}, {Fransson},
  {Sollerman}, {Tartaglia}, {Taddia}, {De}, {Fremling}, {Bagdasaryan},
  {Barbarino}, {Bellm}, {Dekany}, {Dugas}, {Giomi}, {Goobar}, {Graham}, {Ho},
  {Laher}, {Masci}, {Neill}, {Perley}, {Riddle}, {Rusholme}, \&
  {Soumagnac}}]{Karamehmetoglu2021}
{Karamehmetoglu}, E., {Fransson}, C., {Sollerman}, J., {et~al.} 2021, \aap,
  649, A163, \dodoi{10.1051/0004-6361/201936308}

\bibitem[{{Kasen}(2017)}]{Kasen2017}
{Kasen}, D. 2017, in Handbook of Supernovae, ed. A.~W. {Alsabti} \&
  P.~{Murdin}, 939, \dodoi{10.1007/978-3-319-21846-5_32}

\bibitem[{{Kashiyama} \& {Quataert}(2015{\natexlab{a}})}]{2015MNRAS.451.2656K}
{Kashiyama}, K., \& {Quataert}, E. 2015{\natexlab{a}}, \mnras, 451, 2656,
  \dodoi{10.1093/mnras/stv1164}

\bibitem[{{Kashiyama} \& {Quataert}(2015{\natexlab{b}})}]{Kashiyama2015}
---. 2015{\natexlab{b}}, \mnras, 451, 2656, \dodoi{10.1093/mnras/stv1164}

\bibitem[{{Kawana} {et~al.}(2020){Kawana}, {Maeda}, {Yoshida}, \&
  {Tanikawa}}]{2020ApJ...890L..26K}
{Kawana}, K., {Maeda}, K., {Yoshida}, N., \& {Tanikawa}, A. 2020, \apjl, 890,
  L26, \dodoi{10.3847/2041-8213/ab7209}

\bibitem[{{Khatami} \& {Kasen}(2023)}]{Khatami2023}
{Khatami}, D., \& {Kasen}, D. 2023, arXiv e-prints, arXiv:2304.03360,
  \dodoi{10.48550/arXiv.2304.03360}

\bibitem[{{Kleiser} {et~al.}(2018){Kleiser}, {Kasen}, \&
  {Duffell}}]{Kleiser2018CSM}
{Kleiser}, I. K.~W., {Kasen}, D., \& {Duffell}, P.~C. 2018, \mnras, 475, 3152,
  \dodoi{10.1093/mnras/stx3321}

\bibitem[{{Kremer} {et~al.}(2021){Kremer}, {Lu}, {Piro}, {Chatterjee}, {Rasio},
  \& {Ye}}]{Kremer2021}
{Kremer}, K., {Lu}, W., {Piro}, A.~L., {et~al.} 2021, \apj, 911, 104,
  \dodoi{10.3847/1538-4357/abeb14}

\bibitem[{{Kubota} {et~al.}(1998){Kubota}, {Tanaka}, {Makishima}, {Ueda},
  {Dotani}, {Inoue}, \& {Yamaoka}}]{Kubota1998}
{Kubota}, A., {Tanaka}, Y., {Makishima}, K., {et~al.} 1998, \pasj, 50, 667,
  \dodoi{10.1093/pasj/50.6.667}

\bibitem[{{Kuin} {et~al.}(2019){Kuin}, {Wu}, {Oates}, {Lien}, {Emery},
  {Kennea}, {de Pasquale}, {Han}, {Brown}, {Tohuvavohu}, {Breeveld}, {Burrows},
  {Cenko}, {Campana}, {Levan}, {Markwardt}, {Osborne}, {Page}, {Page},
  {Sbarufatti}, {Siegel}, \& {Troja}}]{Kuin2019}
{Kuin}, N. P.~M., {Wu}, K., {Oates}, S., {et~al.} 2019, \mnras, 487, 2505,
  \dodoi{10.1093/mnras/stz053}

\bibitem[{{Laor} \& {Draine}(1993)}]{1993ApJ...402..441L}
{Laor}, A., \& {Draine}, B.~T. 1993, \apj, 402, 441, \dodoi{10.1086/172149}

\bibitem[{{Larsen} \& {Richtler}(1999)}]{Larsen1999}
{Larsen}, S.~S., \& {Richtler}, T. 1999, \aap, 345, 59.
\newblock \doarXiv{astro-ph/9902227}

\bibitem[{{Leung} {et~al.}(2020){Leung}, {Blinnikov}, {Nomoto}, {Baklanov},
  {Sorokina}, \& {Tolstov}}]{Leung2020}
{Leung}, S.-C., {Blinnikov}, S., {Nomoto}, K., {et~al.} 2020, \apj, 903, 66,
  \dodoi{10.3847/1538-4357/abba33}

\bibitem[{{Li}(2022)}]{Li2022}
{Li}, X. 2022, in American Astronomical Society Meeting Abstracts, Vol.~54,
  American Astronomical Society Meeting Abstracts, 114.09

\bibitem[{{Liska} {et~al.}(2018){Liska}, {Hesp}, {Tchekhovskoy}, {Ingram}, {van
  der Klis}, \& {Markoff}}]{Liska2018}
{Liska}, M., {Hesp}, C., {Tchekhovskoy}, A., {et~al.} 2018, \mnras, 474, L81,
  \dodoi{10.1093/mnrasl/slx174}

\bibitem[{{Liu} {et~al.}(2023){Liu}, {Liu}, {Yu}, \& {Zhu}}]{Liu2023}
{Liu}, J.-F., {Liu}, L.-D., {Yu}, Y.-W., \& {Zhu}, J.-P. 2023, \apj, 946, 35,
  \dodoi{10.3847/1538-4357/acbb04}

\bibitem[{{Liu} {et~al.}(2022){Liu}, {Zhu}, {Liu}, {Yu}, \& {Zhang}}]{Liu2022}
{Liu}, J.-F., {Zhu}, J.-P., {Liu}, L.-D., {Yu}, Y.-W., \& {Zhang}, B. 2022,
  \apjl, 935, L34, \dodoi{10.3847/2041-8213/ac86d2}

\bibitem[{{Lyman} {et~al.}(2020){Lyman}, {Galbany}, {S{\'a}nchez}, {Anderson},
  {Kuncarayakti}, \& {Prieto}}]{Lyman2020}
{Lyman}, J.~D., {Galbany}, L., {S{\'a}nchez}, S.~F., {et~al.} 2020, \mnras,
  495, 992, \dodoi{10.1093/mnras/staa1243}

\bibitem[{{Lyutikov}(2022)}]{Lyutikov2022}
{Lyutikov}, M. 2022, \mnras, 515, 2293, \dodoi{10.1093/mnras/stac1717}

\bibitem[{{Lyutikov} \& {Toonen}(2019)}]{Lyutikov2019}
{Lyutikov}, M., \& {Toonen}, S. 2019, \mnras, 487, 5618,
  \dodoi{10.1093/mnras/stz1640}

\bibitem[{{Maeda} \& {Moriya}(2022)}]{Maeda2022}
{Maeda}, K., \& {Moriya}, T.~J. 2022, \apj, 927, 25,
  \dodoi{10.3847/1538-4357/ac4672}

\bibitem[{{Makishima} {et~al.}(1986){Makishima}, {Maejima}, {Mitsuda}, {Bradt},
  {Remillard}, {Tuohy}, {Hoshi}, \& {Nakagawa}}]{Makishima1986}
{Makishima}, K., {Maejima}, Y., {Mitsuda}, K., {et~al.} 1986, \apj, 308, 635,
  \dodoi{10.1086/164534}

\bibitem[{{Marchesi} {et~al.}(2018){Marchesi}, {Ajello}, {Marcotulli},
  {Comastri}, {Lanzuisi}, \& {Vignali}}]{Marchesi2018}
{Marchesi}, S., {Ajello}, M., {Marcotulli}, L., {et~al.} 2018, \apj, 854, 49,
  \dodoi{10.3847/1538-4357/aaa410}

\bibitem[{{Margalit}(2022)}]{Margalit2022b}
{Margalit}, B. 2022, \apj, 933, 238, \dodoi{10.3847/1538-4357/ac771a}

\bibitem[{{Margalit} {et~al.}(2022){Margalit}, {Quataert}, \&
  {Ho}}]{Margalit2022a}
{Margalit}, B., {Quataert}, E., \& {Ho}, A. Y.~Q. 2022, \apj, 928, 122,
  \dodoi{10.3847/1538-4357/ac53b0}

\bibitem[{{Margutti} {et~al.}(2017){Margutti}, {Kamble}, {Milisavljevic},
  {Zapartas}, {de Mink}, {Drout}, {Chornock}, {Risaliti}, {Zauderer},
  {Bietenholz}, {Cantiello}, {Chakraborti}, {Chomiuk}, {Fong}, {Grefenstette},
  {Guidorzi}, {Kirshner}, {Parrent}, {Patnaude}, {Soderberg}, {Gehrels}, \&
  {Harrison}}]{Margutti2017}
{Margutti}, R., {Kamble}, A., {Milisavljevic}, D., {et~al.} 2017, \apj, 835,
  140, \dodoi{10.3847/1538-4357/835/2/140}

\bibitem[{{Margutti} {et~al.}(2019){Margutti}, {Metzger}, {Chornock}, {Vurm},
  {Roth}, {Grefenstette}, {Savchenko}, {Cartier}, {Steiner}, {Terreran},
  {Margalit}, {Migliori}, {Milisavljevic}, {Alexander}, {Bietenholz},
  {Blanchard}, {Bozzo}, {Brethauer}, {Chilingarian}, {Coppejans}, {Ducci},
  {Ferrigno}, {Fong}, {G{\"o}tz}, {Guidorzi}, {Hajela}, {Hurley}, {Kuulkers},
  {Laurent}, {Mereghetti}, {Nicholl}, {Patnaude}, {Ubertini}, {Banovetz},
  {Bartel}, {Berger}, {Coughlin}, {Eftekhari}, {Frederiks}, {Kozlova},
  {Laskar}, {Svinkin}, {Drout}, {MacFadyen}, \& {Paterson}}]{Margutti2019}
{Margutti}, R., {Metzger}, B.~D., {Chornock}, R., {et~al.} 2019, \apj, 872, 18,
  \dodoi{10.3847/1538-4357/aafa01}

\bibitem[{{Matt} \& {Guainazzi}(2003)}]{Matt2003}
{Matt}, G., \& {Guainazzi}, M. 2003, \mnras, 341, L13,
  \dodoi{10.1046/j.1365-8711.2003.06658.x}

\bibitem[{{Mauerhan} {et~al.}(2018){Mauerhan}, {Filippenko}, {Zheng}, {Brink},
  {Graham}, {Shivvers}, \& {Clubb}}]{Mauerhan2018}
{Mauerhan}, J.~C., {Filippenko}, A.~V., {Zheng}, W., {et~al.} 2018, \mnras,
  478, 5050, \dodoi{10.1093/mnras/sty1307}

\bibitem[{{Maund} {et~al.}(2023){Maund}, {H{\"o}flich}, {Steele},
  {Yang(杨轶)}, {Wiersema}, {Kobayashi}, {Jordana-Mitjans}, {Mundell},
  {Gomboc}, {Guidorzi}, \& {Smith}}]{Maund2023}
{Maund}, J.~R., {H{\"o}flich}, P.~A., {Steele}, I.~A., {et~al.} 2023, \mnras,
  521, 3323, \dodoi{10.1093/mnras/stad539}

\bibitem[{{McDowell} {et~al.}(2018){McDowell}, {Duffell}, \&
  {Kasen}}]{McDowell2018}
{McDowell}, A.~T., {Duffell}, P.~C., \& {Kasen}, D. 2018, \apj, 856, 29,
  \dodoi{10.3847/1538-4357/aaa96e}

\bibitem[{{Metzger}(2022)}]{Metzger2022}
{Metzger}, B.~D. 2022, \apj, 932, 84, \dodoi{10.3847/1538-4357/ac6d59}

\bibitem[{{Metzger} \& {Perley}(2023)}]{Metzger2022dust}
{Metzger}, B.~D., \& {Perley}, D.~A. 2023, \apj, 944, 74,
  \dodoi{10.3847/1538-4357/acae89}

\bibitem[{{Metzger} {et~al.}(2008){Metzger}, {Piro}, \&
  {Quataert}}]{Metzger2008}
{Metzger}, B.~D., {Piro}, A.~L., \& {Quataert}, E. 2008, \mnras, 390, 781,
  \dodoi{10.1111/j.1365-2966.2008.13789.x}

\bibitem[{{Metzger} {et~al.}(2014){Metzger}, {Vurm}, {Hasco{\"e}t}, \&
  {Beloborodov}}]{Metzger2014PWN}
{Metzger}, B.~D., {Vurm}, I., {Hasco{\"e}t}, R., \& {Beloborodov}, A.~M. 2014,
  \mnras, 437, 703, \dodoi{10.1093/mnras/stt1922}

\bibitem[{{Milisavljevic} {et~al.}(2015){Milisavljevic}, {Margutti}, {Kamble},
  {Patnaude}, {Raymond}, {Eldridge}, {Fong}, {Bietenholz}, {Challis},
  {Chornock}, {Drout}, {Fransson}, {Fesen}, {Grindlay}, {Kirshner}, {Lunnan},
  {Mackey}, {Miller}, {Parrent}, {Sanders}, {Soderberg}, \&
  {Zauderer}}]{Milisavljevic2015}
{Milisavljevic}, D., {Margutti}, R., {Kamble}, A., {et~al.} 2015, \apj, 815,
  120, \dodoi{10.1088/0004-637X/815/2/120}

\bibitem[{{Mitsuda} {et~al.}(1984){Mitsuda}, {Inoue}, {Koyama}, {Makishima},
  {Matsuoka}, {Ogawara}, {Shibazaki}, {Suzuki}, {Tanaka}, \&
  {Hirano}}]{Mitsuda1984}
{Mitsuda}, K., {Inoue}, H., {Koyama}, K., {et~al.} 1984, \pasj, 36, 741

\bibitem[{{Mohan} {et~al.}(2020){Mohan}, {An}, \& {Yang}}]{Mohan2020}
{Mohan}, P., {An}, T., \& {Yang}, J. 2020, \apjl, 888, L24,
  \dodoi{10.3847/2041-8213/ab64d1}

\bibitem[{{Mor} {et~al.}(2023){Mor}, {Livne}, \& {Piran}}]{Mor2023}
{Mor}, R., {Livne}, E., \& {Piran}, T. 2023, \mnras, 518, 623,
  \dodoi{10.1093/mnras/stac2775}

\bibitem[{{Nayana} \& {Chandra}(2021)}]{Nayana2021}
{Nayana}, A.~J., \& {Chandra}, P. 2021, \apjl, 912, L9,
  \dodoi{10.3847/2041-8213/abed55}

\bibitem[{{Nelson} \& {Papaloizou}(2000)}]{Nelson2000}
{Nelson}, R.~P., \& {Papaloizou}, J. C.~B. 2000, \mnras, 315, 570,
  \dodoi{10.1046/j.1365-8711.2000.03478.x}

\bibitem[{{Nymark} {et~al.}(2006){Nymark}, {Fransson}, \& {Kozma}}]{Nymark2006}
{Nymark}, T.~K., {Fransson}, C., \& {Kozma}, C. 2006, \aap, 449, 171,
  \dodoi{10.1051/0004-6361:20054169}

\bibitem[{{Ofek} {et~al.}(2010){Ofek}, {Rabinak}, {Neill}, {Arcavi}, {Cenko},
  {Waxman}, {Kulkarni}, {Gal-Yam}, {Nugent}, {Bildsten}, {Bloom}, {Filippenko},
  {Forster}, {Howell}, {Jacobsen}, {Kasliwal}, {Law}, {Martin}, {Poznanski},
  {Quimby}, {Shen}, {Sullivan}, {Dekany}, {Rahmer}, {Hale}, {Smith},
  {Zolkower}, {Velur}, {Walters}, {Henning}, {Bui}, \&
  {McKenna}}]{2010ApJ...724.1396O}
{Ofek}, E.~O., {Rabinak}, I., {Neill}, J.~D., {et~al.} 2010, \apj, 724, 1396,
  \dodoi{10.1088/0004-637X/724/2/1396}

\bibitem[{{Paczy{\'n}sky} \& {Wiita}(1980)}]{Paczynski1980}
{Paczy{\'n}sky}, B., \& {Wiita}, P.~J. 1980, \aap, 88, 23

\bibitem[{{Pasham} {et~al.}(2021){Pasham}, {Ho}, {Alston}, {Remillard}, {Ng},
  {Gendreau}, {Metzger}, {Altamirano}, {Chakrabarty}, {Fabian}, {Miller},
  {Bult}, {Arzoumanian}, {Steiner}, {Strohmayer}, {Tombesi}, {Homan},
  {Cackett}, \& {Harding}}]{Pasham2021}
{Pasham}, D.~R., {Ho}, W. C.~G., {Alston}, W., {et~al.} 2021, Nature Astronomy,
  6, 249, \dodoi{10.1038/s41550-021-01524-8}

\bibitem[{{Pellegrino} {et~al.}(2022){Pellegrino}, {Howell}, {Vink{\'o}},
  {Gangopadhyay}, {Xiang}, {Arcavi}, {Brown}, {Burke}, {Hiramatsu},
  {Hosseinzadeh}, {Li}, {McCully}, {Misra}, {Newsome}, {Gonzalez}, {Pritchard},
  {Valenti}, {Wang}, \& {Zhang}}]{Pellegrino2022}
{Pellegrino}, C., {Howell}, D.~A., {Vink{\'o}}, J., {et~al.} 2022, \apj, 926,
  125, \dodoi{10.3847/1538-4357/ac3e63}

\bibitem[{{Perley} {et~al.}(2019){Perley}, {Mazzali}, {Yan}, {Cenko}, {Gezari},
  {Taggart}, {Blagorodnova}, {Fremling}, {Mockler}, {Singh}, {Tominaga},
  {Tanaka}, {Watson}, {Ahumada}, {Anupama}, {Ashall}, {Becerra}, {Bersier},
  {Bhalerao}, {Bloom}, {Butler}, {Copperwheat}, {Coughlin}, {De}, {Drake},
  {Duev}, {Frederick}, {Gonz{\'a}lez}, {Goobar}, {Heida}, {Ho}, {Horst},
  {Hung}, {Itoh}, {Jencson}, {Kasliwal}, {Kawai}, {Khanam}, {Kulkarni},
  {Kumar}, {Kumar}, {Kutyrev}, {Lee}, {Maeda}, {Mahabal}, {Murata}, {Neill},
  {Ngeow}, {Penprase}, {Pian}, {Quimby}, {Ramirez-Ruiz}, {Richer},
  {Rom{\'a}n-Z{\'u}{\~n}iga}, {Sahu}, {Srivastav}, {Socia}, {Sollerman},
  {Tachibana}, {Taddia}, {Tinyanont}, {Troja}, {Ward}, {Wee}, \&
  {Yu}}]{Perley2019}
{Perley}, D.~A., {Mazzali}, P.~A., {Yan}, L., {et~al.} 2019, \mnras, 484, 1031,
  \dodoi{10.1093/mnras/sty3420}

\bibitem[{{Perley} {et~al.}(2021){Perley}, {Ho}, {Yao}, {Fremling}, {Anderson},
  {Schulze}, {Kumar}, {Anupama}, {Barway}, {Bellm}, {Bhalerao}, {Chen}, {Duev},
  {Galbany}, {Graham}, {Gromadzki}, {Guti{\'e}rrez}, {Ihanec}, {Inserra},
  {Kasliwal}, {Kool}, {Kulkarni}, {Laher}, {Masci}, {Neill}, {Nicholl},
  {Pursiainen}, {van Roestel}, {Sharma}, {Sollerman}, {Walters}, \&
  {Wiseman}}]{Perley2021Camel}
{Perley}, D.~A., {Ho}, A. Y.~Q., {Yao}, Y., {et~al.} 2021, \mnras, 508, 5138,
  \dodoi{10.1093/mnras/stab2785}

\bibitem[{{Piro} \& {Lu}(2020)}]{Piro2020}
{Piro}, A.~L., \& {Lu}, W. 2020, \apj, 894, 2, \dodoi{10.3847/1538-4357/ab83f6}

\bibitem[{{Prentice} {et~al.}(2018){Prentice}, {Maguire}, {Smartt}, {Magee},
  {Schady}, {Sim}, {Chen}, {Clark}, {Colin}, {Fulton}, {McBrien}, {O'Neill},
  {Smith}, {Ashall}, {Chambers}, {Denneau}, {Flewelling}, {Heinze}, {Holoien},
  {Huber}, {Kochanek}, {Mazzali}, {Prieto}, {Rest}, {Shappee}, {Stalder},
  {Stanek}, {Stritzinger}, {Thompson}, \& {Tonry}}]{2018ApJ...865L...3P}
{Prentice}, S.~J., {Maguire}, K., {Smartt}, S.~J., {et~al.} 2018, \apjl, 865,
  L3, \dodoi{10.3847/2041-8213/aadd90}

\bibitem[{{Pringle}(1981)}]{Pringle1981}
{Pringle}, J.~E. 1981, \araa, 19, 137,
  \dodoi{10.1146/annurev.aa.19.090181.001033}

\bibitem[{{Pursiainen} {et~al.}(2018){Pursiainen}, {Childress}, {Smith},
  {Prajs}, {Sullivan}, {Davis}, {Foley}, {Asorey}, {Calcino}, {Carollo},
  {Curtin}, {D'Andrea}, {Glazebrook}, {Gutierrez}, {Hinton}, {Hoormann},
  {Inserra}, {Kessler}, {King}, {Kuehn}, {Lewis}, {Lidman}, {Macaulay},
  {M{\"o}ller}, {Nichol}, {Sako}, {Sommer}, {Swann}, {Tucker}, {Uddin},
  {Wiseman}, {Zhang}, {Abbott}, {Abdalla}, {Allam}, {Annis}, {Avila}, {Brooks},
  {Buckley-Geer}, {Burke}, {Carnero Rosell}, {Carrasco Kind}, {Carretero},
  {Castander}, {Cunha}, {Davis}, {De Vicente}, {Diehl}, {Doel}, {Eifler},
  {Flaugher}, {Fosalba}, {Frieman}, {Garc{\'\i}a-Bellido}, {Gruen}, {Gruendl},
  {Gutierrez}, {Hartley}, {Hollowood}, {Honscheid}, {James}, {Jeltema},
  {Kuropatkin}, {Li}, {Lima}, {Maia}, {Martini}, {Menanteau}, {Ogando},
  {Plazas}, {Roodman}, {Sanchez}, {Scarpine}, {Schindler}, {Smith},
  {Soares-Santos}, {Sobreira}, {Suchyta}, {Swanson}, {Tarle}, {Tucker},
  {Walker}, \& {DES Collaboration}}]{2018MNRAS.481..894P}
{Pursiainen}, M., {Childress}, M., {Smith}, M., {et~al.} 2018, \mnras, 481,
  894, \dodoi{10.1093/mnras/sty2309}

\bibitem[{{Quataert} {et~al.}(2019){Quataert}, {Lecoanet}, \&
  {Coughlin}}]{Quataert2019}
{Quataert}, E., {Lecoanet}, D., \& {Coughlin}, E.~R. 2019, \mnras, 485, L83,
  \dodoi{10.1093/mnrasl/slz031}

\bibitem[{{Rest} {et~al.}(2018){Rest}, {Garnavich}, {Khatami}, {Kasen},
  {Tucker}, {Shaya}, {Olling}, {Mushotzky}, {Zenteno}, {Margheim},
  {Strampelli}, {James}, {Smith}, {F{\"o}rster}, \&
  {Villar}}]{2018NatAs...2..307R}
{Rest}, A., {Garnavich}, P.~M., {Khatami}, D., {et~al.} 2018, Nature Astronomy,
  2, 307, \dodoi{10.1038/s41550-018-0423-2}

\bibitem[{{Ricci} {et~al.}(2015){Ricci}, {Ueda}, {Koss}, {Trakhtenbrot},
  {Bauer}, \& {Gandhi}}]{Ricci2015}
{Ricci}, C., {Ueda}, Y., {Koss}, M.~J., {et~al.} 2015, \apjl, 815, L13,
  \dodoi{10.1088/2041-8205/815/1/L13}

\bibitem[{{Rivera Sandoval} {et~al.}(2018){Rivera Sandoval}, {Maccarone},
  {Corsi}, {Brown}, {Pooley}, \& {Wheeler}}]{RiveraSandoval2018}
{Rivera Sandoval}, L.~E., {Maccarone}, T.~J., {Corsi}, A., {et~al.} 2018,
  \mnras, 480, L146, \dodoi{10.1093/mnrasl/sly145}

\bibitem[{{Roychowdhury} {et~al.}(2019){Roychowdhury}, {Arabsalmani}, \&
  {Kanekar}}]{Roychowdhury2019}
{Roychowdhury}, S., {Arabsalmani}, M., \& {Kanekar}, N. 2019, \mnras, 485, L93,
  \dodoi{10.1093/mnrasl/slz035}

\bibitem[{{Sabhahit} {et~al.}(2022){Sabhahit}, {Vink}, {Higgins}, \&
  {Sander}}]{Sabhahit2022}
{Sabhahit}, G.~N., {Vink}, J.~S., {Higgins}, E.~R., \& {Sander}, A. A.~C. 2022,
  \mnras, 514, 3736, \dodoi{10.1093/mnras/stac1410}

\bibitem[{{Scheuer} \& {Feiler}(1996)}]{Scheuer1996}
{Scheuer}, P.~A.~G., \& {Feiler}, R. 1996, \mnras, 282, 291,
  \dodoi{10.1093/mnras/282.1.291}

\bibitem[{{Schlafly} \& {Finkbeiner}(2011)}]{2011ApJ...737..103S}
{Schlafly}, E.~F., \& {Finkbeiner}, D.~P. 2011, \apj, 737, 103,
  \dodoi{10.1088/0004-637X/737/2/103}

\bibitem[{{Shakura} \& {Sunyaev}(1973)}]{ShakuraSunyaev1973}
{Shakura}, N.~I., \& {Sunyaev}, R.~A. 1973, \aap, 24, 337

\bibitem[{{Shen} \& {Matzner}(2014)}]{Shen2014}
{Shen}, R.-F., \& {Matzner}, C.~D. 2014, \apj, 784, 87,
  \dodoi{10.1088/0004-637X/784/2/87}

\bibitem[{{Shivvers} {et~al.}(2016){Shivvers}, {Zheng}, {Mauerhan}, {Kleiser},
  {Van Dyk}, {Silverman}, {Graham}, {Kelly}, {Filippenko}, \&
  {Kumar}}]{2016MNRAS.461.3057S}
{Shivvers}, I., {Zheng}, W.~K., {Mauerhan}, J., {et~al.} 2016, \mnras, 461,
  3057, \dodoi{10.1093/mnras/stw1528}

\bibitem[{{Smartt} {et~al.}(2018){Smartt}, {Clark}, {Smith}, {McBrien},
  {Maguire}, {O'Neil}, {Fulton}, {Magee}, {Prentice}, {Colin}, {Tonry},
  {Denneau}, {Stalder}, {Heinze}, {Weiland}, {Flewelling}, \&
  {Rest}}]{2018ATel11727....1S}
{Smartt}, S.~J., {Clark}, P., {Smith}, K.~W., {et~al.} 2018, The Astronomer's
  Telegram, 11727, 1

\bibitem[{{Smith} {et~al.}(2009){Smith}, {Silverman}, {Chornock}, {Filippenko},
  {Wang}, {Li}, {Ganeshalingam}, {Foley}, {Rex}, \& {Steele}}]{Smith2009}
{Smith}, N., {Silverman}, J.~M., {Chornock}, R., {et~al.} 2009, \apj, 695,
  1334, \dodoi{10.1088/0004-637X/695/2/1334}

\bibitem[{{Soker}(2022)}]{Soker2022}
{Soker}, N. 2022, Research in Astronomy and Astrophysics, 22, 055010,
  \dodoi{10.1088/1674-4527/ac5b40}

\bibitem[{{Soker} {et~al.}(2019){Soker}, {Grichener}, \& {Gilkis}}]{Soker2019}
{Soker}, N., {Grichener}, A., \& {Gilkis}, A. 2019, \mnras, 484, 4972,
  \dodoi{10.1093/mnras/stz364}

\bibitem[{{Stanway} \& {Eldridge}(2018)}]{Stanway2018}
{Stanway}, E.~R., \& {Eldridge}, J.~J. 2018, \mnras, 479, 75,
  \dodoi{10.1093/mnras/sty1353}

\bibitem[{{Stevance} {et~al.}(2020){Stevance}, {Eldridge}, \&
  {Stanway}}]{Stevance2020hoki}
{Stevance}, H., {Eldridge}, J., \& {Stanway}, E. 2020, The Journal of Open
  Source Software, 5, 1987, \dodoi{10.21105/joss.01987}

\bibitem[{{Strubbe} \& {Quataert}(2009)}]{Strubbe2009TDE}
{Strubbe}, L.~E., \& {Quataert}, E. 2009, \mnras, 400, 2070,
  \dodoi{10.1111/j.1365-2966.2009.15599.x}

\bibitem[{{Sun} {et~al.}(2022){Sun}, {Maund}, {Crowther}, \& {Liu}}]{Sun2022}
{Sun}, N.-C., {Maund}, J.~R., {Crowther}, P.~A., \& {Liu}, L.-D. 2022, \mnras,
  512, L66, \dodoi{10.1093/mnrasl/slac023}

\bibitem[{{Sun} {et~al.}(2023){Sun}, {Maund}, {Shao}, \& {Janiak}}]{Sun2022new}
{Sun}, N.-C., {Maund}, J.~R., {Shao}, Y., \& {Janiak}, I.~A. 2023, \mnras, 519,
  3785, \dodoi{10.1093/mnras/stac3773}

\bibitem[{{Suzuki} {et~al.}(2020){Suzuki}, {Moriya}, \&
  {Takiwaki}}]{Suzuki2020}
{Suzuki}, A., {Moriya}, T.~J., \& {Takiwaki}, T. 2020, \apj, 899, 56,
  \dodoi{10.3847/1538-4357/aba0ba}

\bibitem[{{Tampo} {et~al.}(2020){Tampo}, {Tanaka}, {Maeda}, {Yasuda},
  {Tominaga}, {Jiang}, {Moriya}, {Morokuma}, {Suzuki}, {Takahashi}, {Kokubo},
  \& {Kawana}}]{2020ApJ...894...27T}
{Tampo}, Y., {Tanaka}, M., {Maeda}, K., {et~al.} 2020, \apj, 894, 27,
  \dodoi{10.3847/1538-4357/ab7ccc}

\bibitem[{{Tanaka} {et~al.}(2016){Tanaka}, {Tominaga}, {Morokuma}, {Yasuda},
  {Furusawa}, {Baklanov}, {Blinnikov}, {Moriya}, {Doi}, {Jiang}, {Kato},
  {Kikuchi}, {Kuncarayakti}, {Nagao}, {Nomoto}, \&
  {Taniguchi}}]{2016ApJ...819....5T}
{Tanaka}, M., {Tominaga}, N., {Morokuma}, T., {et~al.} 2016, \apj, 819, 5,
  \dodoi{10.3847/0004-637X/819/1/5}

\bibitem[{{Thomsen} {et~al.}(2022){Thomsen}, {Kwan}, {Dai}, {Wu}, {Roth}, \&
  {Ramirez-Ruiz}}]{Thomsen2022}
{Thomsen}, L.~L., {Kwan}, T.~M., {Dai}, L., {et~al.} 2022, \apjl, 937, L28,
  \dodoi{10.3847/2041-8213/ac911f}

\bibitem[{{Tolstov} {et~al.}(2019){Tolstov}, {Nomoto}, {Sorokina}, {Blinnikov},
  {Tominaga}, \& {Taniguchi}}]{Tolstov2019}
{Tolstov}, A., {Nomoto}, K., {Sorokina}, E., {et~al.} 2019, \apj, 881, 35,
  \dodoi{10.3847/1538-4357/ab2876}

\bibitem[{{Tsuna} {et~al.}(2021){Tsuna}, {Kashiyama}, \&
  {Shigeyama}}]{Tsuna2021}
{Tsuna}, D., {Kashiyama}, K., \& {Shigeyama}, T. 2021, \apjl, 922, L34,
  \dodoi{10.3847/2041-8213/ac3997}

\bibitem[{{Uno} \& {Maeda}(2020)}]{Uno2020}
{Uno}, K., \& {Maeda}, K. 2020, \apj, 897, 156,
  \dodoi{10.3847/1538-4357/ab9632}

\bibitem[{{van Velzen} {et~al.}(2019){van Velzen}, {Stone}, {Metzger},
  {Gezari}, {Brown}, \& {Fruchter}}]{VanVelzen2019}
{van Velzen}, S., {Stone}, N.~C., {Metzger}, B.~D., {et~al.} 2019, \apj, 878,
  82, \dodoi{10.3847/1538-4357/ab1844}

\bibitem[{{Virtanen} {et~al.}(2020){Virtanen}, {Gommers}, {Oliphant},
  {Haberland}, {Reddy}, {Cournapeau}, {Burovski}, {Peterson}, {Weckesser},
  {Bright}, {van der Walt}, {Brett}, {Wilson}, {Millman}, {Mayorov}, {Nelson},
  {Jones}, {Kern}, {Larson}, {Carey}, {Polat}, {Feng}, {Moore}, {VanderPlas},
  {Laxalde}, {Perktold}, {Cimrman}, {Henriksen}, {Quintero}, {Harris},
  {Archibald}, {Ribeiro}, {Pedregosa}, {van Mulbregt}, \& {SciPy 1. 0
  Contributors}}]{2020scipy}
{Virtanen}, P., {Gommers}, R., {Oliphant}, T.~E., {et~al.} 2020, Nature
  Methods, 17, 261, \dodoi{10.1038/s41592-019-0686-2}

\bibitem[{{Wang} {et~al.}(2019){Wang}, {Wang}, {Cano}, {Wang}, {Liu}, {Dai},
  {Deng}, {Yu}, {Li}, {Song}, {Qiu}, \& {Wei}}]{Wang2019}
{Wang}, L.~J., {Wang}, X.~F., {Cano}, Z., {et~al.} 2019, \mnras, 489, 1110,
  \dodoi{10.1093/mnras/stz2184}

\bibitem[{{Wang} \& {Gan}(2022)}]{Wang2022}
{Wang}, S.-Q., \& {Gan}, W.-P. 2022, \apj, 928, 114,
  \dodoi{10.3847/1538-4357/ac53aa}

\bibitem[{{Wang} \& {Li}(2020)}]{Wang2020}
{Wang}, S.-Q., \& {Li}, L. 2020, \apj, 900, 83,
  \dodoi{10.3847/1538-4357/aba6e9}

\bibitem[{{Whitesides} {et~al.}(2017){Whitesides}, {Lunnan}, {Kasliwal},
  {Perley}, {Corsi}, {Cenko}, {Blagorodnova}, {Cao}, {Cook}, {Doran},
  {Frederiks}, {Fremling}, {Hurley}, {Karamehmetoglu}, {Kulkarni}, {Leloudas},
  {Masci}, {Nugent}, {Ritter}, {Rubin}, {Savchenko}, {Sollerman}, {Svinkin},
  {Taddia}, {Vreeswijk}, \& {Wozniak}}]{2017ApJ...851..107W}
{Whitesides}, L., {Lunnan}, R., {Kasliwal}, M.~M., {et~al.} 2017, \apj, 851,
  107, \dodoi{10.3847/1538-4357/aa99de}

\bibitem[{{Wiseman} {et~al.}(2020){Wiseman}, {Pursiainen}, {Childress},
  {Swann}, {Smith}, {Galbany}, {Lidman}, {Davis}, {Guti{\'e}rrez},
  {M{\"o}ller}, {Thomas}, {Frohmaier}, {Foley}, {Hinton}, {Kelsey}, {Kessler},
  {Lewis}, {Sako}, {Scolnic}, {Sullivan}, {Vincenzi}, {Abbott}, {Aguena},
  {Allam}, {Annis}, {Bertin}, {Bhargava}, {Brooks}, {Burke}, {Carnero Rosell},
  {Carollo}, {Carrasco Kind}, {Carretero}, {Costanzi}, {da Costa}, {Diehl},
  {Doel}, {Everett}, {Fosalba}, {Frieman}, {Garc{\'\i}a-Bellido}, {Gaztanaga},
  {Glazebrook}, {Gruen}, {Gruendl}, {Gschwend}, {Gutierrez}, {Hollowood},
  {Honscheid}, {James}, {Kuehn}, {Kuropatkin}, {Lima}, {Maia}, {Marshall},
  {Martini}, {Menanteau}, {Miquel}, {Palmese}, {Paz-Chinch{\'o}n}, {Plazas},
  {Romer}, {Sanchez}, {Scarpine}, {Schubnell}, {Serrano}, {Sevilla-Noarbe},
  {Sommer}, {Suchyta}, {Swanson}, {Tarle}, {Tucker}, {Tucker}, {Varga},
  {Walker}, {Walker}, \& {(DES Collaboration)}}]{2020MNRAS.498.2575W}
{Wiseman}, P., {Pursiainen}, M., {Childress}, M., {et~al.} 2020, \mnras, 498,
  2575, \dodoi{10.1093/mnras/staa2474}

\bibitem[{{Xiang} {et~al.}(2021){Xiang}, {Wang}, {Lin}, {Mo}, {Lin}, {Burke},
  {Hiramatsu}, {Hosseinzadeh}, {Howell}, {McCully}, {Valenti}, {Vink{\'o}},
  {Wheeler}, {Ehgamberdiev}, {Mirzaqulov}, {B{\'o}di}, {Bogn{\'a}r}, {Cseh},
  {Hanyecz}, {Ign{\'a}cz}, {Kalup}, {K{\"o}nyves-T{\'o}th}, {Kriskovics},
  {Ordasi}, {P{\'a}l}, {S{\'a}rneczky}, {Seli}, {Szak{\'a}ts}, {Arranz-Heras},
  {Benavides-Palencia}, {Cejudo-Mart{\'\i}nez}, {De la Fuente-Fern{\'a}ndez},
  {Escart{\'\i}n-P{\'e}rez}, {Garc{\'\i}a-De la Cuesta},
  {Gonz{\'a}lez-Carballo}, {Gonz{\'a}lez-Farf{\'a}n},
  {Lim{\'o}n-Mart{\'\i}nez}, {Mantero}, {Naves-Nogu{\'e}s}, {Morales-Aimar},
  {Ru{\'\i}z-Ru{\'\i}z}, {Sold{\'a}n-Alfaro}, {Valero-P{\'e}rez},
  {Violat-Bordonau}, {Zhang}, {Zhang}, {Li}, {Chen}, {Sai}, \&
  {Li}}]{Xiang2021}
{Xiang}, D., {Wang}, X., {Lin}, W., {et~al.} 2021, \apj, 910, 42,
  \dodoi{10.3847/1538-4357/abdeba}

\bibitem[{{Yao} {et~al.}(2022){Yao}, {Ho}, {Medvedev}, {Nayana}, {Perley},
  {Kulkarni}, {Chandra}, {Sazonov}, {Gilfanov}, {Khorunzhev}, {Khatami}, \&
  {Sunyaev}}]{Yao2022}
{Yao}, Y., {Ho}, A. Y.~Q., {Medvedev}, P., {et~al.} 2022, \apj, 934, 104,
  \dodoi{10.3847/1538-4357/ac7a41}

\bibitem[{{Yu} {et~al.}(2019){Yu}, {Chen}, \& {Li}}]{Yu2019}
{Yu}, Y.-W., {Chen}, A., \& {Li}, X.-D. 2019, \apjl, 877, L21,
  \dodoi{10.3847/2041-8213/ab1f85}

\bibitem[{{Yu} {et~al.}(2015){Yu}, {Li}, \& {Dai}}]{Yu2015}
{Yu}, Y.-W., {Li}, S.-Z., \& {Dai}, Z.-G. 2015, \apjl, 806, L6,
  \dodoi{10.1088/2041-8205/806/1/L6}

\bibitem[{{Zhang} {et~al.}(2022){Zhang}, {Shu}, {Chen}, {Sun}, {Shen}, {Tao},
  {Chen}, {Jiang}, {Dou}, {Qin}, {Zhang}, {Zhang}, {Qu}, \& {Wang}}]{Zhang2022}
{Zhang}, W., {Shu}, X., {Chen}, J.-H., {et~al.} 2022, Research in Astronomy and
  Astrophysics, 22, 125016, \dodoi{10.1088/1674-4527/ac9c4b}

\end{thebibliography}
\bibliographystyle{aasjournal}

\end{document}